\documentclass[3p,times,twocolumn]{elsarticle}

\usepackage{ecrc-ac}
\usepackage{booktabs}
\usepackage{amssymb}
\usepackage{amsbsy,color}
\usepackage{subfigure}

\volume{00}

\firstpage{1}

\journalname{Nuclear Physics B Proceedings Supplement}

\runauth{Rainer Sommer and Ulli Wolff}

\jid{nuphbp}

\jnltitlelogo{Nuclear Physics B Proceedings Supplement}

\usepackage{amssymb}

\newcommand{\mathe}{\mathrm{e}}
\newcommand{\Nf}{N_\mathrm{f}}
\newcommand{\fPi}{f_{\pi}}
\newcommand{\fK}{f_\mathrm{K}}
\newcommand{\gbar}{\bar{g}}
\newcommand{\mbar}{\bar{m}}
\newcommand{\gmsbar}{\bar{g}_\mathrm{\overline{MS}}}
\newcommand{\Order}{\mathrm{O}}
\newcommand{\Lmax}{L_\mathrm{max}}
\newcommand{\csw}{c_\mathrm{SW}}
\newcommand{\fP}{f_\mathrm{P}}
\newcommand{\fA}{f_\mathrm{A}}
\newcommand{\cA}{c_\mathrm{A}}
\newcommand{\mhad}{m_\mathrm{had}}

\newcommand{\nf}{N_\mathrm{f}}
\newcommand{\textSF}{Schr\"odinger functional}
\newcommand{\fm}{\mathrm{fm}}
\newcommand{\GeV}{\mathrm{GeV}}
\newcommand{\MeV}{\mathrm{MeV}}

\newcommand{\fig}[1]{figure~\ref{#1}}
\newcommand{\Fig}[1]{Figure~\ref{#1}}
\newcommand{\bes}{\begin{eqnarray}}
\newcommand{\ees}{\end{eqnarray}}
\newcommand{\zp}{Z_\mathrm{P}}
\newcommand{\za}{Z_\mathrm{A}}

\newcommand{\tab}[1]{table~\ref{#1}}
\newcommand{\strange}{\mathrm{strange}}
\newcommand{\fklat}{F_\mathrm{K}}
\newcommand{\msbar}{{\overline{\mathrm{MS}}}}
\newcommand{\fklatphys}{F_{\mathrm{K,phys}}}
\newcommand{\fkphys}{f_{\mathrm{K,phys}}}
\newcommand{\mkphys}{m_{\rm K,phys}}
\newcommand{\mpiphys}{m_{\pi,\mathrm{phys}}}

\newcommand{\yk}{y_\mathrm{K}}
\newcommand{\ypi}{y_\pi }
\newcommand{\rmO}{\mathrm{O}}
\def\babar{\bar b_{\rm A}}
\def\bpbar{\bar b_{\rm P}}
\def\batil{\tilde b_{\rm A}}
\def\bptil{\tilde b_{\rm P}}

\begin{document}

\begin{frontmatter}



\dochead{\small{\flushright{SFB/CPP-14-100 \\ DESY 15-005 \\
                HU-15/02 \\ }}}

\title{Non-perturbative computation of the strong coupling constant on the
lattice}


\author[DE,HU]{Rainer Sommer}
\author[HU]{Ulli Wolff}

\address[DE]{John von Neumann Institute for Computing (NIC), DESY, Platanenallee~6, 15738 Zeuthen, Germany}

\address[HU]{Institut f{\"u}r Physik, Humboldt Universi{\"a}t, 
Newtonstr. 15, 12489 Berlin, Germany}

\begin{abstract}
We review the long term project of the ALPHA collaboration
to compute in QCD the running coupling constant and quark masses
at high energy scales in terms of low energy hadronic quantities.
The adapted techniques required to numerically carry out the
required multiscale non-perturbative calculation with our
special emphasis on the control of systematic errors are
summarized. The complete results in the two dynamical flavor
approximation are reviewed and an outlook is given on the ongoing
three flavor extension of the programme with improved 
target precision.
\end{abstract}


\end{frontmatter}


\section{Introduction}
\label{intro}
Quantum Chromo Dynamics (QCD) is the renormalizable quantum field theory containing gluon and quark fields
that interact in a unique way dictated by SU(3) gauge invariance. It may be seen as arising from the standard model
of elementary particles in a limit where all other fields, including their interactions with quarks and gluons, are stripped away.
Strong interactions and confinement are the characteristics of this sector which hence calls for non-perturbative
evaluations and is in the focus of lattice formulations and simulations.

We here consider QCD with a free number of $\Nf$ color triplets (flavors) of quark species. In Nature
we see the case $\Nf=6$ with the flavors up, down, strange, charm, bottom and top in order of ascending mass. 
The species beyond
light up and down quarks come with characteristic scales of the order of 0.1~GeV, 1~GeV, 4~GeV, 175~GeV.
Therefore it makes sense to consider effective theories with $\Nf < 6$ to describe physics with characteristic
energies significantly below the scales of the dropped degrees of freedom. They then enter only indirectly into
the determination of the free parameters of the effective theory. In lattice simulations the modelling of
the precise flavor content is technically very demanding. Therefore a lot of studies are found and will also be discussed
here that refer to $\Nf=2$ and $\Nf=3$ where the latter number is the minimum to allow for
real applications as an effective theory
\cite{Appelquist:1974tg,Weinberg:1980wa,Weinberg:1978kz,Bernreuther:1981sg,Bruno:2014ufa}. 
In any case generalized QCD has $\Nf+1$ free parameters given
by one quark mass per species and in addition the gauge coupling.

The two light species are in most studies,
including those described here, approximated to be degenerate. The value zero for some or even all $\Nf$
quark masses is theoretically nice as it enhances the chiral symmetry of the model and is thus stabilized
under renormalization. The renormalization of the coupling can be defined in this massless limit and we then 
speak of a massless renormalization scheme.
Such schemes are technically convenient in nontrivial perturbative as well as non-perturbative calculations. 
The renormalization
of the coupling can be left unchanged as quark masses are `turned on later'. To define a renormalized
coupling constant in a massless scheme an additional scale $\mu$ enters via the renormalization
conditions. The resulting scale dependent `running' coupling $\bar{g}(\mu)$ obeys a Callan-Symanzik equation
\begin{equation} \label{CSeq}
 \mu \frac{d}{d \mu} \bar{g}(\mu) = \beta(\bar{g}(\mu)),
\end{equation}
in which the function $\beta$ is determined by the theory once a particular coupling definition
has been adopted. A negative $\beta$-function corresponds to asymptotic freedom.
The free integration constant that arises in solving 
this differential equation can be taken as the free parameter that the bare
coupling has been `traded for' in the process of renormalization. It may be fixed by specifying
$\bar{g}$ for a specific $\mu$ value in GeV. Alternatively one may convert the Callan-Symanzik equation
into the equivalent integral statement that
\begin{eqnarray} \label{Lambda}
\Lambda &=& \mu \left( b_0 \bar{g}^2 (\mu)\right)^{- b_1 / 2 b_0^2} \exp[- 1 / ( 2 b_0 \bar{g}^2 (\mu))] \\
&\times&  \exp \left[ - \int_0^{\bar{g} (\mu)}
   \left\{ \frac{1}{\beta \left( x \right)} + \frac{1}{b_0 x^3} -
   \frac{b_1}{b_0^2 x} \right\} d x \right] \nonumber
\end{eqnarray}
is independent of $\mu$. In this equation $b_0, b_1$ are the leading and scheme independent
coefficients in
the asymptotic expansion
\begin{equation} \label{betaPT}
\beta(x)=-\sum_{n\ge0} b_n x^{2n+3},
\end{equation}
\begin{eqnarray}\label{b0}
b_0&=&\frac1{(4\pi)^2}\left(11-\frac23 \Nf \right), \\ \label{b1}
b_1&=&\frac1{(4\pi)^4}\left(102-\frac{38}3 \Nf \right).
\end{eqnarray}
For asymptotically large $\mu$ it is sufficient to
evaluate (\ref{Lambda}) with the perturbative series for $\beta$
truncated beyond some $n\ge1$. Therefore, in a {\em perturbative} context,
$\Lambda$ is associated with the behavior of $\bar{g}(\mu)$
for $\mu\to\infty$.

\section{Hadronic renormalization scheme and finite size scaling \label{sec_hadronic}}

In lattice simulations also non-perturbative quantities associated with scales
of order one GeV and below can be computed in principle. Examples are the masses
of light hadrons and matrix elements involving their one-particle states, decay
constants like $\fPi, \fK$ for example \cite{Aoki:2013ldr,Fodor:2012gf}. This opens up the possibility to also match
such quantities directly to experiment and in this way determine the free parameters of QCD
which can then be determined with an in principle arbitrary precision\footnote{
This refers to pure QCD. Other interactions are still neglected.
}.
This is not true if perturbation theory at any finite energy is involved, since with an asymptotic
expansion -- even if very high orders were available -- an uncertainty remains. This effect is
expected to be small at the Z-mass, but the situation is much more delicate 
for example for determinations
of $\alpha_s$ in the $\tau$-mass region. 

As a conceptually simple example of a hadronic scheme 
one could imagine to use as input parameters
the mass of the proton and in addition the masses
of $\Nf$ types of stable mesons that are sensitive to the respective quark masses.
In practice one of course has $\Nf+1$ dimensionless parameters at ones disposal
in the lattice theory of which $\Nf$ may be determined by dialing the correct ratios
of meson to proton mass. The remaining degree of freedom allows to tune
the lattice theory to its critical point where the continuum limit is reached.
Due to asymptotic freedom in QCD this is accomplished by sending the bare coupling to zero.
In this limit, all dimensionfull quantities emerge in the form of well-defined multiples of 
appropriate powers of the proton mass which we thus employ to set the scale for all observables.
Equivalently we may say that all that is computed from theories including the lattice and compared with experiment
are dimensionless ratios of observables. The above scheme selects a minimal set of independent
mass ratios and, with these tuned, all other ratios must `fall in place'. We try to be very explicit
on this seemingly trivial issue, as sometimes confusion seems to arise here nevertheless.

In the previous paragraphs we have described a rather idealized situation.
For various technical reasons we will not use this precise hadronic scheme, and
in addition several sources of in practice unavoidable systematic errors have to be
taken into account in lattice computations.

A lattice that is simulated on a computer necessarily has a finite number of sites
and thus finitely many degrees of freedom. This implies a finite extent $L$ and a finite spacing
or resolution $a$ such that one has $(L/a)^4$ sites. In large present day simulations
$L/a \sim 100$ is achieved. If we refer to $\mhad$ as some hadronic mass scale, then $a \mhad>0$
represents a distortion of the physics by an unphysical UV cutoff effect. Details depend on
the chosen lattice discretization, but in practice and employing Symanzik's theory
of cutoff effects 
\cite{Symanzik:1983dc,Symanzik:1983gh,Luscher:1984xn},
we
expect these effects to diminish asymptotically at a rate proportional to $(a \mhad)^2$.
We need to verify that we have reached this asymptotic behavior to estimate the prefactor
by multiple simulations in which the resolution (and nothing else) is varied.
This whole procedure is called continuum extrapolation and, of course, leaves behind
a contribution in the final error budget.

In the same way, unless finite size effects are deliberately looked for (see below),
also a finite product $L \mhad< \infty$ is an unwanted IR cutoff effect to be extrapolated
away or at least bounded.
The situation here is however more benign, as theory \cite{Luscher:1985dn} implies
that the infinite volume limit is reached at an exponential rate proportional to $\exp(-m_{\pi}L)$,
where the pion with mass $m_{\pi}$ enters as the lightest degree of freedom.
In todays large volume lattice simulations we are largely restricted to cutoffs $L\le 6$ fm
and $a^{-1}\le$ 5 GeV, although these extreme values cannot yet quite be realized 
simultaneously and compromises, depending on the physics studied, have to be made.

One would clearly like to confront results extracted by matching perturbation theory
to experiment at high enough energy with those of a hadronic scheme.
This amounts to nothing less than establishing QCD as {\em one} theory at all
these length scales.
An obvious strategy is to compute
the renormalized parameters of a perturbative scheme from `within'
a hadronic scheme. Focussing on the coupling constant this would require
to compute beside a hadron mass $\mhad$ some scale dependent
observable ${\cal O}(\mu)$
that possesses a perturbative expansion
\begin{equation}
{\cal O}(\mu)=\alpha(\mu)+p_1 \alpha^2(\mu)+p_2 \alpha^3(\mu)+\ldots
\end{equation}
where the coupling $\alpha$ refers to some perturbative scheme like $\overline{\mbox{MS}}$.
This evaluation has to proceed at a high enough scale $\mu/\mhad = \rho \gg 1$ to
get sufficient perturbative precision in extracting $\alpha$ at energy $\rho \times \mhad$.
With the help of (\ref{Lambda}) this information may be converted to a value for
$\Lambda/\mhad$. As is well known, the $\Lambda$ parameter of any other scheme follows now
by relating the corresponding couplings at one loop order.

If we imagine to perform such a calculation naively on the lattice, we have to cope with a
multiscale problem where we have to satisfy the string of inequalities
\begin{equation}
a \ll \mu^{-1} \ll \mhad^{-1} \ll L.
\end{equation}
Given the practical constraints on $L/a$ it is clear that such a direct approach
will require severe compromises. An overview over various approaches
including the direct one is found in \cite{Aoki:2013ldr}.

Due to these difficulties strategies have been devised to alleviate the problem
by circumventing one of the required large scale ratios. One idea is to try to
tolerate the scales $a^{-1}$ and $\mu$ to lie in the same range and thus perform
perturbation theory at the scale of the cutoff. Such a calculation, as well as references
to earlier versions, is discussed in \cite{Chakraborty:2014aca}. 
A valid criticism in our opinion is, that
it thus becomes hard to disentangle UV cutoff effects from limitations of (truncated)
perturbation theory. To control lattice artefacts it seems necessary to be able 
to vary the lattice spacing over some range with physical scales $\mhad,\mu$ held fixed.

Our finite size strategy which is in the focus of the remainder of this article may be seen as identifying
$\mu$ and $L^{-1}$ over a major part of the calculation. We will exploit the fact
that finite size effects are universal predictions of quantum field theory.
The Casimir force in QED \cite{Casimir:1948dh} is such a finite size effect which has been experimentally
confirmed as a subtle manifestation of vacuum fluctuations. We extend this concept
to more abstract cases involving periodic or Dirichlet type boundary conditions on finite systems.
They are not realized in the laboratory, but 
universality here means that there again
are unique predictions for effects depending on the size $L$, independent of how the field
theory is regularized and the UV cutoff limit in the finite box is taken.

We now sketch such a computation and come back to details in the later sections.
The key quantity required in our approach is an $L$-dependent finite size observable
$\gbar(L)$ that can function as a non-perturbatively defined coupling constant.
It must exist for arbitrary $L$ and possess a manageable perturbative expansion
that is applicable at small size, for $L^{-1}$ much larger than hadronic scales.
With a suitable normalization $\gbar(L)$ will be related
to other couplings like for example
\begin{equation}
\gbar^2(L)=\gmsbar^2(\mu)+c(\mu L) \gmsbar^4(\mu)+\Order(\gmsbar^6)
\end{equation}
where for good perturbative accuracy one will take $\mu L=\Order(1)$.
Beside perturbation theory $\gbar(L)$ must be easily computable numerically
and a reasonable signal to noise ratio for its estimator is another practical
requirement. Appropriate such couplings are known for
Schr{\"o}dinger functional boundary conditions, see the later sections for details.

With $\gbar(L)$ at hand we now first consider a simulation in a large, i.~e. effectively infinite, volume
where we tune the bare parameters to achieve $a \ll \mhad^{-1} \ll L$. A hadronic scale $\mhad$, or
rather the dimensionless combination $a\mhad$ is the output. Then, {\em for the same bare parameters}
we diminish $L/a$ to the point where $L\mhad\equiv(L/a)(a\mhad)=\rho$ becomes a fixed number of order unity 
(the previous value $a\mhad$ is used).
At this size $\gbar(L)$ will not be small but its value can be computed by simulation.
By repeating these steps for several small $a\mhad$ and extrapolating to the continuum
we thus derive a numerical value $\gbar(L)$ at the scale $L=\rho \mhad^{-1}$ in the
non-perturbative regime. This is a universal number $\gbar(\Lmax=\rho \mhad^{-1})$ in the continuum
where $\Lmax$ can be cited in fermi once $\mhad$ has been related to experiment.

\begin{figure}[ht!]
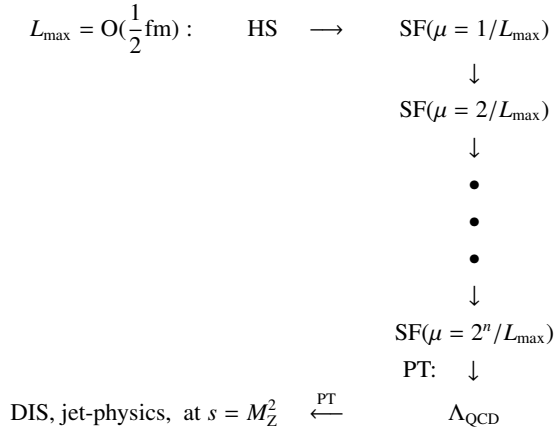

{\small
\bes
 \hspace*{-3mm}{ L_{\rm max}}=\mathrm{O}(\frac{1}{2}\fm): \qquad 
 {\rm HS} \quad \longrightarrow \quad
      &{\rm SF} (\mu=1/{ L_{\rm max}})& \quad 
               \nonumber \\
      &\downarrow&  \nonumber \\
      &{\rm SF} (\mu=2/{ L_{\rm max}})&  \nonumber \\ 
      &\downarrow&  \nonumber \\
      &\bullet&  \nonumber \\
      &\bullet&  \nonumber \\
      &\bullet&  \nonumber \\
      &\downarrow&  \nonumber \\
      &{\rm SF} (\mu=2^n/{ L_{\rm max}})& \nonumber \\
   &\mbox{ \small PT:} \quad  \downarrow \qquad \quad &  \nonumber \\
\hspace*{-3mm}\hbox{DIS, jet-physics, }\mbox{ at } s=M_\mathrm{Z}^2 \quad \stackrel{\rm  PT}{\longleftarrow} 
     \quad   &\Lambda_{\rm QCD} & \nonumber
\ees
}
\vspace{-0.4cm}
\caption{The strategy for a non-perturbative computation of 
         short distance parameters.
         SF refers to the \textSF\ renormalization scheme and HS to 
         a hadronic scheme.
\label{f:strategy}}
\end{figure}

What remains to be done now is to evolve $\gbar(L)$ from the now known
starting point at $\Lmax$ to other $L$ small enough to make contact with couplings
like $\gmsbar$ by perturbation theory. The Callan Symanzik equation
is a differential description of such an evolution guided by the corresponding
$\beta$-function. For a non-perturbative evolution the change of scale by a finite
factor two seems more natural. This leads to the definition of a finite
step-size counterpart of a $\beta$-function, the step scaling function 
\cite{Luscher:1991wu}
\begin{equation}\label{ssfsigmadef}
\sigma(u)=\gbar^2(2L)|_{\gbar^2(L)=u}.
\end{equation}
Note that $\sigma$ does not refer to the lattice at this point but is a universal
continuum quantity, different however for differently defined couplings.
Before we outline the computation of the function $\sigma$ we discuss its application.
We use it to build a sequence $\{u_i, i=0,1,\ldots,n \}$ based on the recursion
$u_i=\sigma(u_{i+1})$ and started at $u_0=\gbar^2(\Lmax)$.
This immediately implies that
\begin{equation}\label{uievol}
u_n=\gbar^2(2^{-n} \Lmax)
\end{equation}
holds and that for sufficiently many steps this value is arbitrarily deep in the
perturbative regime. Here in addition we should also start to see an evolution
that coincides with the one produced by the Callan Symanzik equation with
a perturbative approximation for the appropriate $\beta$ function. In
\fig{f:strategy} a schematic view is given for the example of 
$\gbar$ realized by the  \textSF\
finite volume coupling that will be detailed below.

\begin{figure}[t!]
  \centering
  \includegraphics[width=0.42\textwidth]{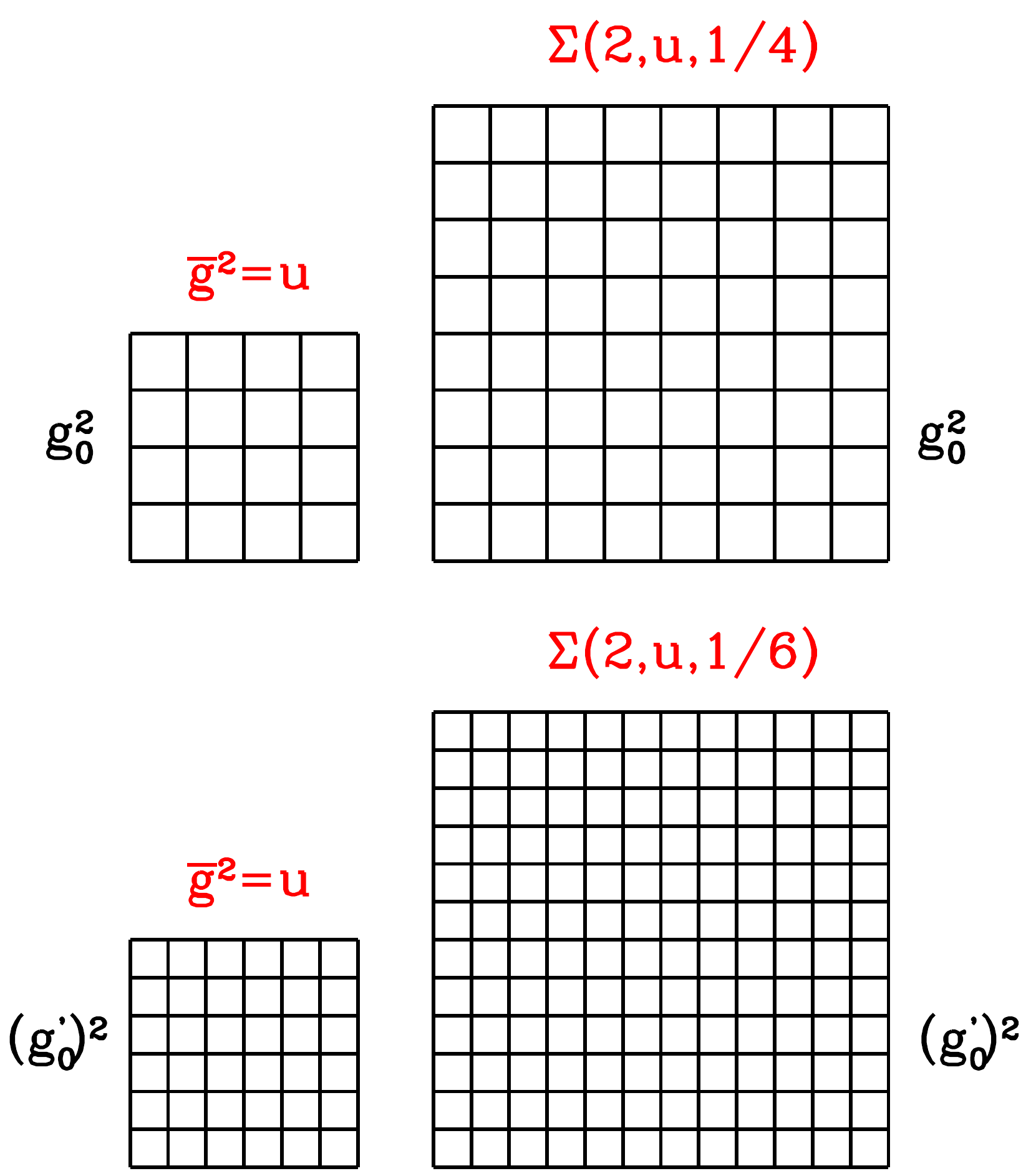}
  \caption{Illustration of the computation of 
  the continuum step scaling function from finer
  and finer lattices. Note that $\Sigma(2,u,a/L)$ in the 
  illustration corresponds to our lattice step scaling function 
  $\Sigma(u,a/L)$.
}
\label{f:latt_ssf}
\end{figure}

It remains to outline the computation of $\sigma$. To this end we pick a resolution
$L/a$ and some value $u$ and tune the bare parameters such that $\gbar^2=u$
results and that the quark masses vanish (massless scheme). Then, keeping the bare
parameters fixed, we double the size to $2L/a$ and determine a value of the lattice
step scaling function $\Sigma(u,a/L)=\gbar^2(2L)$. If this procedure is repeated
for a sequence of $L/a$ we finally extrapolate to
\begin{equation}\label{ssfSigmadef}
\sigma(u)=\lim_{a/L\to0}\Sigma(u,a/L)
\end{equation}
which again is a piece of universal continuum physics. We need to implement
this at a sufficiently dense set of $u$ values to have a sufficiently precise control
over $\sigma(u)$ in the range required for the evolution (\ref{uievol}). The procedure is indicated
in \fig{f:latt_ssf} for one value of $u$.

\section{Lattice discretization}

\subsection{The action}
The gluon vector potential $A_{\mu}(x)$ in Euclidean continuum QCD has its values in the
Lie algebra SU(3) and enters as a parallel transporter over infinitesimal distances into
the covariant derivative $D_{\mu}=\partial_{\mu}+A_{\mu}$. The resulting gauge
covariant curvature $F_{\mu\nu}=[D_{\mu},D_{\nu}]$, also in the Lie algebra,
is the building block of the gauge
invariant action density that is integrated over in the non-negative\footnote{
In our convention $A_{\mu}$ and $F_{\mu\nu}$ are antihermitean.}
 Euclidean Yang-Mills action
\begin{equation}\label{SYMills}
S_{\rm cont} = -\frac1{2g_0^2} \int d^4x \; \mathrm{tr} ( F_{\mu\nu}F_{\mu\nu} ).
\end{equation}

Following Wilson \cite{Wilson:1974sk} we put strong emphasis on gauge invariance
and construct
the discretization on a hypercubic lattice in a way that is
manifestly compatible with this structure.
The smallest separation on the lattice are finite links $(x,\mu)$ starting from any site $x$ into
the $\mu$-direction and ending at its nearest neighbor $x+a\hat{\mu}$
($a$ is the lattice spacing and $\hat{\mu}$ a unit vector). The associated
gauge transporters $U(x,\mu)$ are SU(3) group-valued and represent
the fundamental gluon field on the lattice. In the path integral they are
integrated over with the invariant Haar measure independently on each link.
Curvature on the lattice shows up by the two-step transporters
 $U(x,\mu)U(x+a\hat{\mu},\nu)$ and
$U(x,\nu)U(x+a\hat{\nu},\mu)$ being different or, equivalently,
by the plaquette field
\begin{eqnarray}
&&P(x,\mu,\nu)= \\ \nonumber
&&U(x,\mu)U(x+a\hat{\mu},\nu)[U(x,\nu)U(x+a\hat{\nu},\mu)]^{-1}
\end{eqnarray}
differing from unity. Thus a direct transcription of (\ref{SYMills}) onto the lattice
is given by the Wilson plaquette action
\begin{equation}\label{SWplaq}
S_{\rm W} = \beta \sum_{x,\mu<\nu} \mathrm{Re\, tr}\, [ 1-P(x,\mu,\nu)].
\end{equation}
If we build $U(x,\mu)$ out of slowly varying infinitesimal continuum $A_{\mu}(x)$
we find that $S_{\rm W}$ approaches $S_{\rm cont}$ with $\beta=6/g_0^2$,
i.~e. (\ref{SWplaq}) is a classically valid discretization.

In the same reference \cite{Wilson:1974sk} Wilson has also introduced what today
are called Wilson lattice fermions, which will be the discretization of the quark Dirac field
employed here. Independent quark fields $\psi(x)$ are introduced on the sites of the lattice.
They carry a four-valued Dirac index, the SU(3) color index and an $\Nf$ valued flavor
index which we all suppress in our notation. The covariant forward and backward derivatives 
of quarks are defined by
\begin{equation}\label{Dmu}
\hspace*{-4mm}
(D_{\mu} \psi)(x) =
 \frac1{a}
[U(x,\mu)\psi(x+a\hat{\mu})-\psi(x)]
\end{equation}
\begin{equation}\label{Dmuast}
\hspace*{-4mm}
(D^{\ast}_{\mu} \psi)(x) =
\frac1{a}
 [\psi(x)-U^{-1}(x-a\hat{\mu},\mu)\psi(x-a\hat{\mu})]
\end{equation}
It is well known that the most obvious 
discretized Dirac operator
leads to spurious particle poles, the so-called fermion doubling problem.
Wilson's remedy for this problem is the addition of a
term proportional to a discretized form of the covariant Laplacian, the Wilson term,
which attributes a mass of order $a^{-1}$ to the doubler modes but is otherwise
suppressed by an extra power of $a$.
The total Wilson Dirac operator is now given by
\begin{equation}
D_{\rm W} =\frac12 \gamma_{\mu} (D_{\mu}+D^{\ast}_{\mu}) 
- \frac{a}2 D_{\mu}D^{\ast}_{\mu}.
\end{equation}
The total lattice QCD action is thus given by
\begin{equation}
S_{\rm W;W}[U,\psi,\bar{\psi}]=S_{\rm W}[U] -\sum_x \bar{\psi}(D_{\rm W}+M_0)\psi,
\end{equation}
where $M_0$ is the bare mass matrix with the mass parameters for the various flavors
on its diagonal.
All these ingredients may now be finally assembled to write down the lattice QCD
partition function
\begin{equation}
Z=\int DU D\psi D\bar{\psi} \exp\{ -S_{\rm W;W}[U,\psi,\bar{\psi}]\}
\end{equation}
which beside the group integrations requires additional integrations over
the Grassmann valued quark fields. For numerical simulations
the latter (Gaussian) integrals are carried out and produce the fermion determinant
\begin{equation} \label{ZQCD}
Z=\int DU \exp\{ -S_{\rm W}[U]\} \det(D_{\rm W}+M_0).
\end{equation}
The matrix $D_{\rm W}$ has $U$ fields in its matrix elements and the
determinant represents a complicated nonlocal effective action in $U$
that has to be taken into account for sampling $U$ with the weight given by
this integrand. The standard hybrid Monte Carlo algorithms are able to
cope with this, but nevertheless here is the source of the enhanced complexity of
simulations once dynamical quark degrees of freedom are included.
Some more details on our algorithmic implementation of the
QCD path integral will be given in section \ref{algo}.

Another important point to mention is that the Wilson fermion regularization
breaks chiral symmetry which emerges only in the continuum limit.
Due to the Wilson term, $D_{\rm W}$ does not anticommute
with $\gamma_5$.
As a consequence the masses on the diagonal of $M_0$ undergo additive
renormalization and the physical zero mass condition to set up a massless scheme
has to be enforced as a nontrivial renormalization condition 
-- usually some chiral Ward identity -- 
which will force $a M_0$ to approach
a $g_0$ dependent nontrivial critical value $a m_c(g_0)$. Note that
in perturbation theory one finds
$a m_c(g_0)=c_1 g_0+\mathrm{O}(g_0^2)$
which amounts
to a linearly diverging bare mass parameter.

\subsection{Symanzik improved action}

A further consequence of missing chiral symmetry in the regularized theory
is the appearance of cutoff effects that are linear in the lattice spacing $a$
(multiplied by powers of logarithms).
In Symanzik's approach the structure of cutoff effects can be studied
by describing the lattice theory {\em including leading cutoff effects} 
by an effective theory {\em in the continuum}. This requires additional terms
in the action of the latter beyond the combination of renormalizable terms 
of dimension up to four that we have discretized before. In asymptotically
free theories the dimension of terms can be used to organize the additional 
contributions to the action:
extra terms of dimension five represent lattice artefacts that vanish linearly
in $a$, dimension six those that are quadratic and so on. The second essential
criterion is to only admit terms that are invariant under the (reduced) symmetry
that is still present in the lattice theory. Here chiral symmetry is missing
for Wilson fermions. The interplay between dimension
and symmetry leads to a finite number of additional couplings that allows to
match all cutoff effects up to a given order in $a$. This is the standard situation
in effective theories, with a rapid proliferation of the number of terms as the
order is increased. In addition, all that one can hope for is an asymptotic
expansion that is relevant close to the continuum limit
where combinations like $a\mhad$ are already small. In addition to the
enlarged action also in renormalized observables extra
mixings with higher dimensional terms of the same symmetry have to be
taken into account to achieve a complete representation of cutoff effects in correlation
functions. The whole concept may be seen as an
extension of the renormalization
programme that normally just focuses on divergences ($\log a$ and possibly inverse powers)
to small positive powers of $a$.

The dimension five terms relevant for the linear order in $a$ have been classified and listed
in \cite{Luscher:1996sc} for mass degenerate quarks. In this case there are five operators
of which three can be absorbed into modifications of the bare coupling and the bare mass.
The remaining two operators are
the Pauli term
\begin{equation}
{\cal O}_1=\bar{\psi}\sigma_{\mu\nu}F_{\mu\nu} \psi
\end{equation}
and
\begin{equation}
{\cal O}_2=\bar{\psi}D_{\mu}D_{\mu} \psi+
\bar{\psi}\overleftarrow{D}_{\mu}\overleftarrow{D}_{\mu} \psi.
\end{equation}
Note that within the Symanzik effective theory we systematically expand in $a$.
Therefore the dimension five terms in the action with an explicit factor $a$ are
expanded down from the exponent and appear as operator insertions.

The description and ultimately elimination of a class of cutoff effects
becomes much more manageable if we restrict ourselves to what is
called on-shell improvement \cite{Luscher:1984xn}. The restriction refers to correlations
with a finite number of local operators which all reside at physical
separations from each other, finite multiples of $\mhad^{-1}$ for example.
Then we can deform the integration
variables of the path integral at all points without inserted operators to
derive the so-called equations of motion as operator identities. They can
be used to transform between the higher dimension terms classified before
and to reduce their number while still matching cutoff effects in on-shell
correlations. For the case at hand the equation of motion is just the Euclidean
Dirac equation
\begin{equation}
(\gamma_{\mu}D_{\mu}+M_0) \psi =0, \quad
\bar{\psi}(\gamma_{\mu}\overleftarrow{D}_{\mu}-M_0)=0
\end{equation}
where only the renormalizable terms are considered as
we want to only modify the order five terms  and neglect yet higher
dimensional terms. This equation is usually employed to eliminate
${\cal O}_2$ and then the only new bulk term needed to describe O($a$)
on-shell cutoff effects in the effective theory is the Pauli term ${\cal O}_1$.
One more complication has to be mentioned. The insertions generated by the
$a$-expanded improved action appear integrated over 
Euclidean space-time and thus are not strictly
separated from the observables as the use of the equations of motion would require.
A more detailed analysis confirms however that these violations of the
equations of motion due to overlapping insertions (`contact terms')
can be compensated in the observable improvement terms
that were mentioned before.

\begin{figure}[ht!]
  \centering
  \includegraphics[width=0.25\textwidth]{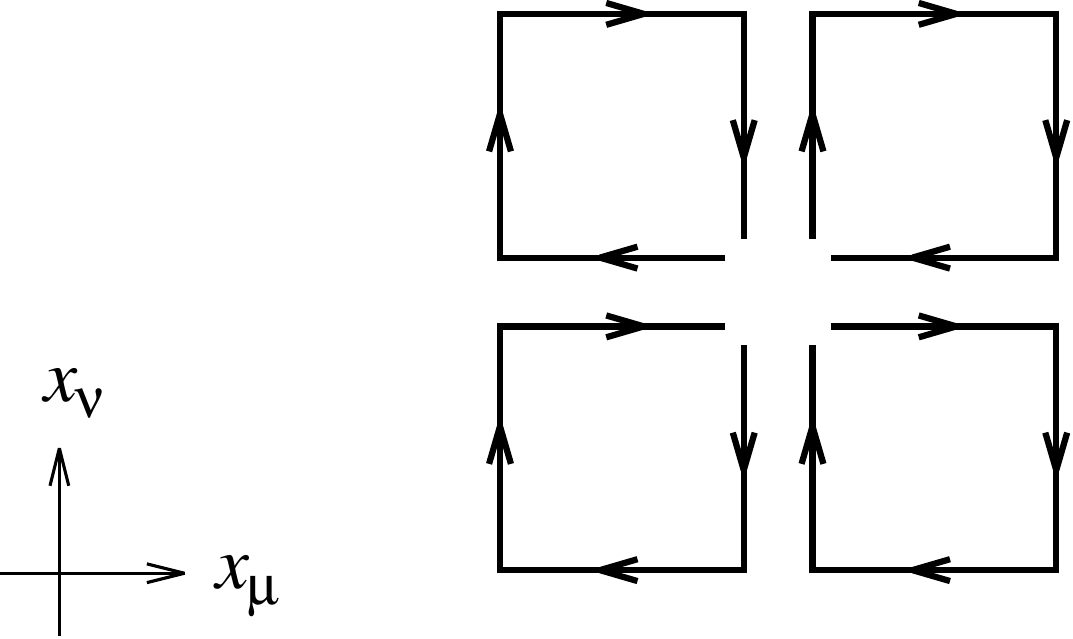}
  \caption{The clover leaf representation  $\hat{F}_{\mu\nu}(x)$ of  ${F}_{\mu\nu}(x)$.
}
\label{f:clover}
\end{figure}

The Symanzik description of cutoff effects can be used in a next step
to eliminate these contributions by what is called Symanzik improvement. To that end
a discretized version of the extra term(s) is added to the original action
with coefficients that are tuned such that the corresponding couplings in the
effective action vanish. This may then be seen as an alternative discretization
without the leading artefacts that have been systematically canceled in this way.
For the operator ${\cal O}_1$ above this is the so called clover term first proposed
in \cite{Sheikholeslami:1985ij}. We then have the Sheikholeslami-Wohlert (SW)
improved lattice action
\begin{equation} \hspace{-1.3em}
S_{\rm W;SW}=S_{\rm W;W}+\csw  a^5 \sum_x 
\bar{\psi}(x)\frac{i}4 \sigma_{\mu\nu}\hat{F}_{\mu\nu}(x) \, \psi(x)
\end{equation}
where the name clover is owed to the lattice representative of the field
strength $\hat{F}_{\mu\nu}(x)$ from four co-planar plaquettes (`leafs')
\begin{equation}
\hat{F}_{\mu\nu}(x)=\frac1{8a^2}\{ Q_{\mu\nu}(x)-Q_{\nu\mu}(x)\},
\end{equation}
\begin{eqnarray}
Q_{\mu\nu}(x)&=&
P(x,\mu,\nu)+\mbox{three more}\\ \nonumber
&&
\mbox{rotated } 1\times 1 \mbox{ loops opened at }x,
\end{eqnarray}
see \fig{f:clover}. An improvement condition (generalized renormalization condition)
that has to hold in the continuum theory has to be enforced to determine
$\csw(g_0)$.

The number of terms and coefficients required at leading order
is small enough that we can and shall implement complete on-shell O($a$) improvement.
This is not practicable any more at the next order $a^2$. 
If one implements however only some part of the
improvement terms with some prescription for the coefficients
one still obtains a legal variant discretization different from the one without extra terms.
Numerical experience suggests that the addition of a rectangle term to the gluon
plaquette action with a strength suggested by improving at tree level of perturbation theory
leads to a variant action with better properties than the pure plaquette form
although artefacts are O($a^2$) in both cases.
This action, called tree level improved L{\"u}scher-Weisz action, 
generalizes (\ref{SWplaq}) and reads
\begin{equation}\label{SLW}
S_{\rm LW}=\beta \sum_{i=0}^1 c_i \sum_{{\cal C}\in {\cal S}_i}
\mathrm{Re\; tr}\, [ 1-P({\cal C})].
\end{equation}
Here ${\cal S}_0$ is the set of all different (unoriented) plaquette ($1\times 1$) loops on
the lattice and $P({\cal C})$ a parallel transporter around it. Hence, for $c_0=1, c_1=0$
this action would coincide with $S_{\rm W}$. The second term involves the set ${\cal S}_1$
of all different planar $1\times 2$ loops (rectangles) and for the tree level improved 
L{\"u}scher-Weisz action the weights assume the values
\begin{equation}
c_0=\frac53,\quad c_1=-\frac1{12}.
\end{equation}

\subsection{Improved currents and renormalization}

Quark currents are observables of primary importance in QCD.
In particular the isovector axial current formed from the two light quarks
\begin{equation}
A_{\mu}^a(x)=\bar{\psi} \gamma_{\mu} \gamma_5 \tau^a \psi(x)
\end{equation}
and the pseudo-scalar density
\begin{equation}\label{Pdef}
P^a(x)=\bar{\psi} \gamma_5 \tau^a \psi(x)
\end{equation}
enter into the discussion of chiral symmetry. They are here given first as
bare currents in terms of bare fields at the same lattice site
and Pauli matrices $\tau^a$ operate on the up and down quarks.
As discussed in the previous subsection for O($a$) improvement these
dimension three operators can mix with dimension four terms of the right
symmetry. It turns out that there is no such term for $P^a$, but the improved
axial current can mix with the gradient of $P^a$ and is hence given by
\begin{equation}
(A_\mathrm{I})_{\mu}^a(x)=A_{\mu}^a(x)+a c_\mathrm{A} \tilde{\partial}_{\mu} P^a.
\end{equation}
Here $\tilde{\partial}_{\mu}=(\partial_{\mu}+\partial_{\mu}^{\ast})/2$ is the symmetrized
lattice derivative and $c_\mathrm{A}(g_0)$ is an improvement coefficient that
has to be fixed by another improvement condition. It will turn out that its perturbative
expansion starts at O($g_0^2$).

In \cite{Luscher:1996sc} the nontrivial interplay between the use of a massless renormalization scheme
and improvement is discussed in some detail. In such a scheme all renormalization
conditions are formulated at a normalization scale $\mu$. 
For nonzero, but for simplicity
degenerate, quark masses $m_0$ the relation between bare and renormalized
coupling must be taken as
\begin{equation}
g_\mathrm{R}^2=\tilde{g}_0^2 Z_g(\tilde{g}_0^2,a\mu),\quad \tilde{g}_0^2=g_0^2(1+b_g a m_q).
\end{equation}
Here $b_g(g_0)$ is an improvement constant that eliminates O($a$) effects at nonzero $m_q$
which in turn is the subtracted quark mass
 \begin{equation}
m_q=m_0-m_c(g_0)
\end{equation}
such that $m_q=0$ implies a vanishing physical mass.
The term with $b_g$ reflects a dimension five term
in the Symanzik effective action
 proportional to
$m {\rm tr}(F_{\mu\nu}^2)$.
It is only with this term (and the correct $b_g$) that in the process
of expressing physical observables in terms of $g_\mathrm{R}$
not only divergences but also linear lattice artefacts are eliminated.

In a similar way the usual multiplicative mass renormalization
must be replaced by
\begin{equation}
m_\mathrm{R}=\tilde{m}_q Z_m(\tilde{g}_0^2,a\mu),\quad \tilde{m}_q=m_q(1+b_m a m_q).
\end{equation}
Quite similar formulas follow for the current renormalizations
\begin{eqnarray}\label{ZAdef}
(A_\mathrm{R})_{\mu}^a&=&Z_\mathrm{A}(1+b_\mathrm{A} a m_q) (A_\mathrm{I})_{\mu}^a,\\
(P_\mathrm{R})^a&=&Z_\mathrm{P}(1+b_\mathrm{P} a m_q) P^a. \label{ZPdef}
\end{eqnarray}

If in the continuum limit the chiral symmetry group SU($\Nf$) $\times$ SU($\Nf$)
is recovered up to finite mass effects, then we expect the PCAC relation
\begin{equation}\label{PCACrel}
\partial_{\mu} (A_\mathrm{R})_{\mu}^a = 2 m_\mathrm{R} (P_\mathrm{R})^a
\end{equation}
to emerge as an operator relation that can be inserted into matrix elements.
It is very advantageous to focus on this relation to actually {\em define} the
renormalized mass in the improved theory, once the improved renormalized
currents have been introduced already. The finite size
Schr{\"o}dinger functional scheme
that we introduce next is a very convenient setting to do so.

Non-perturbative techniques to determine the improvement coefficients
were developed in the quenched approximation \cite{impr:pap3,impr:babp,impr:pap4}, and later applied for the two and three flavor theories 
in \cite{impr:csw_nf2,impr:ca_nf2,impr:babp_nf2,impr:csw_nf3}. 
Some coefficients such as $b_\mathrm{P}$ 
remain unknown non-perturbatively but 
can be taken from one-loop perturbation 
theory \cite{impr:pap2,impr:pap5}. 

\section{The Schr{\"o}dinger functional\label{theSF}}

\begin{figure}[t!]
  \centering
  \includegraphics[width=0.25\textwidth]{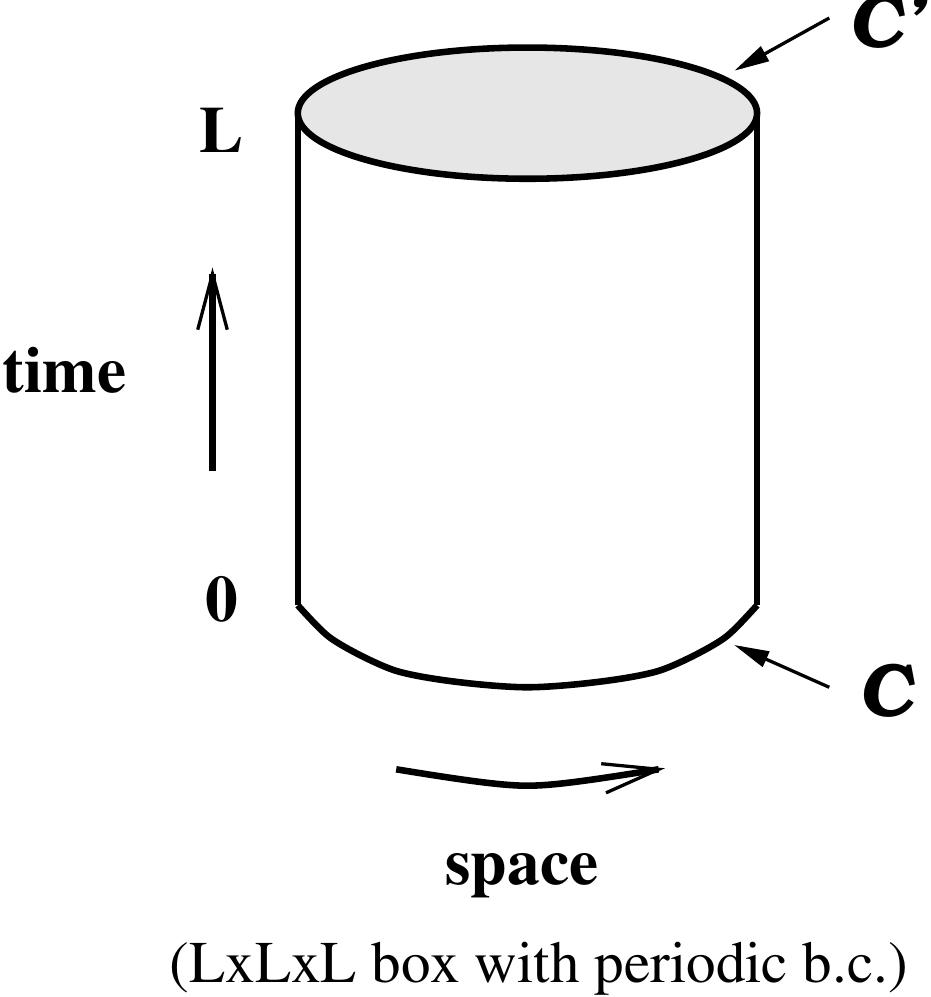}
  \caption{The Schr{\"o}dinger functional geometry. 
}
\label{f:can}
\end{figure}

For the Schr{\"o}dinger functional (SF) \cite{Luscher:1992an}
-- and later the associated renormalization scheme --
we consider a finite portion of
Euclidean space time with spatial extent $L$ and temporal size $T$ as depicted in \fig{f:can}.
In the spatial directions $\mu=k=1,2,3$ we impose periodic boundary
conditions $x\equiv x\pm L\hat{k}$ while Dirichlet boundary conditions
fix certain field components at $x_0=0,T$ to externally given values.
The SF can be defined in the continuum and is in fact studied 
by dimensionally regularized 1-loop
perturbation theory in \cite{Luscher:1992an}.
As we here want to mainly discuss non-perturbative computations
we prefer to immediately start on the lattice, where some features
of the SF even become simpler to discuss. This implies, of course, that
both $L/a$ and $T/a$ must be integer.

\subsection{Gauge sector}

In the standard form of the SF 
the Dirichlet conditions for the gluon field $U(x,\mu)$ are
\begin{eqnarray}
U(x,k)|_{x_0=0}&=&\exp(aC_k), \\
U(x,k)|_{x_0=T}&=&\exp(aC'_k),
\end{eqnarray}
where $C_k,C'_k$ are constant Abelian vector potentials
in the form of diagonal traceless imaginary matrices.
Note that temporal links 
$U(x,0)$ exist as integration variables for $x_0=0,a,\ldots,T-a$
and are not subject to any boundary conditions.

A possible interpretation of this Euclidean path integral
with boundaries is in the Hamiltonian or transfer matrix
formalism. It goes with a Schr{\"o}dinger representation
of states in Hilbert space as wave function(al)s on (three dimensional) 
spatial configurations
$U({\bf x},k)$. We denote a state concentrated on a
fixed field configuration $U({\bf x},k)\equiv \exp(aC_k)$ 
by a ket $|C \rangle$
(like $|x\rangle$-states in quantum mechanics).
Then the SF partition function ${\cal Z}$ is equal to the matrix
element
\begin{equation}
{\cal Z}(C';C)=\langle C'| \mathe^{-T\mathbb{H}}\mathbb{P}|C\rangle,
\end{equation}
where $\mathe^{-a\mathbb{H}}$ is the transfer matrix and  $\mathbb{P}$
is a projector to gauge invariant states \cite{Seiler:1982pw}.

From the Euclidean point of view
in the SF setup the time direction is distinguished from the others
and there is no translation invariance in the time-direction. Therefore
in this case we generalize the Wilson action to
\begin{equation}\label{Wsf}
S_{\rm Wsf}=\beta \sum_{{\cal C}\in {\cal S}_0}
w({\cal C})\mathrm{Re tr}\, [ 1-P({\cal C})].
\end{equation}
Here the novelty is the plaquette dependent
weight $w$ for which the transfer matrix formalism
suggests to take $w({\cal C})=1$ for all plaquettes
except the purely spatial ones on the boundary 
where $w({\cal C})=1/2$. This is so because it
is natural to symmetrically distribute these contributions
to the two adjacent transfer matrix factors
(as for the potential $V(x)$ in quantum mechanics).

In the Symanzik effective action additional terms 
representing cutoff effects are possible due to
the SF geometry. In the continuum they are given by three
dimensional integrals over boundary planes with the
dimension four densities $\mathrm{tr} ( F_{0k}F_{0k} )$, 
$\mathrm{tr} ( F_{kl}F_{kl} )$ as integrands. These
contributions are associated with
O($a$) boundary cutoff effects.
For Symanzik improvement these terms are discretized
and included in the lattice action with adjustable coefficients.
For $S_{\rm Wsf}$ this is incorporated by two different
nontrivial weights for space-time and for space-space plaquettes
at the boundary. With our Abelian boundary fields the latter
type does not contribute and we set
\begin{equation}
w({\cal C})=c_t(g_0)=1+c_t^{(1)}g_0^2+\ldots
\end{equation}
for 0k plaquettes touching the boundary and $w({\cal C})=1$
for all others. In principle $c_t(g_0)$ has to be fixed by yet
another improvement condition. In practice these terms
can at present only be set to perturbative values 
as already indicated above. See \cite{nara:rainer} for a  
discussion of the relevance of this approximation.
For the tree-level L{\"u}scher Weisz action similarly modified
weights are required for both plaquettes and rectangles
close to the boundary which are discussed in
 \cite{Aoki:1998qd} or more recently in \cite{Bulava:2013cta}.
In the following we always assume that these O($a$) improvement
terms are included in lattice actions for the SF. 

To not introduce further scales the boundary fields
 $C,C'$ are taken as multiples
of $L^{-1}$ and an often used standard  choice is
\begin{eqnarray}
C_k&=&\frac{i}{L}\mbox{diag}(\eta-\pi/3,-\eta/2,-\eta/2+\pi/3)\\
\nonumber
C'_k&=&\frac{i}{L}\mbox{diag}(-\eta-\pi,\eta/2+\pi/3,\eta/2+2\pi/3).
\end{eqnarray}
The dimensionless parameter $\eta$ allows to vary the boundary
values and is set
to zero after taking derivatives with respect to it.
In \cite{Luscher:1992an} it is shown that
these particular boundary values (for not too small $L/a,T/a$)
lead to a stable minimum of $S_\mathrm{Wsf}$. 
This minimum is unique up to gauge transformations which
(at the boundaries) we restrict to the subgroup that preserves
the boundary values.
A representative $U(x,\mu)=\exp(aB_{\mu})$ for this 
gauge orbit of minima is
\begin{equation}\label{Bfield}
B_0=0, B_k=[x_0C'_k +(T-x_0)C_k]/T
\end{equation}
which linearly interpolates between the boundaries.

We are now in a position to introduce the renormalized SF
coupling. We start from the effective action or free energy
\begin{equation}\label{GammaB}
\Gamma[B]=-\ln {\cal Z}(C';C).
\end{equation}
In perturbation theory a saddle point expansion around $B$
requires the usual Fadeev-Popov gauge fixing and yields a
regular expansion
\begin{equation}
\Gamma [B]=S_{\rm Wsf} [B] + \Gamma_1 [B]+g_0^2 \Gamma_2 [B]+\ldots,
\end{equation}
where we note that the classical or tree level term $S_{\rm Wsf}[B]$
in (\ref{Wsf}) is proportional $g_0^{-2}=\beta/6$.
The definition of the coupling associated with the scale $L$  finally reads
\begin{equation}\label{gSFdef}
\gbar_\mathrm{SF}^2(L)=g_0^2 \left.\frac{\partial S_{\rm Wsf}/\partial\eta}
{\partial\Gamma/\partial\eta}\right|_{\eta=0}.
\end{equation}
In $\partial\Gamma/\partial\eta$ we differentiate the logarithm of a partition function
with respect to a parameter entering into the (boundary terms of the) action.
It immediately leads to an expectation value that is independent of
unphysical factors in the path integral measure and can be estimated as a mean value
of a well defined observable when field configurations are sampled with the probability
$\exp(-S_{\rm Wsf})$. The normalization factor $g_0^2\partial S_{\rm Wsf}/\partial\eta$
can easily be computed in closed form for (\ref{Bfield}).

\subsection{Quark sector}

The extension of the SF to fermions has been first
presented in \cite{Sint:1993un}. 

We generalize the periodic boundary conditions in space to
a periodicity up to a phase\footnote{
It has become customary to call this twisted boundary
conditions, although it should not be confused
with  t'Hooft type twisted boundary conditions
referring to planes of a torus.
}
for the quark fields
\begin{equation}
\psi(x+L\hat{k})=\mathe^{i\theta_k}\psi(x), \quad
\bar{\psi}(x+L\hat{k})=\mathe^{-i\theta_k}\bar{\psi}(x),
\end{equation}
which leaves all bilinear densities in the action and elsewhere
strictly periodic. The angles $\theta_k$ allows us to vary
the finite size kinematics in useful ways.

As for the Dirichlet boundary conditions in time, it turns out that only half
of the components of the independent Grassmann fields $\psi$
and $\bar{\psi}$ have to be fixed. Formally this is a consequence
of the first order nature of the Dirac equation and the correspondingly
modified boundary value problem as was already noted in connection
with the bag model \cite{Chodos:1974pn}. Alternatively, and more
laboriously, one may argue on the basis of the transfer matrix for Wilson fermions
\cite{Luscher:1976ms}. The result for the SF in any case are Dirichlet conditions
\begin{equation}
P_+ \psi|_{x_0=0}=\rho({\bf x}),\quad 
P_- \psi|_{x_0=T}=\rho'({\bf x})
\end{equation}
and
\begin{equation}
\bar{\psi}P_-|_{x_0=0}=\bar{\rho}({\bf x}),\quad 
\bar{\psi}P_+|_{x_0=T}=\bar{\rho}'({\bf x})
\end{equation}
with projectors
\begin{equation}
P_{\pm}=\frac12 (1\pm\gamma_0).
\end{equation}
The spatial fields $\rho,\rho',\bar{\rho},\bar{\rho}'$
are formal Grassmann valued sources which, after possible
differentiations will always be set to zero. The partition function
(\ref{GammaB}) now depends on all boundary fields
${\cal Z}={\cal Z}(C',\bar{\rho}',\rho';C,\bar{\rho},\rho)$
and is given by the finite volume path integral
\begin{equation}
{\mathcal Z} =\int DU D\psi D\bar{\psi} \exp[-S_\mathrm{W;Wsf}(U,\psi,\bar{\psi})]
\end{equation}
which as an example we have written with the plaquette
action and plain Wilson fermions and the (here suppressed) boundary fields
enter into the action.

As mentioned before we want to eliminate O($a$) artefacts throughout.
We know already that this requires the clover term to be added to the bulk
quark action. But, as for the gluons, the presence of boundaries allows for
new terms in the Symanzik action that have to be canceled by corresponding
improvement terms. A discussion of the relevant dimension four densities
integrated over the $x_0=0,T$ time slices is found in \cite{Luscher:1996sc}
together with their reduction due to the on-shell conditions. The upshot is a rather
simple modification of the quark action by a contribution at $x_0=a$ given by
\begin{equation}
(\tilde{c}_t-1) a^4\sum_{\bf x}
[\bar{\psi}P_+ D_0^{\ast} \psi+
\bar{\psi} \overleftarrow{D}_0^{\ast}P_- \psi]
\end{equation}
and another such term with the same coefficient
at the other boundary. Again $\tilde{c}_t$
has to be tuned and possesses a perturbative expansion
\begin{equation}
\tilde{c}_t(g_0)=1+\tilde{c}_t^{(1)}g_0^2+\ldots .
\end{equation}
Additional terms involving spatial boundary terms do not
contribute for constant sources $\rho,\ldots$,
to which our applications will be restricted,
or can be absorbed into rescaling the sources
which undergo multiplicative renormalization anyway.
If in the following we refer to the Sheikholeslami
Wohlert improved quark action $S_\mathrm{SWsf}$
for the SF, we assume the $\tilde{c}_t$ terms to be included, too.

\subsection{Boundary quark operators}
\begin{figure}[t!]
  \centering
  \includegraphics[width=0.4\textwidth]{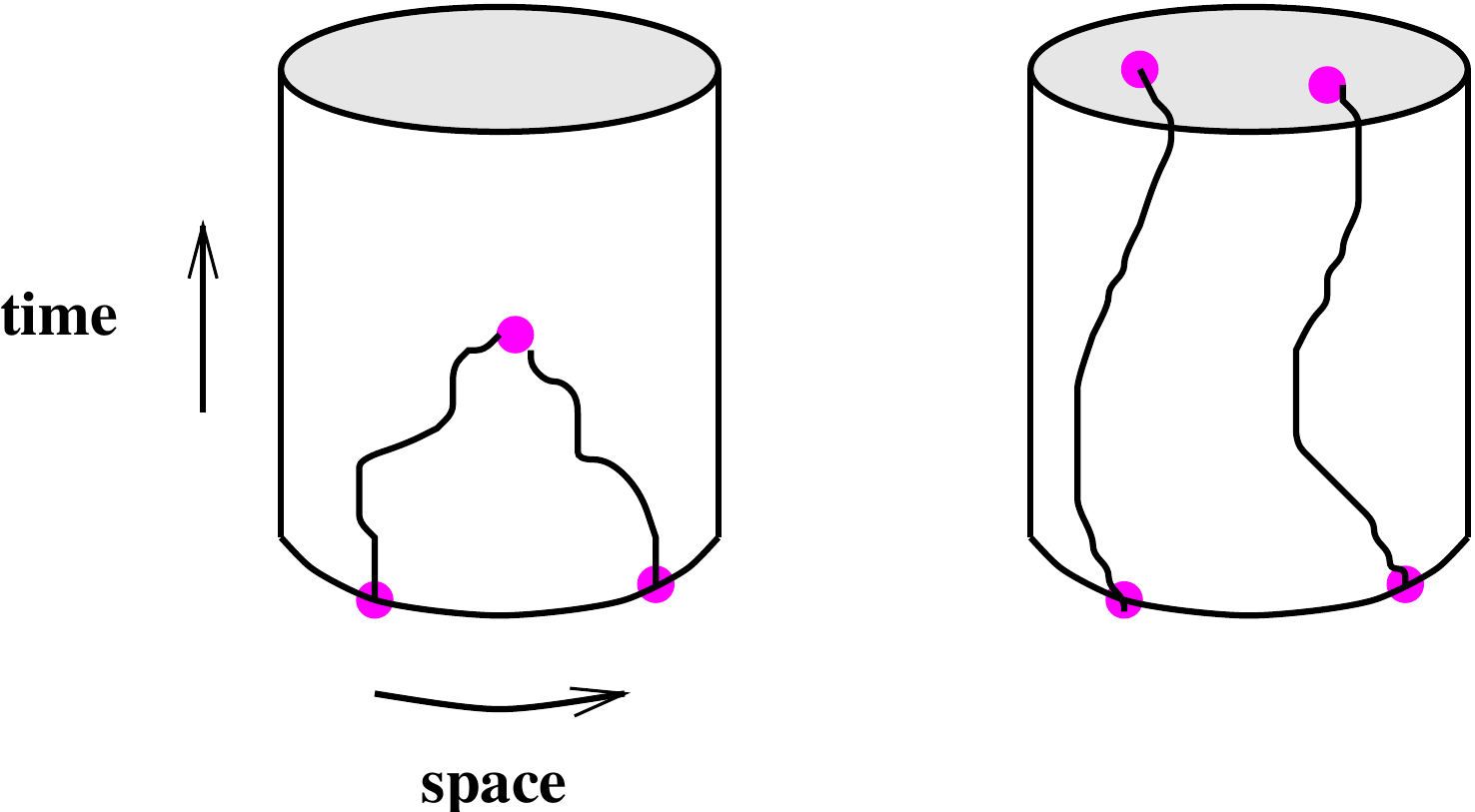}
  \caption{Correlation functions in
  the Schr{\"o}dinger functional. On the left we show
  boundary-to-bulk correlation functions 
  such as $f_\mathrm{P}$ and on the right the
  boundary-to-boundary correlation function $f_1$.   
}
\label{f:matr}
\end{figure}

We introduce `functional' differentiation operators for the boundary sources
\begin{eqnarray}
\zeta({\bf x})&=&a^{-3}\frac{\partial}{\partial\bar{\rho}({\bf x})},\quad
\bar{\zeta}({\bf x})=-a^{-3}\frac{\partial}{\partial\rho({\bf x})}\\
\zeta'({\bf x})&=&a^{-3}\frac{\partial}{\partial\bar{\rho'}({\bf x})},\quad
\bar{\zeta}'({\bf x})=-a^{-3}\frac{\partial}{\partial\rho'({\bf x})}.\nonumber
\end{eqnarray}
If they contribute to an observable ${\cal O}$ expectation values are meant in the sense
\begin{equation}
\quad \langle {\cal O}\rangle=
\end{equation}
\begin{displaymath}
\left\{ \frac1{\cal Z} \int DU D\psi D\bar{\psi}\,{\cal O}\, \exp[-S(U,\psi,\bar{\psi})]
\right\}_{\bar{\rho}'=\rho'=\bar{\rho}=\rho=0}
\end{displaymath}
where $S$ is any of our lattice actions and it contains the boundary values.
Inspection of any of these actions shows that differentiation with respect to
$\rho({\bf x})$ for example leads to single insertions of the dynamical quark
fields with suitable parallel transporters. They transform contragrediently to
$\rho({\bf x})$, i.~e. like $\bar{\rho}({\bf x})$. Hence the various $\zeta,\ldots$
can be contracted to form boundary currents like for example
\begin{equation}
{\cal O}^a = a^6 \sum_{\bf u,v}  \bar{\zeta}({\bf u})
 \gamma_5 \frac12 \tau^a\zeta({\bf v})
\end{equation}
and the corresponding primed operator at the other boundary.
They allow to form the SF standard correlation functions
\begin{equation}
\fP(x_0) = -\frac13 \langle P^a(x) \, {\cal O}^a \rangle
\end{equation}
and the boundary to boundary constant
\begin{equation}
f_1= \langle {\cal O}'^a {\cal O}^a \rangle,
\end{equation}
which are illustrated in \fig{f:matr}.
We may use these correlation functions to define a convenient
normalization condition for $Z_{\rm P}$ in (\ref{ZPdef}) by postulating
\begin{equation}\label{ZPnorm}
Z_{\rm P}=\mbox{const.} \sqrt{f_1}/\fP(T/2)
\end{equation}
where multiplicative renormalization factors of the boundary fields cancel.
Some choice has to be adopted for the kinematical parameters:
aspect ratio $T/L$, the sources, angles $\theta_k$ and for $x_0/T$.
The constant is then chosen such that $Z_{\rm P}=1$
holds at tree level. As emphasized before we would like to establish the SF
as a massless renormalization scheme. We therefore want to tune the bare
masses $m_0$ to their critical value. At least in perturbation theory this is possible
as the SF supplies an infrared regulator by providing a mass gap of order $L^{-1}$.

A similar standard matrix element involving the axial current is defined by
\begin{equation}
\fA(x_0) = -\frac13 \langle A_0^a(x) \, {\cal O}^a \rangle
\end{equation}
which allows us to define a bare improved PCAC mass
\begin{equation}\label{mPCACI}
m=\frac{
\tilde{\partial}_0 \fA(x_0) +\cA a \partial_0^{\ast}  \partial_0 \fP(x_0)
}{
2\fP(x_0)
}.
\end{equation}
In the next step this leads to the renormalized running mass
\begin{equation}\label{mbardef}
\mbar(L)=m \frac{
Z_\mathrm{A}(1-b_\mathrm{A} am_q)
}{
Z_\mathrm{P}(1-b_\mathrm{P} am_q)
}.
\end{equation}
If all improvement and renormalization factors assume their correct values
$\mbar$ is well defined up to O($a^2$) uncertainties. This is visible as small variations
under changes of kinematic parameters ($x_0$ for example) on which the lattice but not the continuum 
results depend.


\section{Gradient flow}
\label{flow}
The finite volume SF coupling defined in the previous section has been the
key tool for the ALPHA collaboration's effort to relate low and high energies in
QCD for many years. Recently the technique of the gradient flow \cite{Luscher:2010iy}
has given access to a large class of alternative observables
 that may serve in the same function.
It has turned out in the meantime that it is a good strategy to keep
SF boundary conditions, typically with zero boundary fields, to define
massless finite size schemes based on gradient flow (GF) couplings \cite{Fritzsch:2013je}. 
They will be found superior in practise
to the SF coupling at intermediate energy scales, but not at the perturbative
end of the scale evolution. In addition, perturbation theory is more developed
and also easier
for the traditional SF coupling where the three-loop term of the $\beta$-function
as well as some two loop results for improvement coefficients like $c_t$
are available (for $S_\mathrm{W;SWsf}$) \cite{Bode:1999sm}. Therefore 
a good way forward is to use a particular optimized
GF coupling for the lower energy
part of the scale evolution and, at an intermediate energy,
to match it to the SF coupling which is then run up
to perturbatively high energies. This promises the best over-all precision and is thus
a good reason to also exploit gradient flow couplings and review them here.

\subsection{Gradient flow in the continuum theory}

In the continuum Yang Mills theory we follow \cite{Luscher:2010iy} and
grow a one parameter ($t$) family of gauge potentials $B_{\mu}(t,x)$ starting in an
arbitrary four dimensional potential $A_{\mu}(x)$ which is taken as
an initial value
\begin{equation}
B_{\mu}(t=0,x)=A_{\mu}(x).
\end{equation}
The trajectory is defined by imposing
the first order flow equation
\begin{equation}\label{floweq}
\frac{\partial B_{\mu}(t,x)}{\partial t}= -\frac{\delta S_\mathrm{YM}[B]}{\delta B_{\mu}(t,x)}
\end{equation}
where $S_\mathrm{YM}$ is the usual Euclidean Yang Mills action evaluated
for potentials at any $t$ value, which will be referred to as flow `time'.
The action is lowered with growing $t$ as we move along the steepest descent direction
of the action.
Thus along the trajectory the initial field $A_{\mu}$ gets smoothed.

It is most remarkable that correlation functions at finite
$t$ are finite without
additional renormalizations beyond those of the four dimensional theory.
In \cite{Luscher:2011bx} this has been shown to all orders of perturbation theory by
performing an analysis based on Feynman rules of a
five dimensional theory  where $t$ is an extra coordinate with the dimension
of a squared length. The five dimensional action is constructed with additional
Lagrange multipliers
in such a way that
the flow equation appears as a constraint in the functional integral.
Similar techniques have been known for some time in the
related field of stochastic quantization \cite{ZinnJustin:1987ux}.

A particularly striking new finite observable is
the local action density
\begin{equation}
E(t)=-\frac12  \mathrm{tr}
( G_{\mu\nu}(t,x)G_{\mu\nu}(t,x) ),
\end{equation}
where $G_{\mu\nu}(t,x)$ is the SU(3) curvature
tensor of the potential $B_{\mu}(t,x)$,
which has a  finite expectation value
at positive flow time.
In \cite{Luscher:2010iy} the leading order perturbation theory result
\begin{equation}\label{Econt}
\langle t^2 E(t) \rangle =\frac3{16 \pi^2} \gmsbar^2(\mu) +\mathrm{O}(\gmsbar^4),
\quad \mu=\frac1{\sqrt{8t}}
\end{equation}
has been derived. To leading order the length $\sqrt{8t}$ is the radius over which
the initial values $A_{\mu}(x)$ are smoothed by solving (\ref{floweq}) and is thus
a natural scale for the $\overline{\mathrm{MS}}$ coupling to expand in.
This shows that up to a computable normalization factor the combination
$t^2\langle E(t) \rangle$ can be regarded as a renormalized coupling constant
which -- in contrast to $\gmsbar$ -- is however defined also beyond perturbation theory.
In section \ref{sec_hadronic} we have explained the advantages for our purposes in defining
coupling constants running with the finite size. This remains true for GF couplings.
We want however to stay with couplings running with only a {\em single scale} and therefore
tie together $t,L$ by setting
\begin{equation}\label{c_def}
\sqrt{8t}=c L
\end{equation}
with a fixed proportionality
constant $c$ of order unity. Its value and the details of the finite size boundary conditions
together form part of the coupling definition and each choice represents a different scheme.
These different variants are in finite relations with each other and a 
particular choice will be determined by
practical considerations. In QCD with $\Nf$ quark species the flow (\ref{floweq}) is unchanged
with only $S_\mathrm{YM}$ appearing. Also the leading order result (\ref{Econt}) is
$\Nf$ independent. 

We remark that the result (\ref{Econt}) indicates that GF couplings are not
technically simple in perturbation theory. The normalization factor in defining
a renormalized coupling constant via the bare coupling is usually 
given by a trivial tree level result.
For GF couplings a diffusion process corresponding to the free flow equation has to be solved 
to normalize $E(t)$
and this calculation is already comparable to the
evaluation of one-loop diagrams. The next correction would resemble a two loop
calculation etc.

\subsection{Gradient flow on a finite lattice}

In principle the transcription to the lattice is straight forward,
but many choices have to be made that differ in their lattice artefacts.
First of all, a lattice discretization for the action in the 
four dimensional path integral
has to be chosen as before, with which configurations $U(x,\mu)$ are
generated. In addition, the action whose gradient is taken in the flow
equation must be discretized, and this need not necessarily be the same discretization.
Finally, a lattice substitute for the action density must be picked.

On the lattice the gluon field is Lie group rather than algebra valued.
This requires standard changes to (\ref{floweq}). Instead of $B_{\mu}(t,x)$
we now introduce a family of group valued link field $V(t,x,\mu)$ by 
\begin{equation}
V(t=0,x,\mu)=U(x,\mu)
\end{equation}
and
\begin{equation}\label{latflow}
a^2 \frac{\partial V(t,x,\mu)}{\partial t} V(t,x,\mu)^{-1}
= -g_0^2 \partial_{x,\mu} S[V].
\end{equation}
Here $S$ is any lattice gluon action (e.~g. $S_\mathrm{W}$ or  $S_\mathrm{LW}$). 
The Lie algebra valued left derivative
gradient of a scalar function on the group 
is given by (suppressing here the link index $x,\mu$)
\begin{equation}\label{Liederi}
\partial f[U]=\left.\sum_a  T^a \frac{d}{ds} f[\exp (sT^a) U] 
\right|_{s=0}
\end{equation}
with $T^a$ being a basis of group generators. Such a lattice version of the gradient flow
based on $U(x,\mu)$ fields has been first written down in \cite{Luscher:2010iy}.

In \cite{Ramos:2014kka} a careful
assessment of lattice artefacts of gradient flow observables
has been reported. We do not review any details which proceed
via an analysis of the five dimensional Symanzik effective theory.
The remarkably simple result is that the major part of all
O($a^2$) coming from the flow equation is cancelled
if we invoke
the tree level improved L{\"u}scher Weisz action 
for the gradient flow with just
one additional term and replace (\ref{latflow}) by 
\begin{equation}\label{latflowi}
a^2 \frac{\partial V(t,x,\mu)}{\partial t} V(t,x,\mu)^{-1}=
\end{equation}
\begin{displaymath}
 -g_0^2(1+\frac{a^2}{12} D^{\ast}_{\mu}D_{\mu}) \partial_{x,\mu} S_\mathrm{LW}[V]\,,
\end{displaymath}
where the covariant derivatives defined in (\ref{Dmu}), (\ref{Dmuast})
have to be taken with $V(t,x,\mu)$ here. The discretized
energy density to form $E$ has to be
read off from the density in $S_\mathrm{LW}$ (\ref{SLW}).
It has to be remembered however that the usual O($a^2$) effects
of the four dimensional theory are unchanged, and it is just the additional
source from the flow that one tries to control here. More numerical experiments
with the rather recent proposal (\ref{latflowi}) are required.

To evaluate flow observables in simulations some solver
for first order equations in time has to be employed.
This will require also a discretization of flow time, but this step-size
error can be kept at a negligible level. In appendix C of \cite{Luscher:2010iy}
a third order Runge Kutta integrator on Lie groups has been proposed
in a fully explicit form that is easy to implement. It is reported that
a step-size $\epsilon=0.01$ in $t/a^2$ is accompanied by errors of $10^{-6}$
in link variables. A more sophisticated variant in \cite{Fritzsch:2013je}
automatizes the choice of $\epsilon$ by combining the previous Runge Kutta
integrator with adaptive step-size control.
 
To define a finite size scheme based on a GF coupling given
by the expectation value of $\langle E(t)\rangle$ boundary conditions have to
be specified. The first attempt to do so was made for simple
periodic boundary conditions in \cite{Fodor:2012td}. It has been known
for a long time that Yang Mills theory on a small torus is complicated
due to the presence of non-Gaussian fluctuation modes 
(see refs. in \cite{Fritzsch:2013je}). This leads to a
small coupling expansion of torus based finite size couplings
that is non-analytic in $g_0^2$.
As discussed in section \ref{theSF}
the SF in contrast has a unique minimum with
purely Gaussian fluctuations up to exact gauge modes that have
to be fixed in the usual way and then no non-analyticity arises.
Moreover, zero field SF Dirichlet boundary conditions provide an
infrared regulator and hence the additional bonus is that
the quark masses can be set to zero and we continue to have
a massless scheme as with the SF coupling before.

To achieve the exact normalization on the lattice  
${\cal N}^{-1}t^2\langle E(t,x_0) \rangle=g_0^2+\mathrm{O}(g_0^4)$
the factor ${\cal N}$ has to be calculated for the finite lattices in use by taking
into account all details like choice of discretization, boundary conditions, the ratio
$c$ in (\ref{c_def}) etc. Note that in the SF there also is the indicated $x_0$ 
dependence and the obvious standard choice is to take $x_0=T/2$ here.
Depending on the action in use ${\cal N}$ is calculated in closed form
or numerically, involving a finite momentum sum, and the required values may be
tabulated once and for all. The effort in any case is negligible.

\begin{figure*}[t!]
  \centering
  \includegraphics[width=0.74\textwidth]{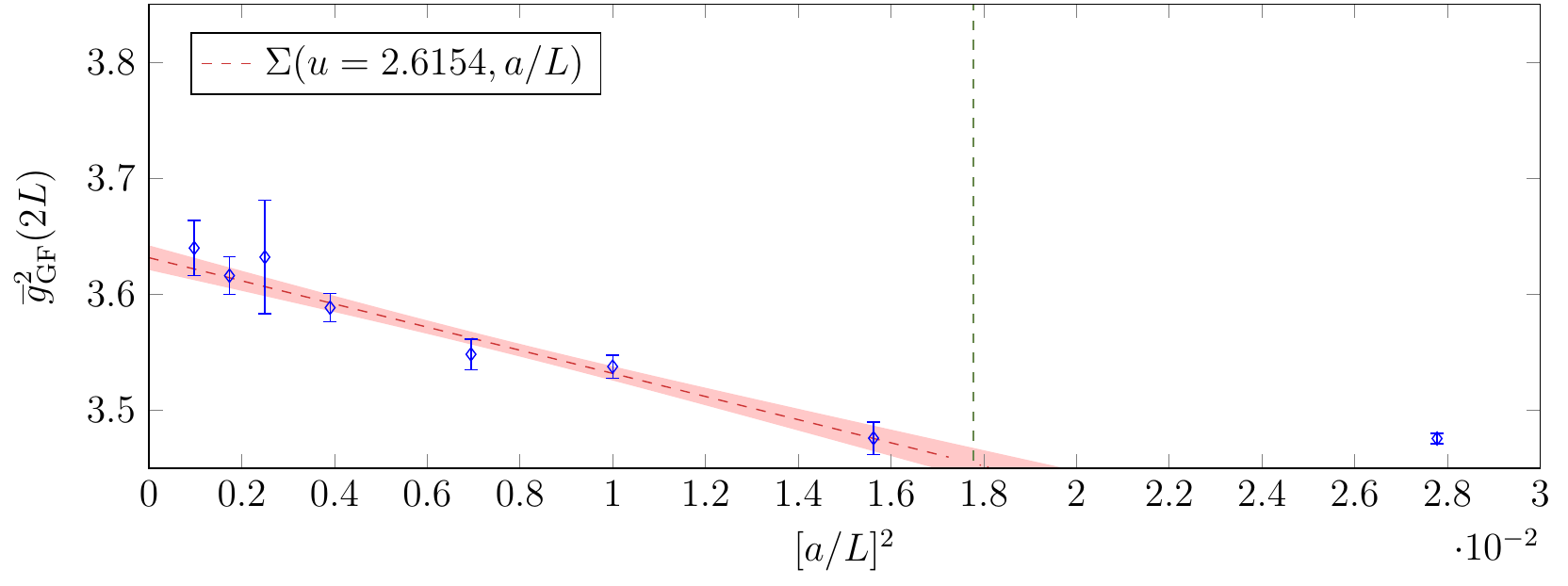}
  \caption{Step scaling function of a gradient flow coupling for $\nf=0$.
}
\label{f:ssfGF}
\end{figure*}

In \fig{f:ssfGF} we have a first look at the cutoff dependence of a step scaling
function of a GF coupling a la \cite{Fritzsch:2013je}. Some details of this experiment are:
$\Nf=0, T=L, c=0.3$ with the actions $S_\mathrm{LW}$ for the $U$ sampling,
$S_\mathrm{W}$ in the flow equation (\ref{latflow}) and a clover-discretized
$E$. We see a nice extrapolation to the 
continuum limit but also a non-negligible slope 
and a clear break off from an approximate asymptotic $a^2$ 
behavior beyond $a/L \ge 1/8$ (vertical dashed line).
More detailed investigations of the cutoff effects
of GF couplings are underway and the improved flow equation (\ref{latflowi})
is presently simulated.

\section{Some details on simulation algorithms \label{algo}}

The whole physics programme reviewed in this article depends
on the availability of appreciable (parallel) computer resources and
their optimal exploitation with efficient algorithms (see \cite{Luscher:2010ae} for a review). 
While the
resources mainly have to be provided by outside supercomputer centers,
algorithmic research and development is a major occupation of the
physics collaborations. Usually useful ideas can only be advanced if
the physics that one tries to extract is understood in detail. This section
is therefore meant to reflect some of these efforts without being able
to cover the whole field.

\subsection{Hybrid Monte Carlo}
Algorithms of the Hybrid Monte Carlo (HMC) class are at present
the only known method that is able to cope with QCD at $\Nf > 0$,
although for $\Nf = 0$ and for other bosonic lattice field theories much more
efficient algorithms exist. The problem is the nonlocal contribution of
the fermion determinant in (\ref{ZQCD}).

The HMC \cite{Duane:1987de,Gottlieb:1987mq} method is based
on molecular dynamics. There for each link $U(x,\mu)$ a conjugate momentum
$\pi (x,\mu)$ in the Lie algebra is introduced and a Hamiltonian is considered
\begin{equation}
{\cal H}[\pi,U]=-\sum_{x \mu}{\rm tr}(\pi^2 (x,\mu))+S[U]
\end{equation}
with some lattice action $S[U]$ that plays the role of a potential here.
We now consider an enlarged ensemble with a partition function
given by the path integral
\begin{equation}
\tilde{Z}=\int D\pi DU \mathe^{-{\cal H}[\pi,U]}
\end{equation}
where the $\pi$ fields are integrated with the obvious measure
over their 8 dimensional real vector space for each link.
It is clear that observables depending only on $U$ assume the
desired expectation values in this ensemble as the Gaussian
$\pi$ integrations simply factorize out.

The HMC update procedure for the enlarged ensemble is now given by
the following sequence of steps:
\begin{itemize}
\item a complete field of independently Gaussian distributed $\pi (x,\mu)$
is drawn at random,
\item with the given configuration $(\pi,U)$ as initial values at $t=0$ Hamilton's equations
are solved in molecular dynamics time\footnote{
This $t$ has nothing to do with the flow time before.
} $t$
\begin{equation}
\dot{\pi}(x,\mu)=-\partial_{x,\mu}S[U]
\end{equation}
\begin{equation}
\dot{U}(x,\mu)=\pi (x,\mu) U (x,\mu)
\end{equation}
up to some trajectory length $t=\tau$. In practice this is done
with some inexact discretized integrator with a finite step size
that has to be exactly {\em time reversible},
\item only due to step size errors of the integrator the Hamiltonian
between $t=0$ and $t=\tau$ will change ${\cal H}\to{\cal H}+\Delta{\cal H}$.
The end configuration is taken as a successor of the initial one
with the Metropolis acceptance probability $\min(1,\exp(-\Delta{\cal H}))$.
In the case of rejection the old configuration remains unchanged.
\end{itemize}
An important point here is that no zero step size limit is needed
to prove detailed balance for the HMC. In practice however the step size
$\tau/N_\mathrm{step}$ has to be small enough to make the rejection rate
small, around 10\% for example in practice.

With two dynamical quark species for $S$ an effective action like
\begin{equation}
S_\mathrm{eff}[U]=S_\mathrm{W}[U]-\ln \det (M^2)
\end{equation}
should be taken that includes the fermion determinant.
We have introduced here the usual lattice parameterization for the Dirac matrix
(for one flavor)
\begin{equation}
M=2\kappa a(D+m_0)
\end{equation}
with the hopping parameter
\begin{equation}\label{kappadef}
\kappa = \frac1{8+2 a m_0},
\end{equation}
where $D$ will be $D_\mathrm{SW}$  in most cases here.
Its non-negative
squared determinant corresponds to two degenerate quark species
as in the $\Nf=2$ theory. Other cases complicate simulations
further and will only be mentioned later.
The advantage of HMC in this context is that in $\partial_{x,\mu}S_\mathrm{eff}$
only the response of $S_\mathrm{eff}$ to infinitesimal rather than finite moves
in $U$, which enters into matrix elements of $M$, is required. The
derivative of the $\ln\det$ contribution in $S_\mathrm{eff}$ leads
to the necessity to control arbitrary matrix elements of the non-sparse
matrix $M^{-1}$. In four dimensional QCD on large lattices also
this is impractical. Therefore the additional trick of introducing pseudofermions is
needed that we discuss next. With them we shall only have to solve systems
of linear equations with the Dirac matrix as coefficients for which highly
efficient tools exist.

\subsection{Pseudofermions}

A pseudofermion is a complex field $\phi(x)$ which carries the same Dirac and color indices
as one quark species. The squared determinant can now be represented by a Gaussian path
integral
\begin{equation}
|\det(M)|^2=\int D[\phi] \exp(-S_\mathrm{pf}[\phi])
\end{equation}
with
\begin{equation}
S_\mathrm{pf}[U,\phi]=
\sum_x |M^{-1}\phi|^2.
\end{equation}
A number of comments are in order:
\begin{itemize}
\item $D[\phi]$ means independent integration over
the real and imaginary part of all components and some 
(later irrelevant) normalization,
\item due to the inversion of $M$ the action of $\phi(x)$
is highly nonlocal,
\item for a pair of degenerate improved SW quarks there is
no sign problem although
the Dirac operator $D_\mathrm{SW}$ has complex eigenvalues.
This is so since the determinant is real because of the relation
\begin{equation}
M^{\dagger}=\gamma_5 M \gamma_5.
\end{equation}
\end{itemize}

With the pseudofermion representation of the determinant
the HMC Hamiltonian becomes
\begin{equation}
{\cal H}[\pi,U,\phi]=\frac12 (\pi,\pi)+S_\mathrm{W} [U]
+S_\mathrm{pf}[U,\phi]
\end{equation}
with a scalar product notation for the $\pi$ part and Wilson gluons 
as an example. Note that we have not introduced momenta conjugate 
to $\phi$. As they are Gaussian a correctly distributed $\phi$ configuration
can be trivially drawn (for given $U$) by applying once the Dirac operator in
\begin{equation}
\phi = M\eta.
\end{equation}
Here $\eta$ has the same indices (color, Dirac) as $\phi$ and each component
is an independent Gaussian random number with zero mean and unit variance.
Now each HMC trajectory starts with a global choice of $\pi$ and $\phi$ and
then an evolution in molecular dynamics time $t=0,\ldots,\tau$ and the 
accept/reject step.  The force during the evolution stems from the gluon
action and from $S_\mathrm{pf}$. The derivative (\ref{Liederi}) of the gluon
action for a given link is given by a sum of small Wilson loops, `staples' for
$S_\mathrm{W}$, that require small computation time. The derivative of 
$S_\mathrm{pf}$ derives from  the variation of the discretized Dirac operator
$M\to M+\delta M$ as the $U(x,\mu)$ entering its matrix elements change.
It is given by
\begin{equation}
\delta S_\mathrm{pf}=-\sum_x \chi^{\dagger} M^{-1}\delta M \chi +c.c.
\end{equation}
with
\begin{equation}
M\chi= \phi.
\end{equation}
For each computation of the pseudofermion force we thus have to solve
two linear systems involving the Dirac matrix. It is this step -- solving the
lattice Dirac equation with given right hand sides -- which consumes by
far the most time in QCD simulations with the HMC algorithm.

Obviously it is worthwhile to optimize the linear equation solvers used with
HMC as far as possible and a corresponding effort has been devoted to this issue by
the lattice community. As the Dirac matrix in any of the discretizations discussed
here is sparse, iterative Krylov space solvers are the method of choice. At first
the simple and robust conjugate gradient method \cite{Press:1058314} has been widely used.
A large number of improvements have been developed over the years
with incremental small speed-up factors. A kind of breakthrough with regard
to the slowing down for small quark masses has been achieved 
more recently with
low mode deflation \cite{Luscher:2007es,Luscher:2007se}
and with multigrid methods \cite{Frommer:2013fsa}.

Also the integrators used for Hamilton's equations inside of HMC allow for improvement.
One example is the multiple time step size technique \cite{Sexton:1992nu}.
It is based on the observation that the pseudofermion force, that is
expensive to compute, is usually much smaller than the gluonic force.
The former would therefore admit larger time steps leading to fewer evaluations.
In \cite{Sexton:1992nu} a modified leapfrog integrator is proposed
that allows for different step sizes for the two components while still
maintaining the required reversibility. Other possibilities are to minimize
the higher order step size errors and a popular integrator in this respect
has been proposed in \cite{Omelyan2003272}.

 \subsection{Hasenbusch preconditioning\label{Hasenbusch}}
 Hasenbusch  \cite{Hasenbusch:2001ne} has proposed a 
 simple factorization of the two flavor determinant 
\begin{equation}
|\det(M)|^2=|\det(\tilde{M})|^2 \times
|\det(M\tilde{M}^{-1})|^2
\end{equation}
which has been further investigated in  \cite{Hasenbusch:2002ai}. Above,
$\tilde{M}$ contains a larger mass ($\tilde{\kappa} < \kappa$).
One pseudofermion is introduced now for each factor
with an action
\begin{equation}
S_\mathrm{pf}=(\tilde{M}^{-1}\phi_1,\tilde{M}^{-1}\phi_1)+
                           (\tilde{M}M^{-1}\phi_2,\tilde{M}M^{-1}\phi_2).
\end{equation}
With some tuning of $\tilde{\kappa}$ the condition numbers of both
factors can be lower than the one of the original $M$.
As a consequence the forces are of smaller magnitude
with smaller  fluctuations
and the step size can be increased by a factor two or so.
It turned out that it was important to tune $\tilde{\kappa}$
properly \cite{Urbach:2005ji} and 
also more than two factors can be introduced.


\subsection{Simulation of nondegenerate quarks, $\Nf>2$}

All simulation techniques and improvements discussed so far
have exploited that the a fermion determinant was squared
for two degenerate fermions. Its weight could not go negative and
the factorization was useful for introducing a pseudofermion.
To model QCD more precisely the inclusion of the strange
(and ultimately also charmed) quarks becomes necessary.

The strange quark implies an additional weight $\det(M_s)$ 
to be included in the Boltzmann factor where the
Dirac matrix $M_s$ includes the strange mass. Following
\cite{Luscher:2010ae} and \cite{Luscher:2012av} we
start from the trivial factorization
\begin{equation}
\det(M_s)=W_s \det(R^{-1}),\quad W_s=\det(M_s R).
\end{equation}
The goal is to construct $R$ such that
\begin{itemize}
\item
$R\approx (M_s^{\dagger}M_s)^{-1/2}$ and $W_s\approx 1$,
\item
$R$ can be represented by one or several pseudofermions.
\end{itemize}
A well known solution is the rational approximation of degree $m$
\cite{Clark:2006fx}
\begin{equation}
R=C\prod_{k=0}^{m-1}\frac{M_s^{\dagger}M_s+\omega_k^2}
{M_s^{\dagger}M_s+\nu_k^2}.
\end{equation}
If one decides on a spectral interval $(\epsilon^2,r^2)$ of
$M_s^{\dagger}M_s$ on which $R$ approximates
the inverse square root then there is the
well defined Zolotarev algorithm to construct 
those $C,\{\omega_k\},\{\nu_k\}$ that optimize the
approximation quality in a certain norm. In addition
the shifts can be taken real, positive and ordered
\begin{equation}
0<\nu_0<\omega_0<\nu_1<\cdots<\omega_{m-1}.
\end{equation}
It is now not difficult to see that, using the rational
factorization,
 $\det(R^{-1})$ can be represented
by one or several pseudofermions. By attaching
the subsets of the factors with similar shifts
to several pseudofermions,
even something similar
to mass preconditioning can be achieved.

In spite of a good rational approximation (at moderate degree $m$)
we still have to take care of the correction factor $W_s$.
Due to the lack of chiral symmetry the Wilson type lattice Dirac
operator has no rigorous spectral cutoff and may 
in principle even develop zero eigenvalues
for some gauge fields occurring in the path integral (Monte Carlo).
In such cases $W_s$ may not be close to one and therefore important.
Such fluctuations are known to become extremely rare once the mass is not
too small and additional light quarks are present. They are
further suppressed for large volume. For the strange quark $W_s$
should be monitored to have only moderate fluctuations around unity.
In precisely this case it can be estimated stochastically by
\begin{equation}
W_s=\langle \mathe^{-(\eta,(R^{-1}M_s^{-1}-1)\eta} \rangle
\end{equation}
with Gaussian random fields $\eta$.
Its estimator, possibly averaging over several
independent $\eta$ fields, may then be included in
observables as usual for reweighting.

Clearly, stochastic strange quark reweighting can not account
for fluctuations to negative $\det(M_s)$. As indicated this is however not
expected to happen for large enough mass and volume. In massless
$\Nf=3$ simulations for renormalization purposes the finite volume
SF boundary conditions have to be monitored via $W_s$
to sufficiently stabilize the spectrum. In the applications reviewed 
here this is indeed the case.

\subsection{Topological freezing\label{topfreeze}}

The efficiency of any Monte Carlo algorithm depends on the degree
in how far it is able to produce at reasonable cost {\em statistically independent} lattice field
configurations distributed according to the 
action of the field theory under study. The statistical error of any observable
is given by
\begin{equation}
\sigma^2({\cal O})=v({\cal O}) \frac{2 \tau_{\rm int,{\cal O}}}{N}.
\end{equation}
In this relation $v({\cal O})$ is the variance of ${\cal O}$ which
is another expectation value. The number of generated estimates $N$
is divided by the integrated autocorrelation time  $\tau_{\rm int,{\cal O}}$
which summarizes the decorrelation power of the algorithms for a given observable.
There also is a maximal autocorrelation time $\tau_{\rm exp}$ that is related to
the spectral gap in the transition probability (Markov) operator of an algorithm.
More details on autocorrelations and error analysis can be found in \cite{Wolff:2003sm}
for example.

Each smooth configuration in a continuum Yang Mills theory on a four dimensional torus
carries an integer valued topological quantum number $Q$. In the path integral all configurations
are summed over and all $Q$ contribute even if large $|Q|$ may have a small relative weight
in some situations depending on the physical size of the torus and the observable.
The famous self-dual instanton
configurations, for example, are classical fields with $Q=\pm 1$. Quantization and in particular
the lattice regularization
eliminates in the first place all naive arguments based on continuity and topology.
It has been argued however \cite{Luscher:1981zq} 
that topological winding numbers do effectively imply a decomposition
of the space of $U(x,\mu)$ configurations into sectors that are separated
by barriers with large Euclidean action.
The numerical problem now is that known Monte Carlo algorithms for QCD, including HMC, change
configurations in small steps and the probability to change sectors by sequences
of such moves gets very small when the continuum limit is approached. 
As one lowers $a$ this is first seen in steeply growing autocorrelation times.
Once these are very large, they are hard to even discover in the statistical error
analysis and there is a risk of unnoticed systematic errors. Clearly observables
formed in terms of (a lattice approximation) of $Q$ are optimal diagnostic tools
to control this problem of topological freezing.

The seriousness of the freezing problem was noticed during the physics programme
reviewed here and led to quite some delay. A precise analysis of the problem was given in 
\cite{Schaefer:2010hu}. Here a special strategy to sharpen the error analysis was proposed
that allows for a reliable error estimation in the range of lattice spacings down to
$a\approx 0.05$ fm where with our standard algorithms and actions freezing starts
to become a problem. The idea of the method is to extract the longest autocorrelation time
from measuring $Q^2$ and $E(t)$ at finite flow time
and feed this information into the error analysis of other
observables that couple to topology only weekly. Slightly different strategies can be used
for the running GF coupling with SF boundary conditions at intermediate volume \cite{Fritzsch:2013yxa}
which however also suffers from
freezing.

Finally in \cite{Luscher:2011kk} a fundamental solution to the problem was given.
Open boundary conditions in Euclidean time -- while keeping periodicity in the three
spatial directions -- abolishes the quantization of topological charge and thus the 
barriers in field space. Charges can so to speak flow in and out of the boundary
hyperplanes. In view of the seriousness of the freezing problem, the price of a somewhat
more complicated data analysis in the large volume simulations due to 
the lost time translation invariance seems
reasonable to accept. In addition Symanzik improvement has to be modified due to the
presence of boundaries in a way that is somewhat familiar from the SF. 

For a recent review covering lattice QCD algorithms in more detail than possible here we refer to
\cite{Schaefer:2012tq}.

\section{Running of coupling and masses}

In this section we review numerical results that have been achieved
for the running of the non-perturbatively defined coupling and mass
with the finite system size as renormalized scale in the \textSF.

\subsection{Step scaling function of the SF coupling at $\Nf=2$.}

\begin{figure}[t!]
  \centering
  \includegraphics[width=0.49\textwidth]{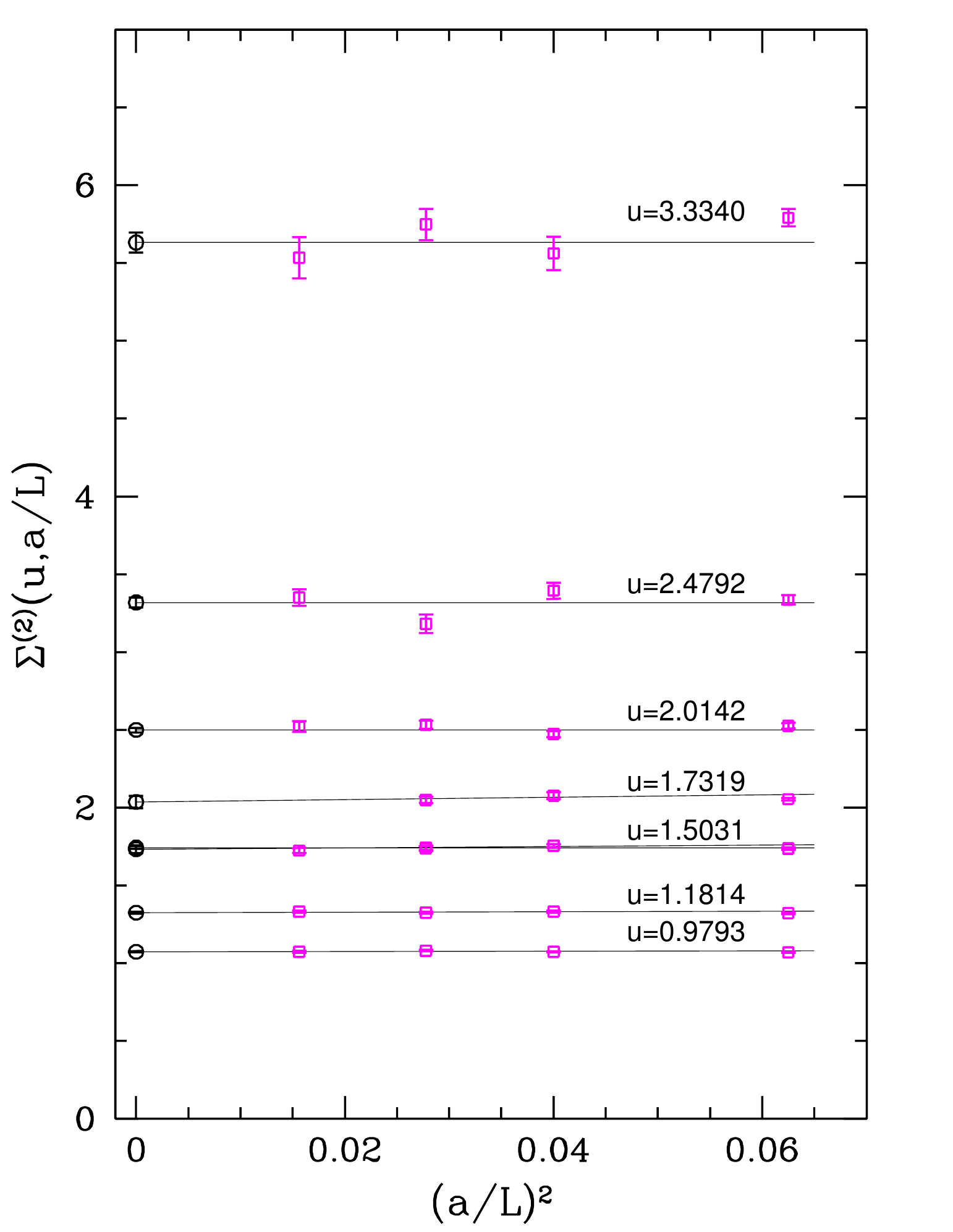}
  \caption{Continuum extrapolations of the \textSF\
  coupling with $\nf=2$ dynamical quark flavors. Figure from \cite{DellaMorte:2004bc}.
}
\label{f:Sigma_nf2}
\end{figure}
The SF coupling has been defined in (\ref{gSFdef}) and its evolution is
analyzed by computing the lattice step scaling function (SSF)
(\ref{ssfSigmadef}) and by extrapolating it to $\sigma$. 
In the numerical investigations it has turned out
that it is highly profitable to accelerate the continuum limit of $\Sigma$.
We consider the perturbative expansion of the deviation
\begin{equation} \hspace*{-4mm}
\frac{\Sigma(u,a/L)-\sigma(u)}{\sigma(u)}=
\delta_1(a/L)u+\delta_2(a/L)u^2+\ldots.
\end{equation}
For the simulations on which we report here the SF was implemented with the Wilson
plaquette gluon action $S_\mathrm{Wsf}$ with SW (`clover') improved quarks 
$S_\mathrm{SWsf}$. For this total action the two loop perturbation theory
has been studied on the lattice \cite{Bode:1999sm}
and hence $\delta_1,\delta_2$ are known for the relevant values of $L/a$.
We use them to define a perturbatively two loop improved SSF
\begin{equation}
\Sigma^{(2)}(u,a/L)=\frac{\Sigma(u,a/L)}{1+\delta_1(a/L)u+\delta_2(a/L)u^2}\,.
\end{equation}
The very flat and well controlled
continuum extrapolation of this quantity is shown in \fig{f:Sigma_nf2}.
Several other technical issues had to be mastered to produce these data.
The two bare parameters $g_0$ and the bare quark mass (\ref{kappadef}) $\kappa$
for each $L/a$ have to be tuned to the $u$ values of the series shown
together with a vanishing quark mass. This can be achieved only to some limited precision
and small corrections have to be applied based on perturbative as well as numerical
information. For the quark mass a particular definition $m$ based on a PCAC relation
(\ref{mPCACI}) is adopted
\cite{DellaMorte:2004bc} and it is estimated that a tuning up to $|m L| < 0.05$ suffices for
the attempted precision. 

\begin{figure}[t!]
  \centering
  \includegraphics[width=0.46\textwidth]{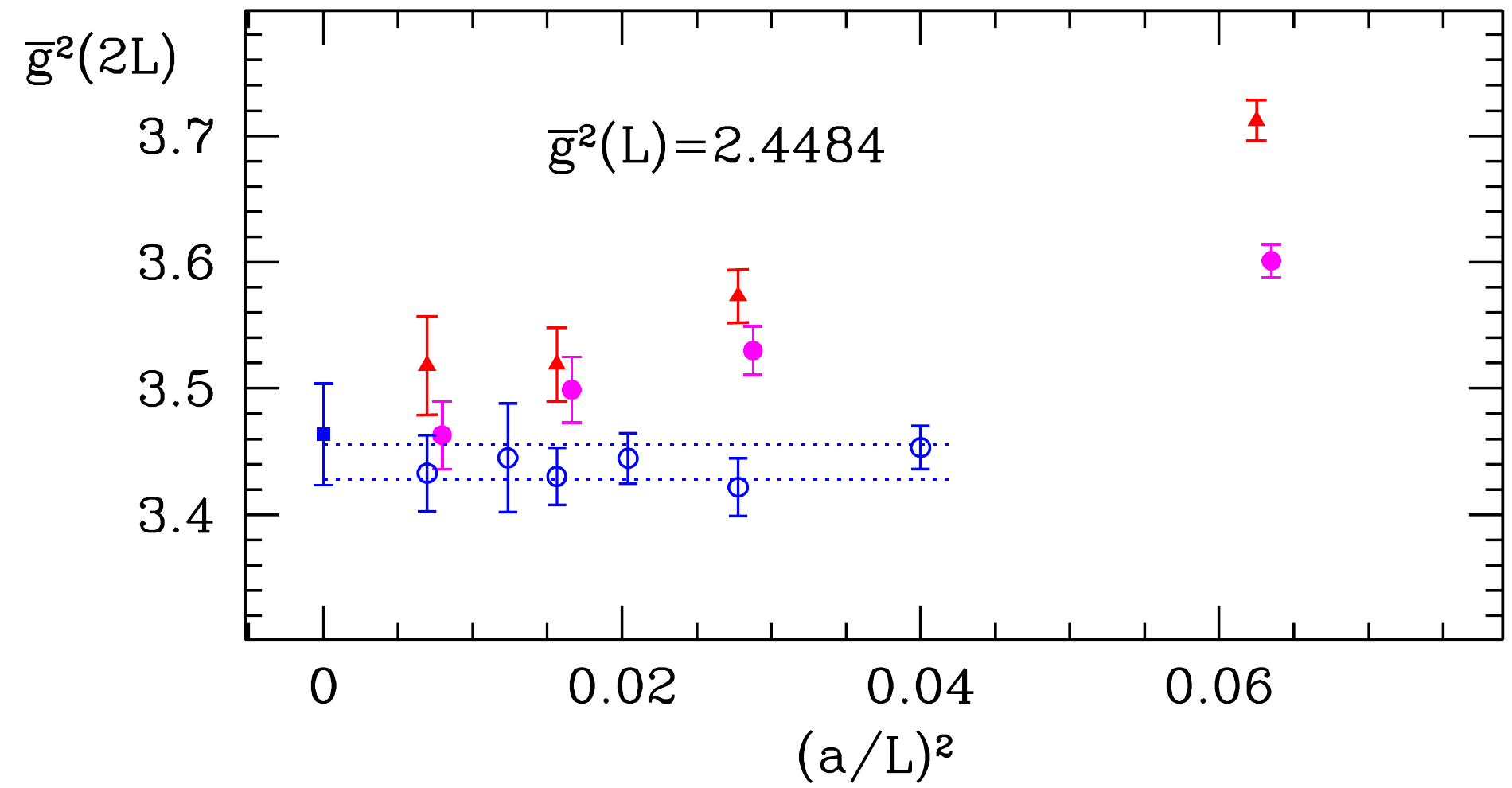}
  \caption{A test of the continuum extrapolations 
  with different actions for $\nf=0$. 
  The data from 
top (triangles) to 
bottom (open circles) are for the Iwasaki, the tree level L{\"u}scher Weisz 
and the Wilson gauge action. Both the 
boundary improvement of the action and the improvement of the observables
have been included. At present this is possible at the  2-loop level for the 
Wilson gauge action  only,
and at the 1-loop level in the two other cases.
 Figure from \cite{nara:rainer} based on data from 
\protect\cite{Takeda:2004xha,Bode:2001jv}.
}
\label{f:Universality1}
\end{figure}
\begin{figure}[t!]
  \centering
  \includegraphics[width=0.45\textwidth]{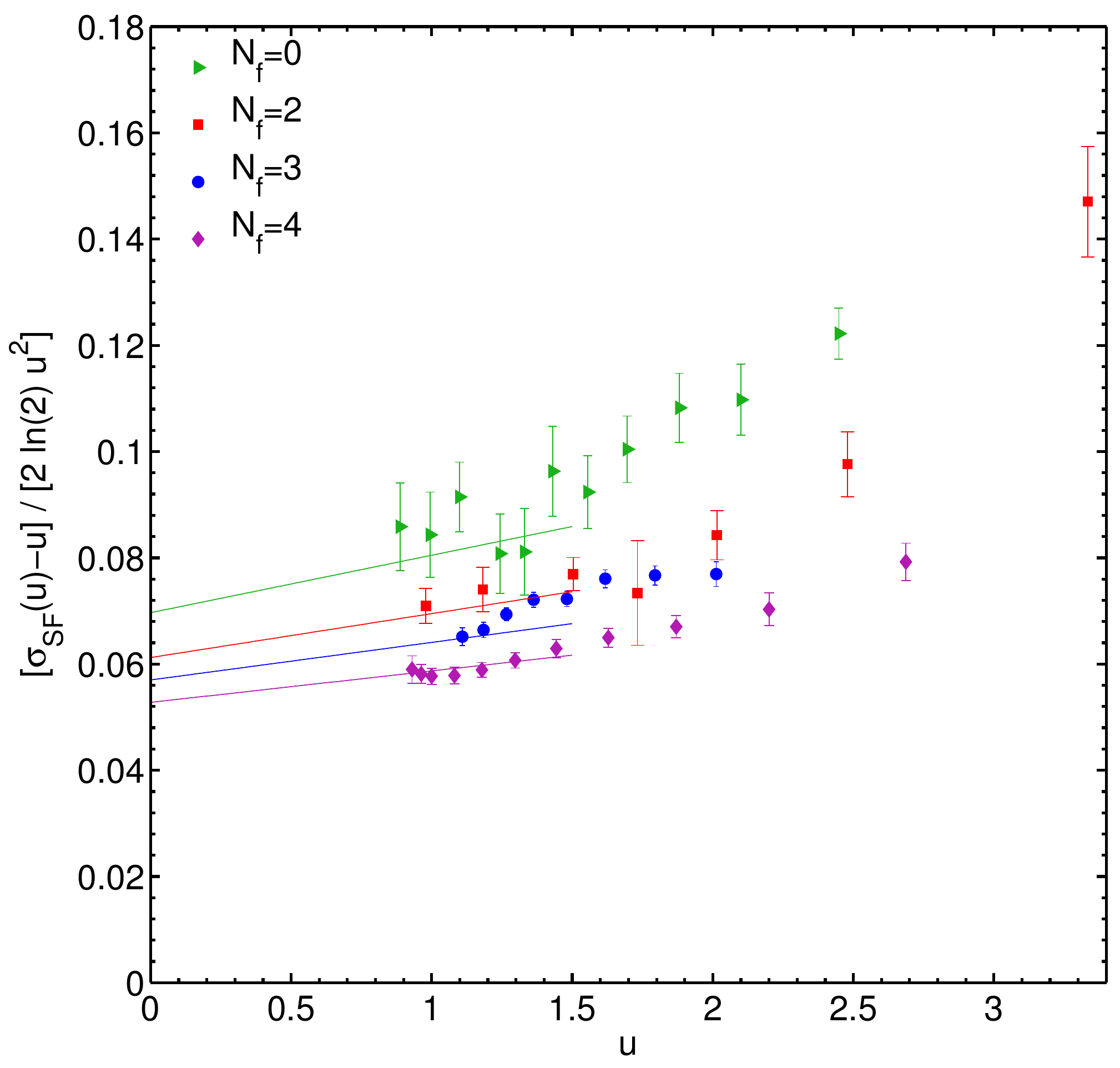}
  \caption{ The $\nf$-dependence of the step scaling
  function of the \textSF\ coupling \cite{Patrick_lat14}. Non-perturbative results 
  are shown together with the two-loop curves.
}
\label{f:nfdep}
\end{figure}

In \fig{f:Universality1} we find an additional demonstration that the values of $L/a=6 \ldots 12$ 
have very small discretization errors, 
at least after our 2-loop improvement of the observable
and with the Wilson plaquette gauge action.  
One can therefore carry out a precise continuum
limit with these rather small lattices.
On the other hand the figure shows that one 
cannot take this for granted for any action. Care to take the continuum limit is the most important 
requirement for a trustable determination of the 
step scaling functions and ultimately also the $\Lambda$ parameter.

\begin{figure*}[t!]
  \centering
  \includegraphics[width=0.35\textwidth]{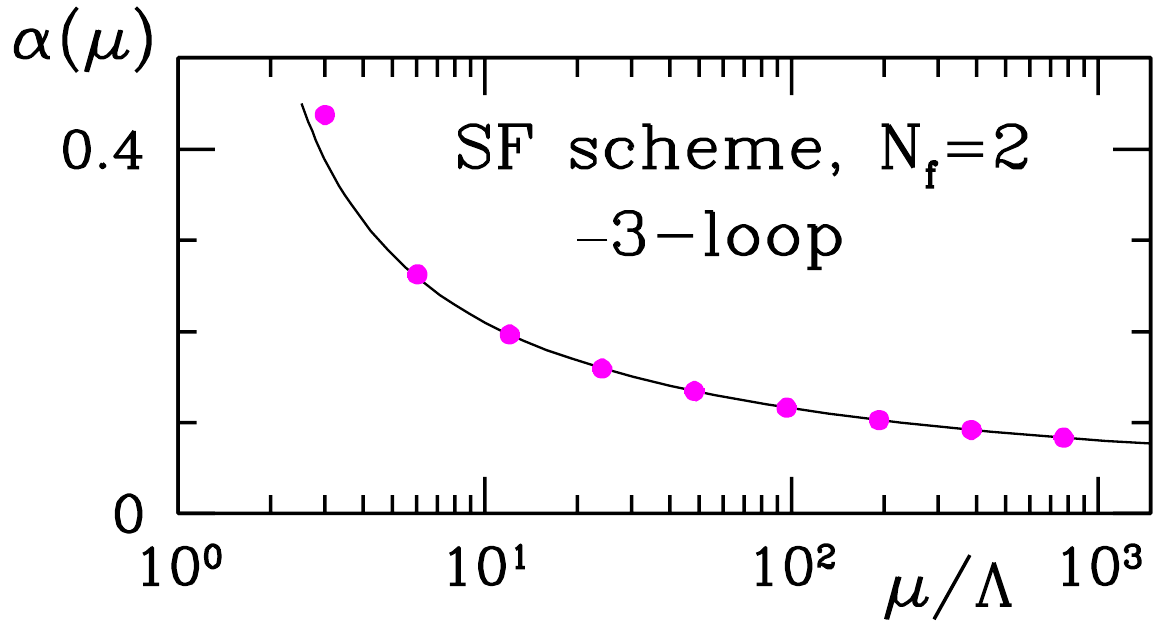}
  \includegraphics[width=0.35\textwidth]{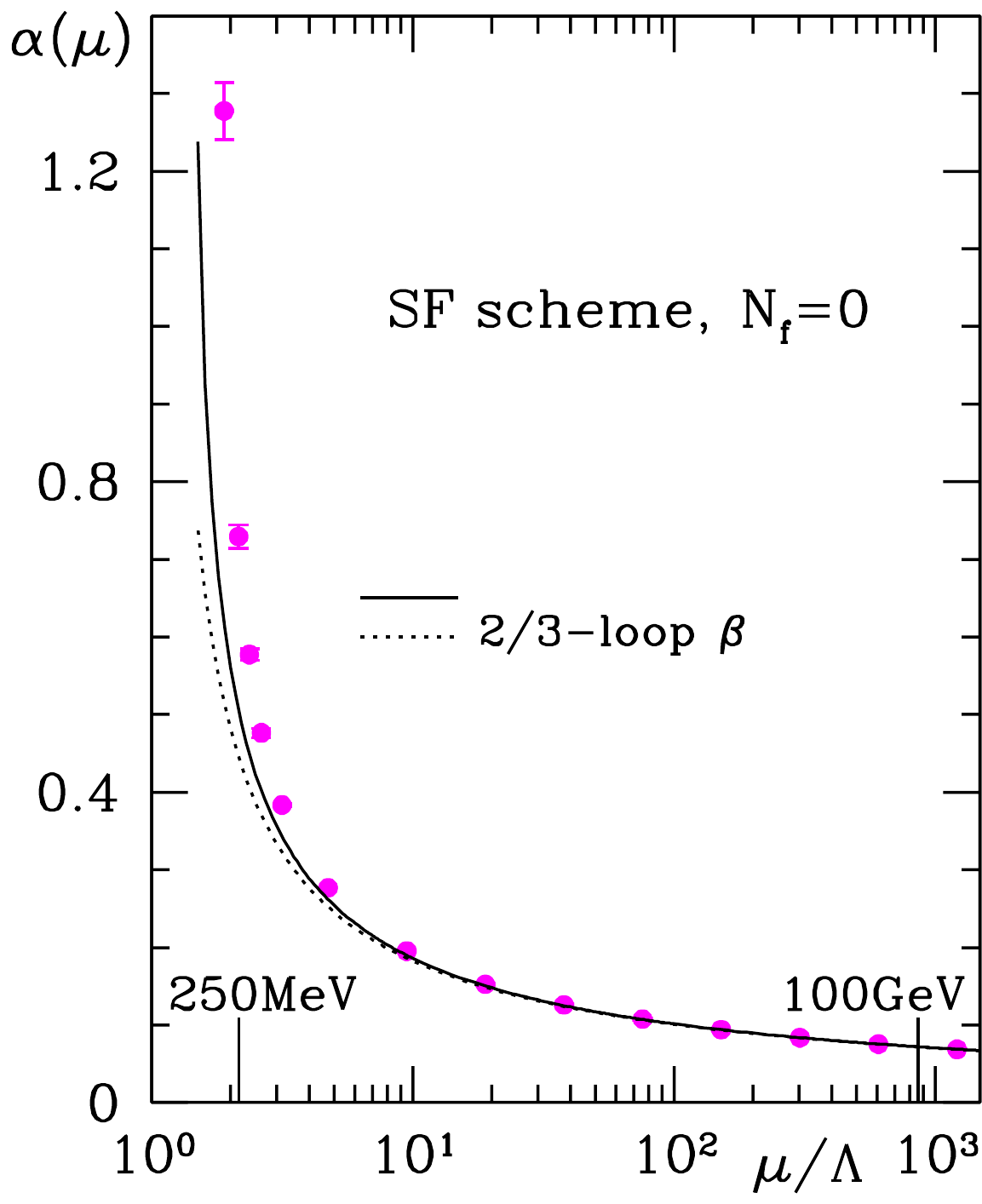}
  \caption{Running coupling for $\nf=2$ compared to 
  $\nf=0$. Figure from \cite{nara:rainer} 
  based on results of  
\protect\cite{Luscher:1993gh,Capitani:1998mq,Heitger:2001hs,Bode:2001jv,DellaMorte:2004bc}.
}
\label{f:running02}
\end{figure*}
The continuum extrapolated step scaling function can finally be iterated to construct
the non-perturbative running coupling for a number of scale arguments in \fig{f:running02}.
We will see that changing the number of quark flavors from $\nf=0$ to $\nf=2$ 
does not induce any qualitative changes. 
The connection from low to high energies is rather
smooth and perturbation theory can be trusted in
the \textSF\ schemes rather precisely at energies
$\mu \gtrsim 50\,\GeV$ or larger.
For the purpose of the comparison we show side 
by side $\nf=0$ results from \cite{Capitani:1998mq}
where the methods were developed and $\nf=2$ results.

With the continuum step scaling function $\sigma(u)$ under our control
for a range $u\in(0,u_\mathrm{max})$ we can connect the associated scales
$\Lambda_\mathrm{SF}$ and $\Lmax$ implicitly defined by the condition
\begin{equation}
\gbar^2(\Lmax)=u_\mathrm{max}=4.484.
\end{equation}
The result, including a complete error analysis, is \cite{Fritzsch:2012wq}
\begin{equation}
\Lmax \Lambda_\mathrm{SF}^{(2)}=0.264(15).
\end{equation}
In terms of the semi-phenomenological scale 
$r_0\approx 0.49$fm the estimate $\Lmax \approx 0.39$fm can be given.
This scale is defined \cite{Sommer:1993ce}  
in terms of the force $F$ between static quarks by solving
\begin{equation}\label{r0def}
r_0^2 F(r_0)=1.65.
\end{equation}
It is loosely connected to quarkonia models and has in addition been
related to other observables in previous lattice simulations.
In the next section more direct connections of $\Lmax$ to phenomenology
will be cited.

\subsection{Verification of asymptotic freedom}

Asymptotic freedom is
normally taken for granted for QCD with any $0 \le \Nf \le 6$
and it is certainly a self-consistent
property in perturbation theory.
Once we have some control beyond perturbation theory
we should however remember this situation and analyze
our data under this aspect.

From the continuum limit SSF we may form
the combination
\bes
  b_0^\mathrm{eff}(u)  =  \frac{\sigma(u) - u}{2 \log 2\,u^2 }\,.
\ees
It is expected to extrapolate
to the perturbative coefficient $b_0$  at small
$u$, with an asymptotic approach that is linear in $u$. 
In \fig{f:nfdep}, where we include
additional data
for $\nf=0,2,3,4$, we do see the expected behavior.
This provides a clear non-perturbative confirmation of
asymptotic freedom including the universal perturbative $\Nf$
dependence at small coupling.

In perturbation theory asymptotic freedom is lost 
beyond $\nf=16$ because $b_0$ changes its sign. 
We just remark in passing that we see a considerable effort 
to determine SSFs for $\nf=8 \dots 12$ where one
expects an almost vanishing $\beta$-function and approximate
conformal invariance that may be of phenomenological interest
for models of the technicolor variety.
Many of these
investigations use the methods described in this article,
but in a much more difficult situation.

\subsection{Running quark mass\label{runmass}}
\begin{figure*}[t!]
  \centering
  \includegraphics[width=0.51\textwidth]{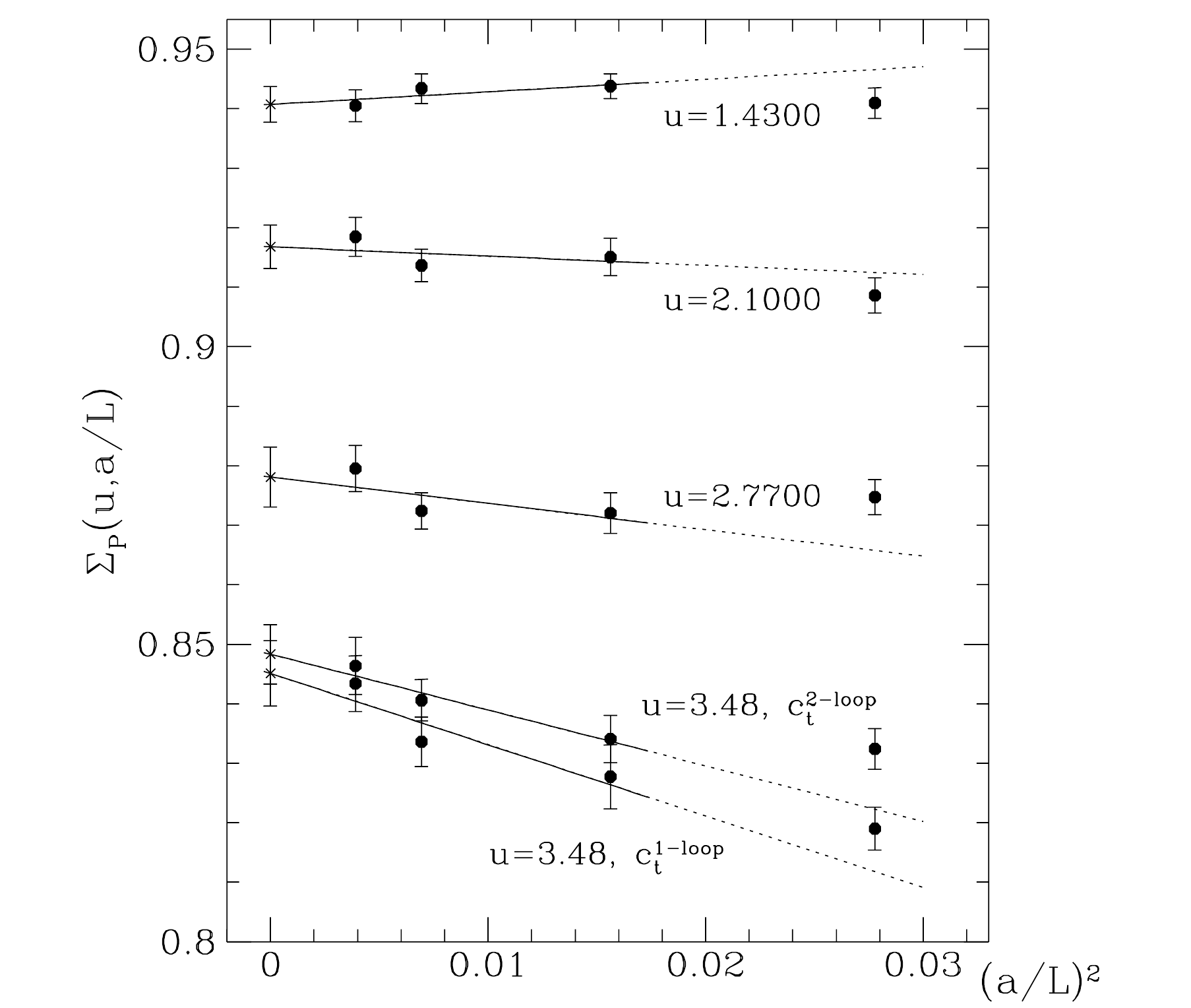} \hfill
  \includegraphics[width=0.45\textwidth]{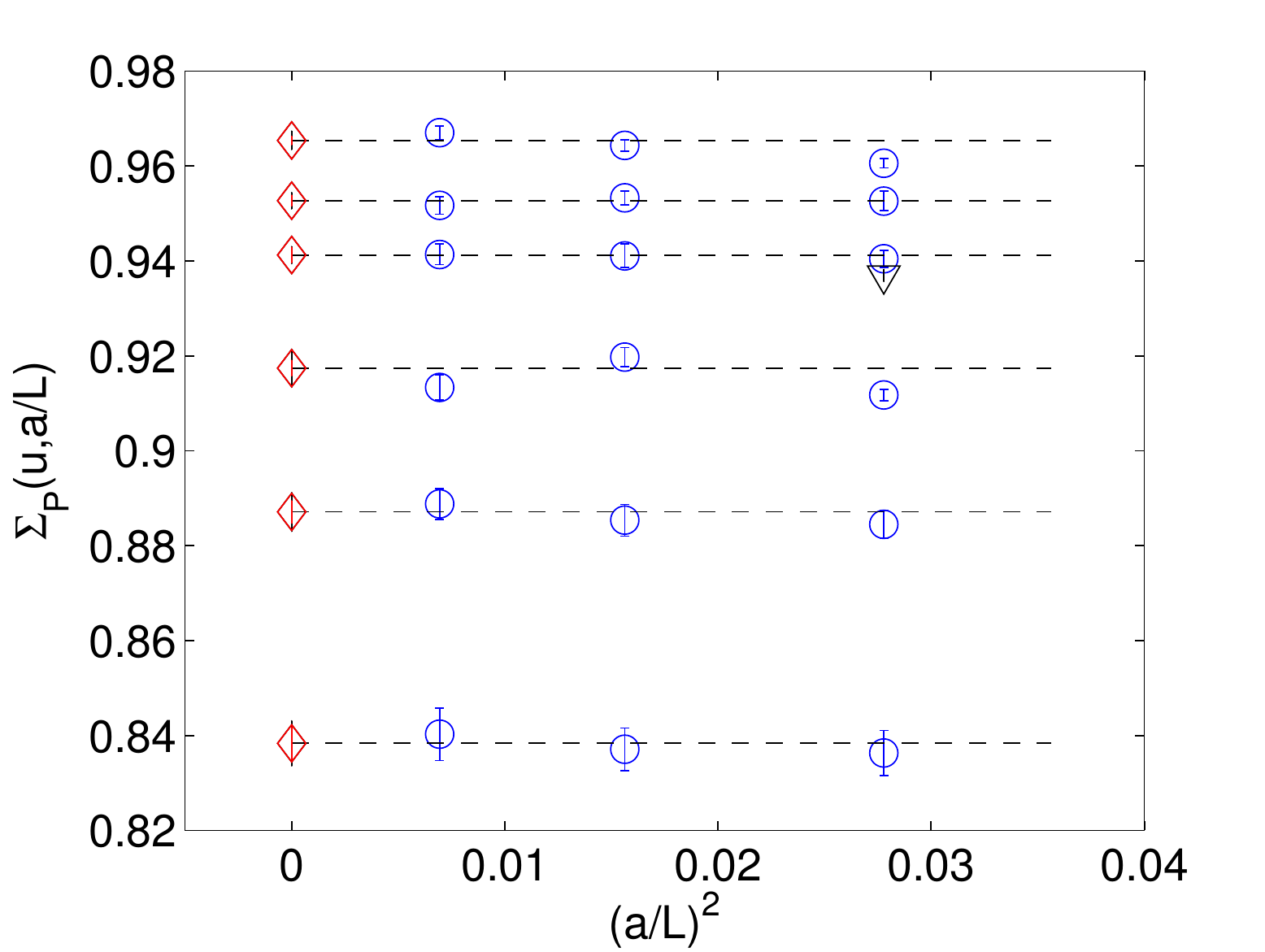}
  \caption{Continuum extrapolations of the step scaling function $\Sigma_\mathrm{p}$ in
  the quenched approximation (left)~\cite{Capitani:1998mq} and 
   with $\nf=2$ dynamical quark flavors~\cite{DellaMorte:2005kg}. In the right graph 
the coupling $u$ ranges from $u=0.979$ to $u=3.33$. 
}
\label{f:Sigmap}
\end{figure*}
We now mention also some result for the running of the quark mass
which is actually intertwined with the evolution of the coupling constant.
The scale dependence of the running mass $\mbar(L)$ in (\ref{mbardef}) derives form the
$L$ dependence of $\zp$ in (\ref{ZPnorm}).
The renormalization group equation for the mass reads
\begin{equation}
\mu \frac{d\mbar}{d \mu} = \tau(\gbar) \mbar,\quad \mu=L^{-1}
\end{equation}
with $\tau(g)=-8g^2/(4\pi)^2+\mathrm{O}(g^4)$ in perturbation theory.
Note that this scale evolution is coupled with the one of $\gbar(\mu)$ in (\ref{CSeq}).
To have non-perturbative control also here, we define another step scaling
function
\begin{equation}
\Sigma_\mathrm{P}(u,a/L)=\left. \frac{\zp(g_0,2L/a)}{\zp(g_0,L/a)}\right|_{u=\gbar^2(L)}.
\end{equation}

\begin{figure*}[t!]
  \centering
  \includegraphics[width=0.47\textwidth]{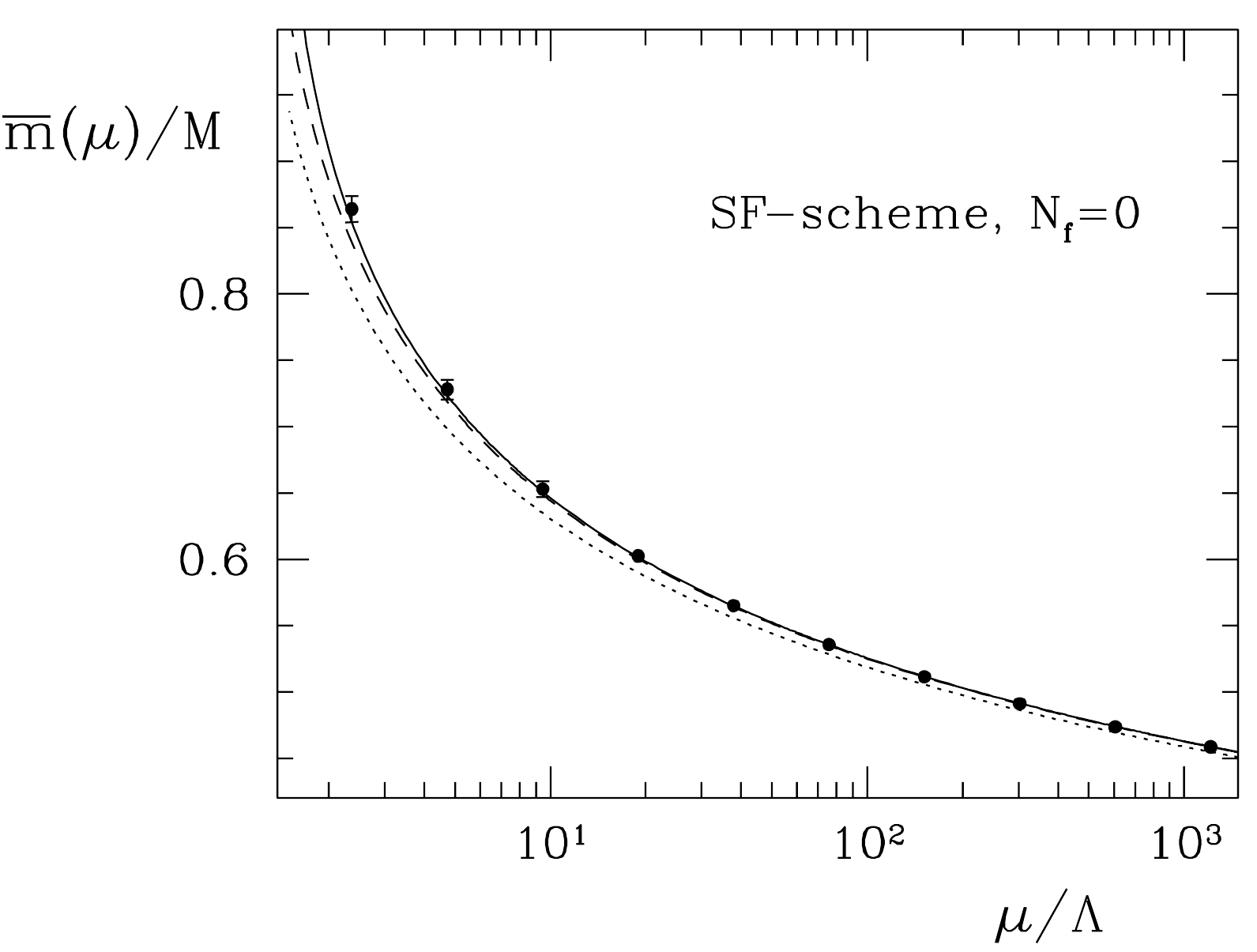}\hfill
  \includegraphics[width=0.45\textwidth]{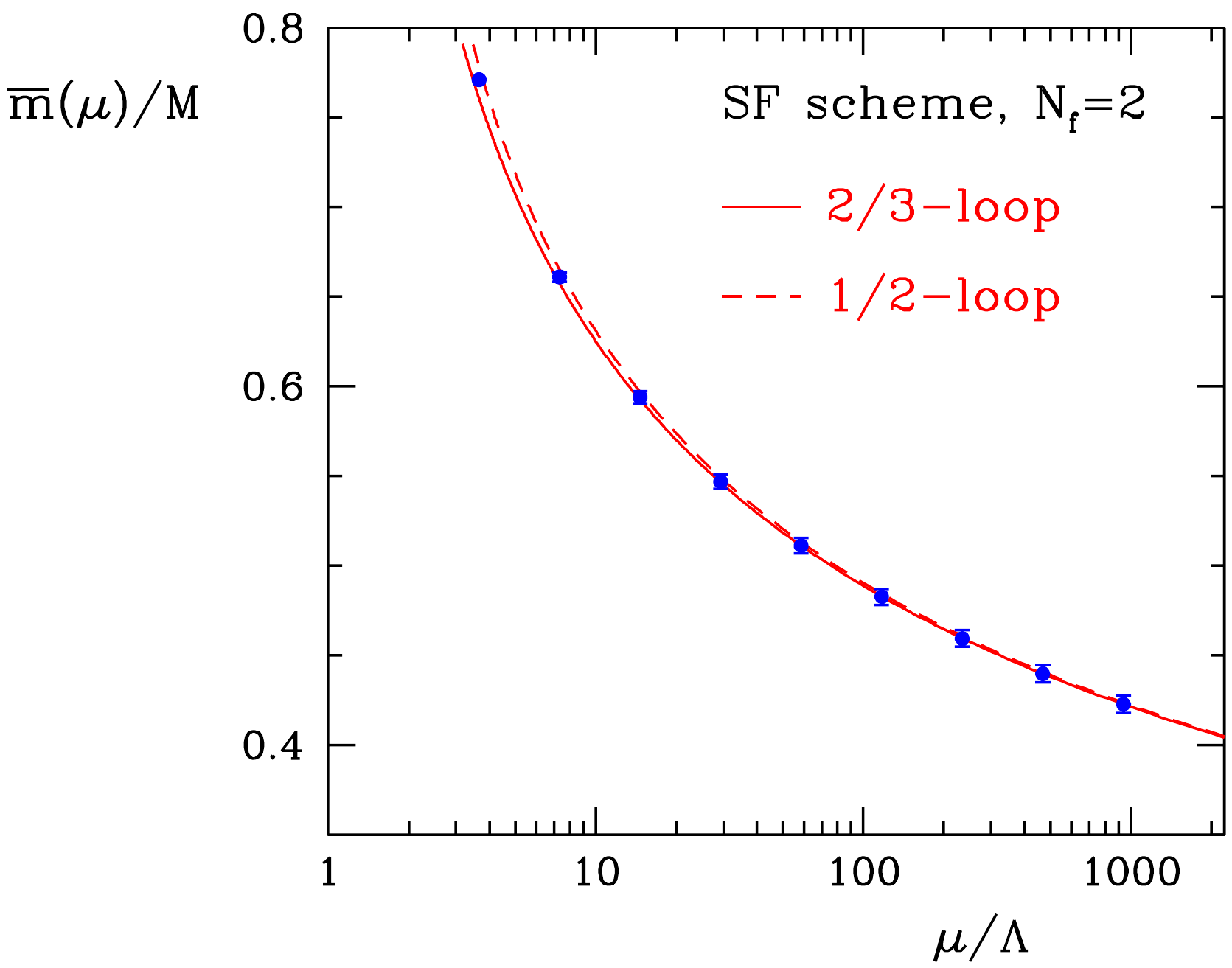}
  \caption{Scale dependence of the quark mass $\mbar$
   in the quenched approximation (left)\cite{Capitani:1998mq} and 
   with $\nf=2$ dynamical quark flavors\cite{DellaMorte:2005kg}.
   For $\nf=0$ the dotted, dashed and solid curves are 
obtained from eqs.~(\ref{Lambda}) and
(\ref{M}) using the $1/2$-, $2/2$- 
and $2/3$-loop expressions
for the $\tau$- and $\beta$-functions respectively.
}
\label{f:runningmass}
\end{figure*}

Also the renormalization group equation for the mass can be converted to an equivalent
integral equation
\bes
\label{M}
  M &=& \mbar\,(2 b_0\gbar^2)^{-d_0/2b_0} 
  \times
  \\&& 
  \exp\left\{-\int_0^{\gbar}\mathrm{d} g 
  \left[{\tau(g)\over\beta(g)}-{d_0\over b_0g}\right]\right\}, \nonumber
\ees
where $M$ is the scale independent RGI mass, that is the analogue to the $\Lambda$
parameter in the case of the coupling. The perturbative approximations of $\tau$ and $\beta$
may be used at very high energy (small $L$ in the SF) to compute $M,\Lambda$ from
$\mbar(L)$ and $\gbar(L)$. After extrapolating to the continuum limit
\begin{equation}
\sigma_\mathrm{P}(u)=\lim_{a/L\to 0} \Sigma_\mathrm{P}(u,a/L)\,,
\end{equation}
see \fig{f:Sigmap},
the continuum running mass was constructed. 
It is shown in \fig{f:runningmass}.

\subsection{Optimized strategy}
\label{s:opt}

\begin{figure}[t!]
  \centering
  \includegraphics[width=0.49\textwidth]{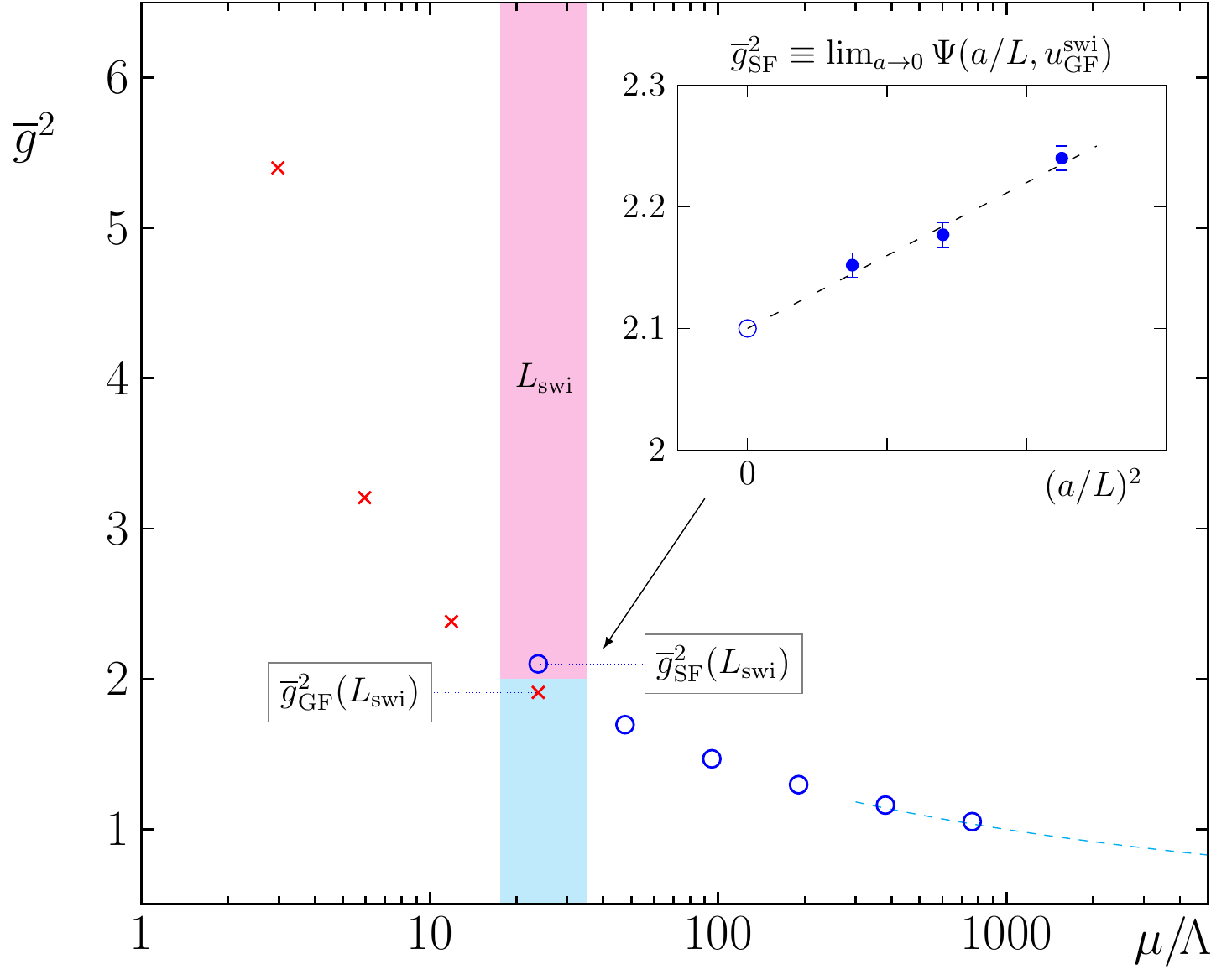}
  \caption{Sketch (no real data) of the overall strategy being employed
  for $\nf=3$ \cite{Patrick_lat14}.
}
\label{f:sketch}
\end{figure}

In \fig{f:nfdep} we have already shown some results for the
running of the SF coupling with $\nf=3$.
The goal in the present stage of the project is to improve the precision 
at the same time as having this more realistic number of flavors.
As already mentioned at the beginning of section \ref{flow}
the combination of SF and GF coupling is promising for the precision issue. 
In \fig{f:sketch} we show a sketch
of the overall strategy to use their complementarity and
combine the two couplings. We note however,
that here we do not yet see real data!

The presently quite different size of the discretization errors in the two observables are seen
by comparing \fig{f:Universality1} and
\fig{f:ssfGF}. We hope that the improved flow equation
(\ref{latflowi}) and the other optimizations reviewed in that subsection
will still considerably accelerate the continuum limit of the GF coupling.

\begin{figure}[t!]
  \centering
  \includegraphics[width=0.49\textwidth]{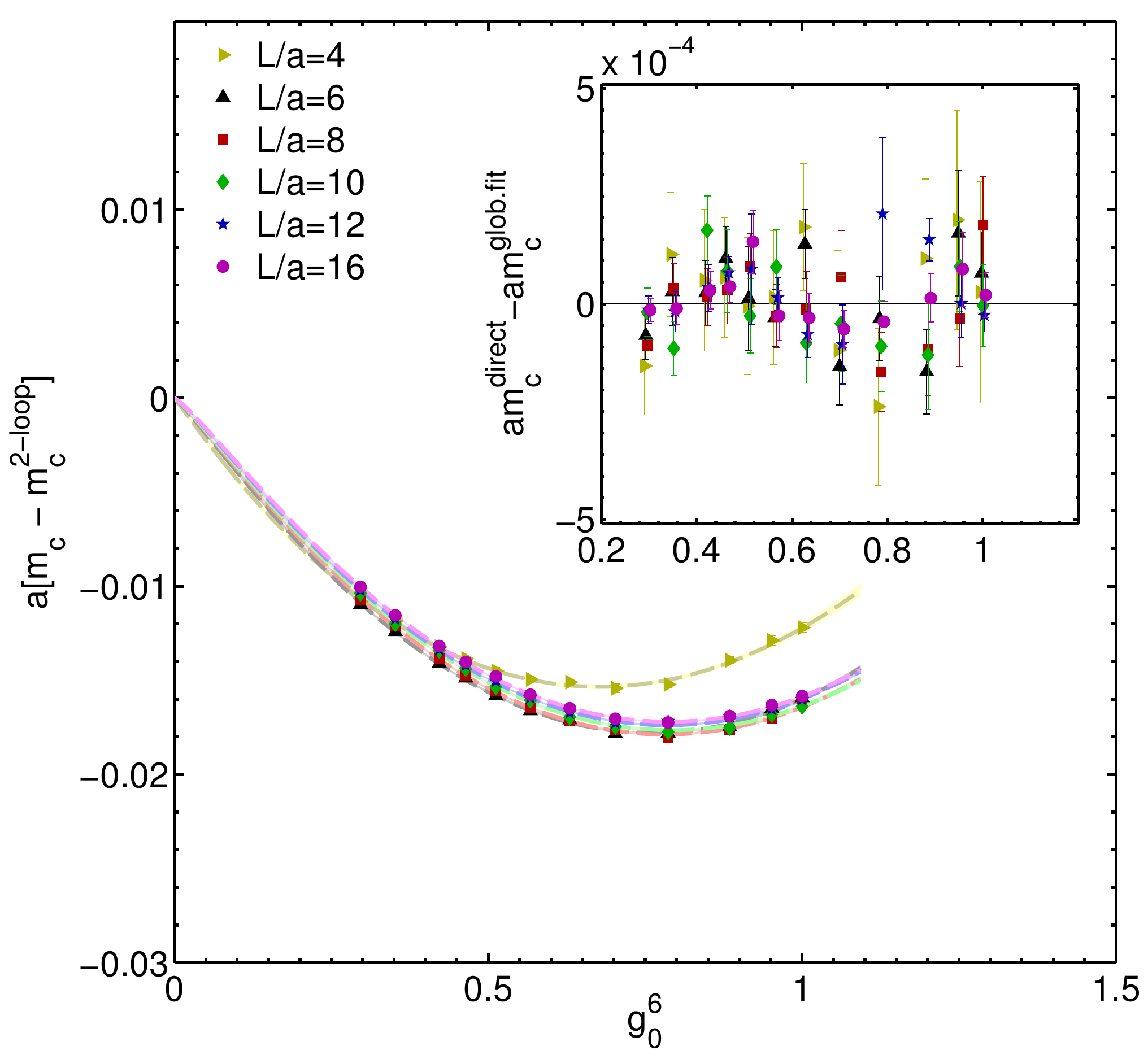}
  \caption{Determination of the critical lines for
  various values of $L/a$. Note that a big part, the 2-loop expression, is subtracted from the data. Figure from \cite{Patrick_lat14}.
}
\label{f:mcrit}
\end{figure}

Increasing the precision also requires a more precise tuning
to the massless theory than before. This is at the same time desirable for
other projects such as HQET, where the precise knowledge
of the relation between $\gbar$ and $L$ is used to 
perform simulations in a volume of fixed physical size.
For SW improved Wilson fermions that we use this means
that the critical mass in lattice units $a m_c(g_0)$ has to be
determined to a many digit precision in the relevant range of couplings. 
For the Wilson gauge action such a
result is given in \fig{f:mcrit}. For future simulations these data are
represented by a smooth interpolating fit formula
shown as lines in \fig{f:mcrit}.



\section{Large volume simulations}
In this section we review the completed results for $\Nf=2$
and some data of the ongoing $\Nf=3$ simulations.
Quenched computations from the nineties were an extremely useful
preparation but will be skipped here.

\subsection{Algorithmic issues}
Here simulations are described that have been run for $\nf=2$ 
improved Wilson quarks
within the Coordinated Lattice Simulation (CLS) 
consortium.

The algorithmic framework is mostly as discussed in section \ref{algo}
where it has become clear that the frequent inversion of the lattice Dirac
matrix is the dominant numerical task. It is in particular required
to evaluate the fermionic driving force for the molecular dynamics evolution 
in HMC.

\begin{figure*}[t!]
\subfigure{}{
\includegraphics[scale=0.29,angle=270]{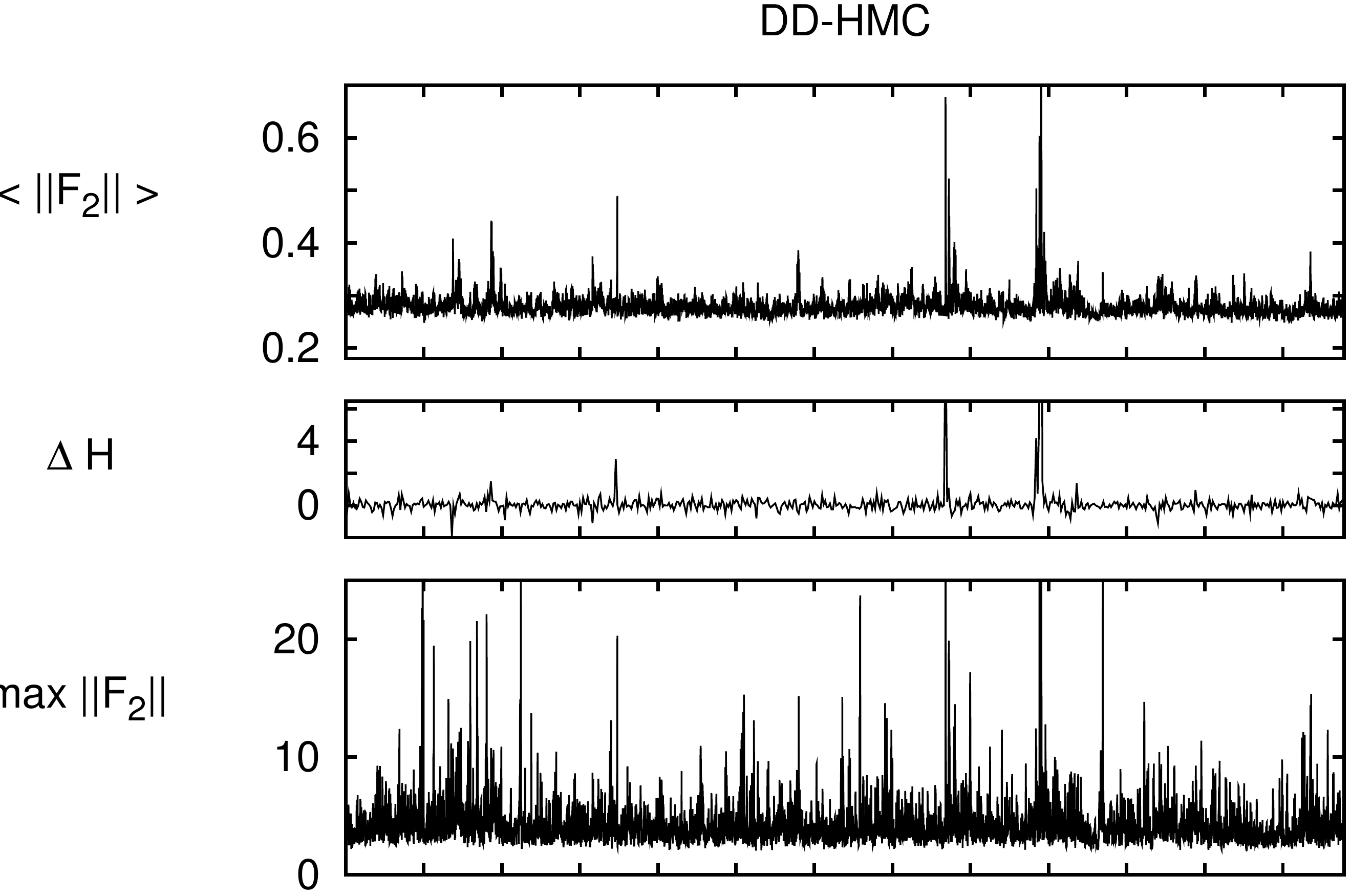}
}
\subfigure{}{
\includegraphics[scale=0.29,angle=270]{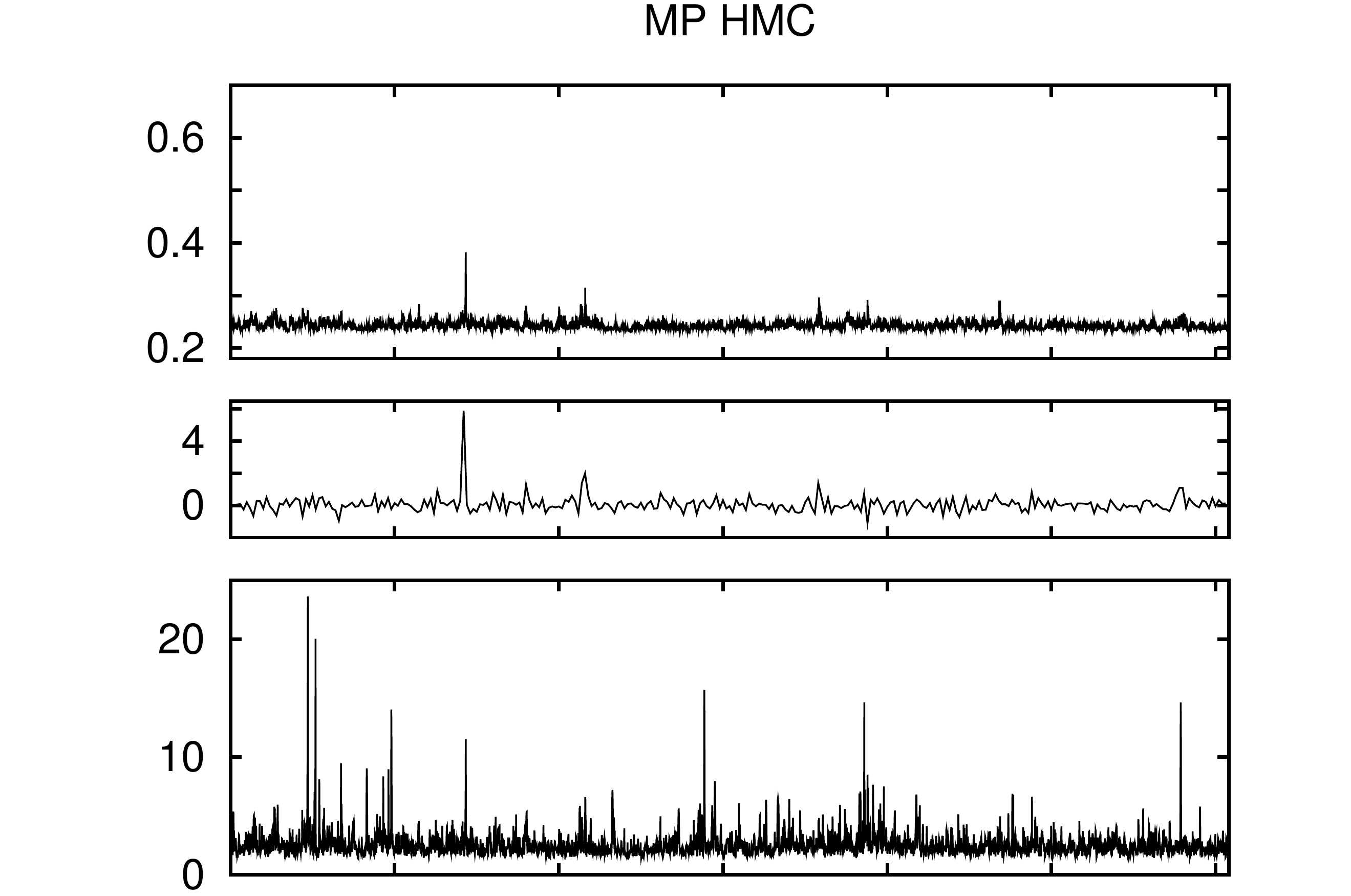}
}

\subfigure{}{
\includegraphics[scale=0.29,angle=270]{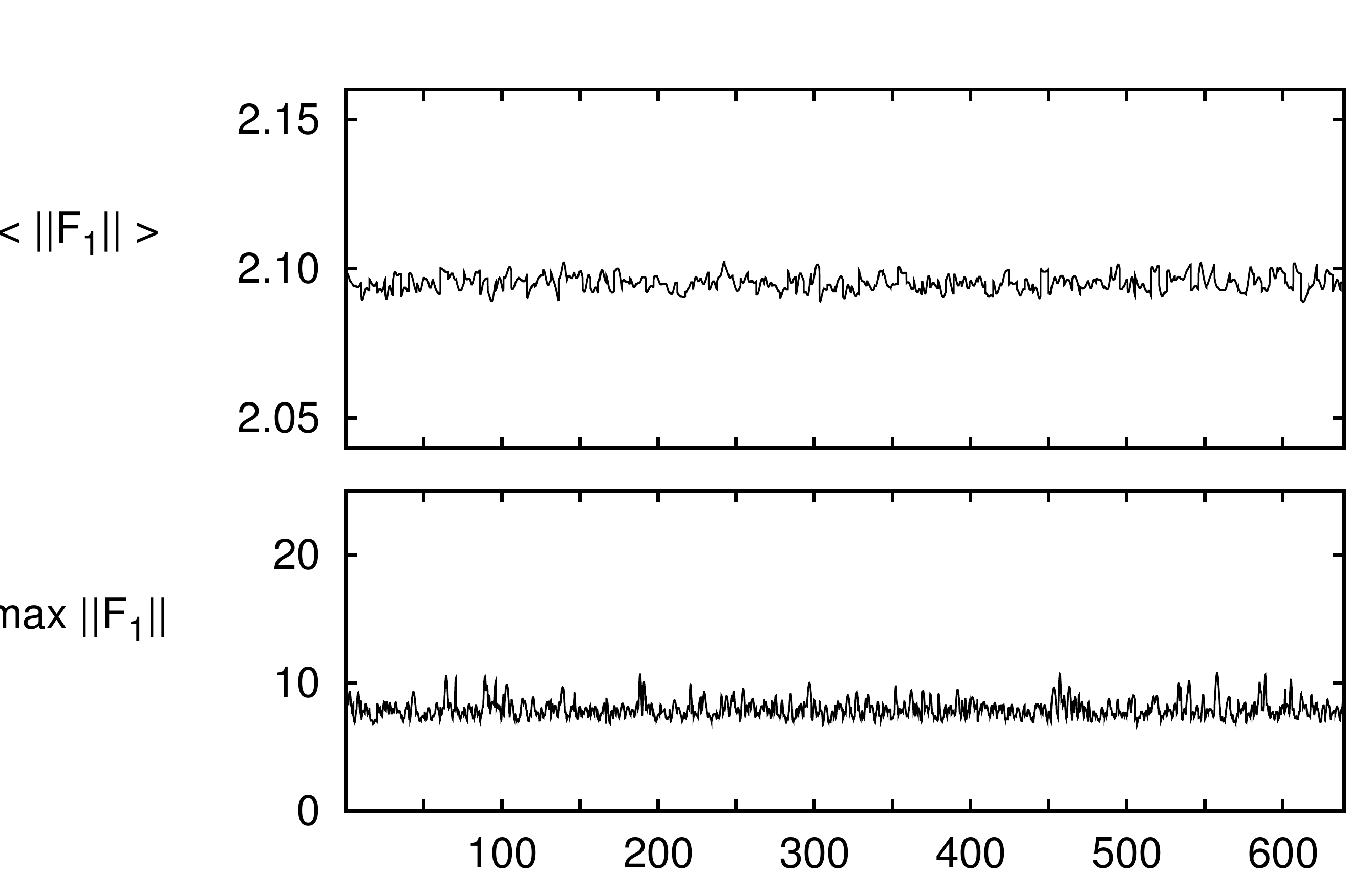}
}
\subfigure{}{
\includegraphics[scale=0.29,angle=270]{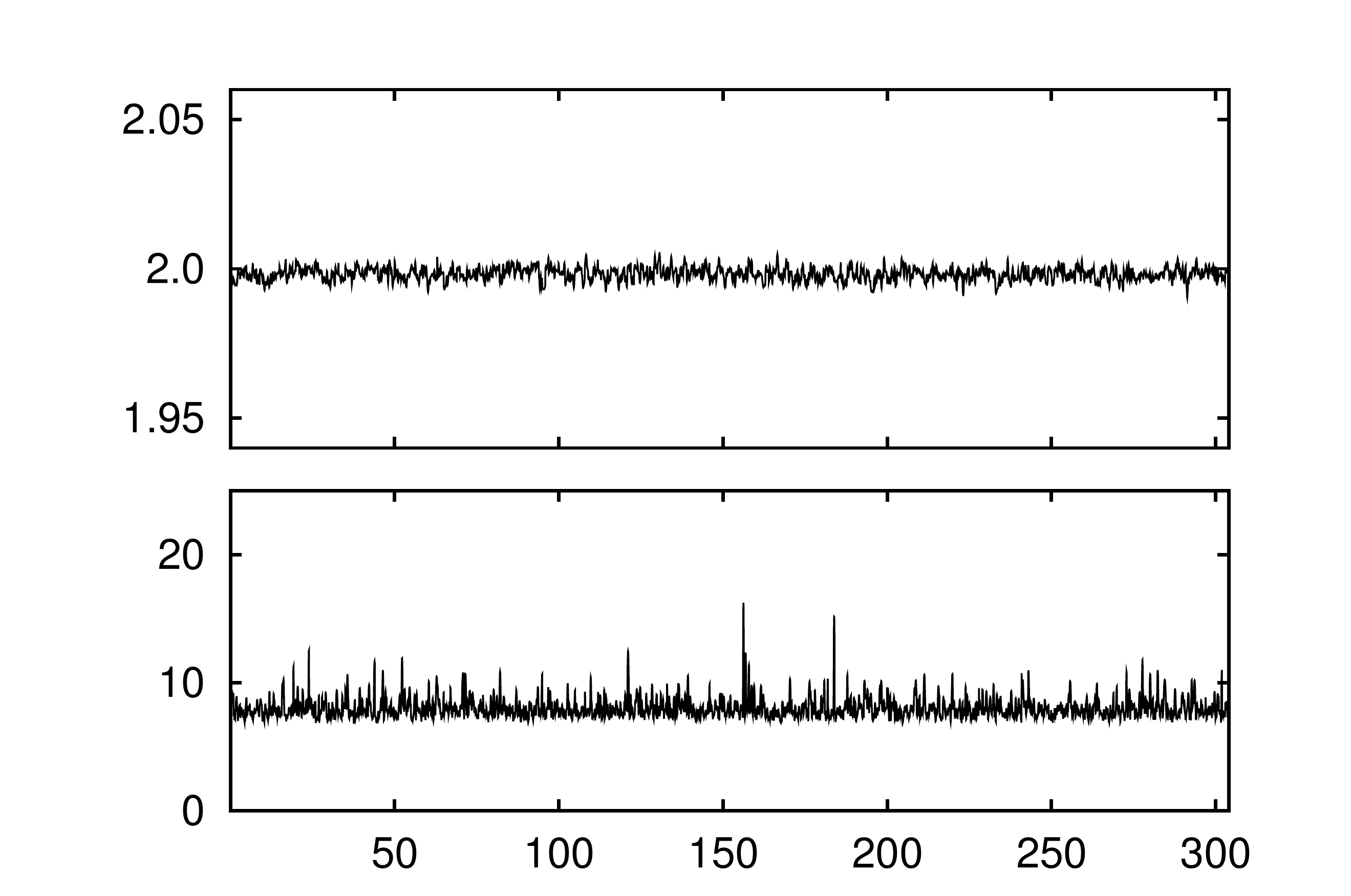}
}
\caption{Histories of the energy violation $\Delta H$, as well as maximum and average forces $F_2$ and $F_1$, for each
force update, plotted as a function of the trajectory number. Values corresponding to the DD-HMC 
algorithm are shown to the left and the integration step-sizes for the two forces relate as 
$\Delta t_2 : \Delta t_1 = 1:6$. The values for MP-HMC are shown in the two right panels and 
the corresponding ratio of the integration steps is $\Delta t_1 : \Delta t_2 = 1:10$. The lattice 
size is $48\times24^3$ and $\kappa_\mathrm{sea}=0.13625$.}
\label{f:forces}
\end{figure*}
Before we have switched to the more efficient
Hasenbusch preconditioning (see subsection \ref{Hasenbusch}),
the domain decomposed DD-HMC algorithm \cite{Luscher:2005rx} had been used initially.
This at first very promising variant was ultimately dismissed
because of inferior autocorrelations and critical slowing down,
but we here still look at some data produced with DD-HMC for
comparison and for historical reasons.
In  \fig{f:forces} we see several histories in molecular dynamics time,
where the left plots refer to DD-HMC simulations
and the right ones (MP-HMC) to Hasenbusch mass preconditioning.
Components of the fermion forces from two separate pseudofermions are
shown and the Hamiltonian violation $\Delta {\cal H}$ that enters
into the accept step. The spike-like fluctuations that one sees point 
to algorithmic problems due to nearly singular Dirac matrices.
With them present, one is restricted to small step sizes
and finds large iteration numbers for the Dirac solvers which
both damage the efficiency. The plots demonstrate the
smoother running of the MP-HMC after tuning its parameters
in a reasonably close to optimal way \cite{Marinkovic:2010eg}.
After detailed studies MP-HMC was adopted as the method of choice.

\begin{figure*}[t!]
\subfigure{}{
\includegraphics[scale=0.2,angle=270]{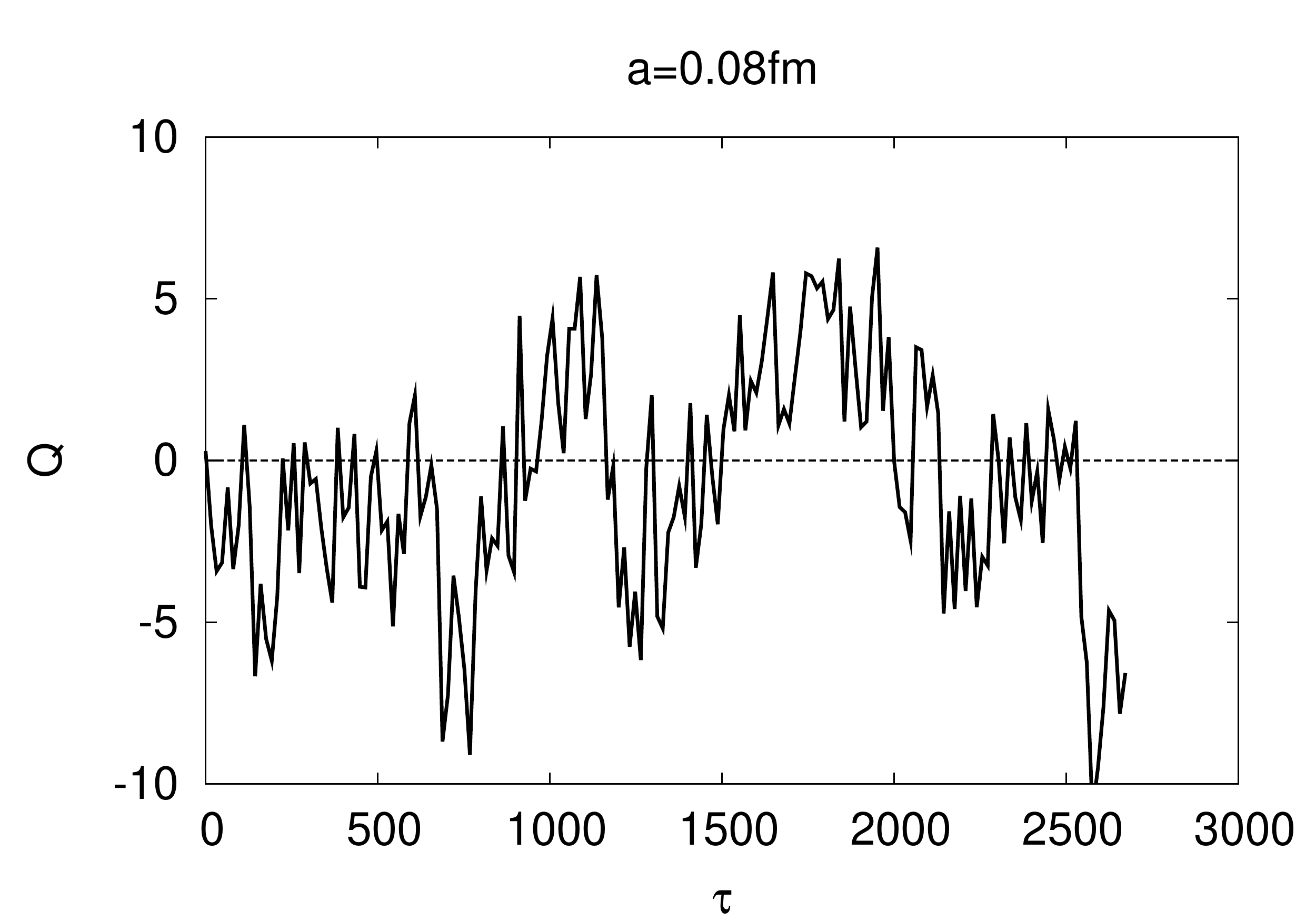}
}
\subfigure{}{
\includegraphics[scale=0.2,angle=270]{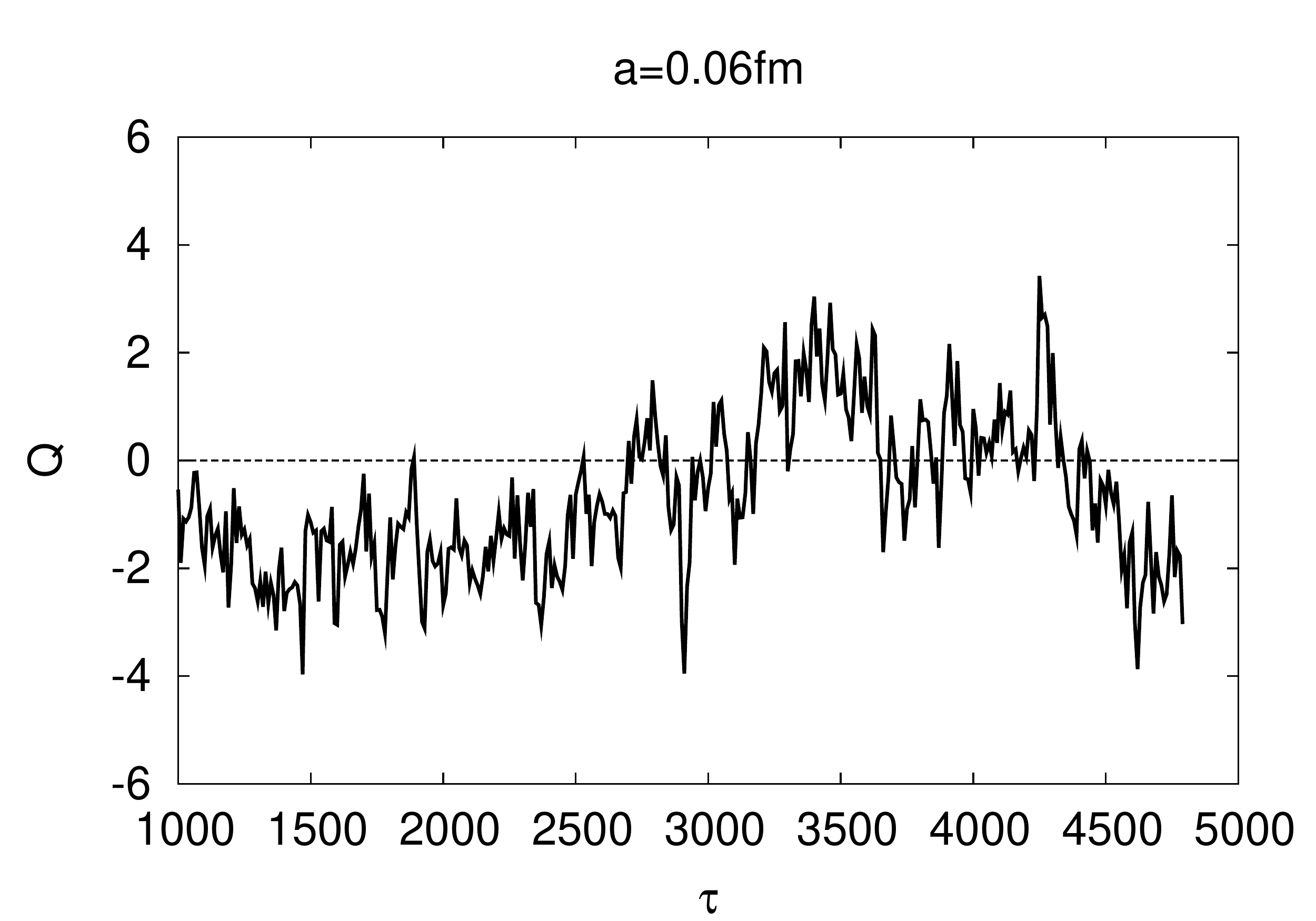}
}
\subfigure{}{
\includegraphics[scale=0.2,angle=270]{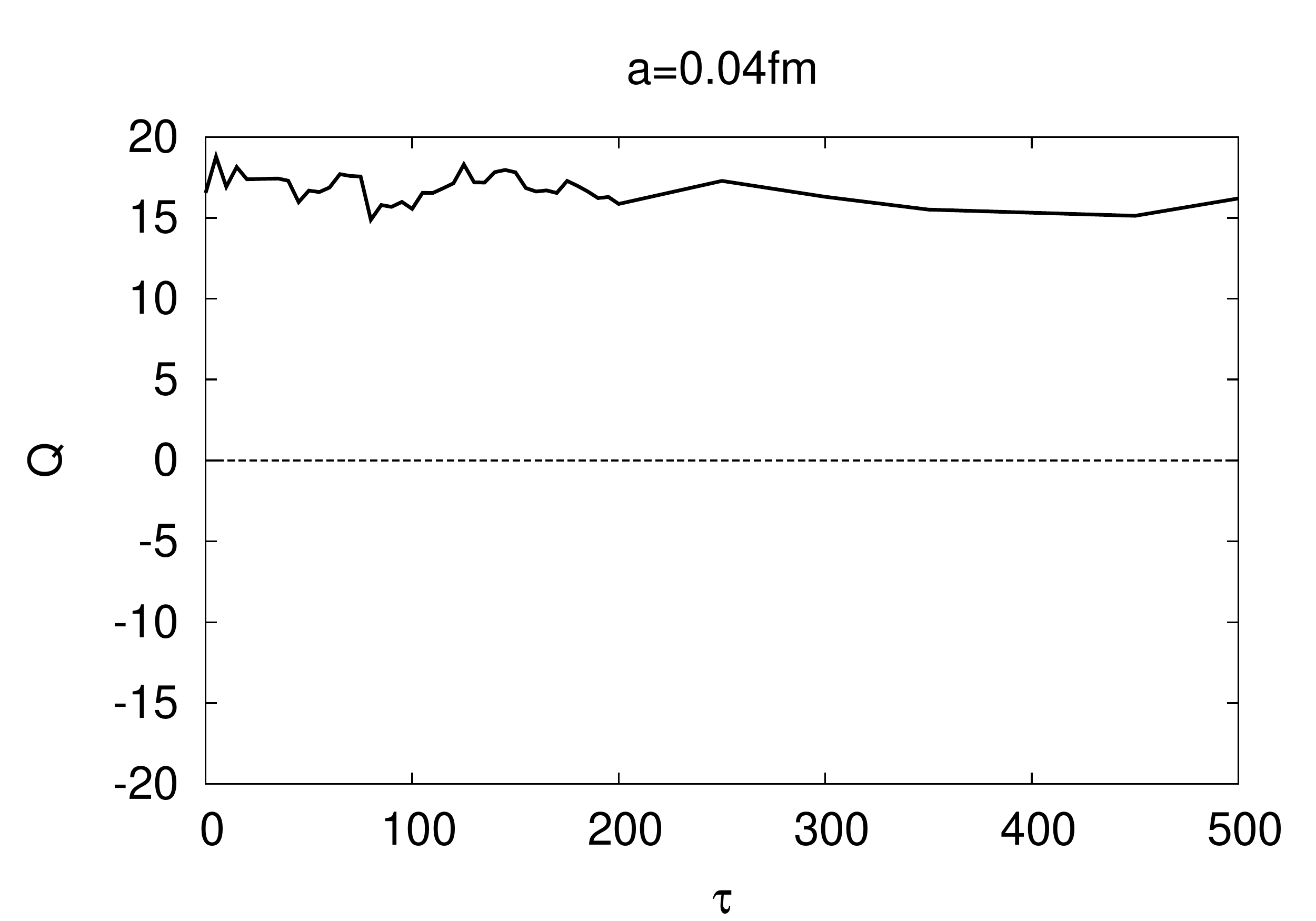}
}
\caption{Histories of an estimate of the topological charge on runs with $\nf=2$~\cite{Schaefer:2009xx}.}
\label{f:slowingdown}
\end{figure*}

Topological freezing (see subsection \ref{topfreeze}) 
was the other problem that significantly held back the completion
of large volume $\nf=2$ simulations.
In the history of the topological charge in
\fig{f:slowingdown}
one sees that the simulations are slowing down dramatically as the lattice spacing is reduced.
The observed structures extend over a non-negligible fraction of the length of a typical run.
On the other hand, studying the effect of the slow modes of the 
Monte Carlo algorithm on typical hadronic correlation
functions, it was found that these receive only 
suppressed contributions to their autocorrelation functions. 
Therefore, for $a\approx 0.05$ fm the
standard error computation could still be adapted to the situation 
\cite{Schaefer:2010hu,Schaefer:2009xx}. 
Conservative  error estimates are still possible 
with MC histories which are of the order of 20-100 
times the slowest relaxation time of the system, $\tau_\mathrm{exp}$.

%
%
%
%
\subsection{General considerations for scale setting }

\label{s:scale}

\newcommand{\bi}{\begin{itemize}}
\newcommand{\ei}{\end{itemize}}

\def\rmd{\mathrm{d}}
\def\tr{\mathrm{tr}}
\def\Rho{{\cal R}}
\newcommand{\mprot}{m_\mathrm{p}}
\newcommand{\Mprot}{M_\mathrm{p}}
\def\prot{\mathrm{p}}
\def\mlat{m^\mathrm{lat}}
\def\quark{\mathrm{quark}}
\newcommand{\sect}[1]{Sect.~\ref{#1}}
\newcommand{\Sect}[1]{Section~\ref{#1}}
\def\sm{S^\mathrm{m}}
\def\sa{S^\mathrm{a}}
\def\metas{m_{\eta_s}}
\def\texp{\tau_\mathrm{exp}}
\def\tauexp{\tau_\mathrm{exp}}
\newcommand{\mpi}{m_{\pi}}
\newcommand{\fpi}{f_{\pi}}
\newcommand{\fk}{f_\mathrm{K}}
\newcommand{\mk}{m_\mathrm{K}}
\def\vud{V_\mathrm{ud}}
\def\vus{V_\mathrm{us}}
\def\expe{{\cal E}}
\def\tauexp{\tau_\mathrm{exp}}
\def\tauint{\tau_\mathrm{int}}
\def\tmc{t_\mathrm{MC}}
\newcommand{\simas}[1]{\raisebox{-.1ex}{
            $\stackrel{\small{#1}}{\sim}$}}
\newcommand{\eq}[1]{eq.~(\ref{#1})}
\newcommand{\Eq}[1]{Eq.~(\ref{#1})}
\newcommand{\eqs}[1]{eqs.~(\ref{#1})}
\newcommand{\Eqs}[1]{Eqs.~(\ref{#1})}



\begin{figure*}[ht!]
\centering
   \includegraphics[width=0.83\textwidth]{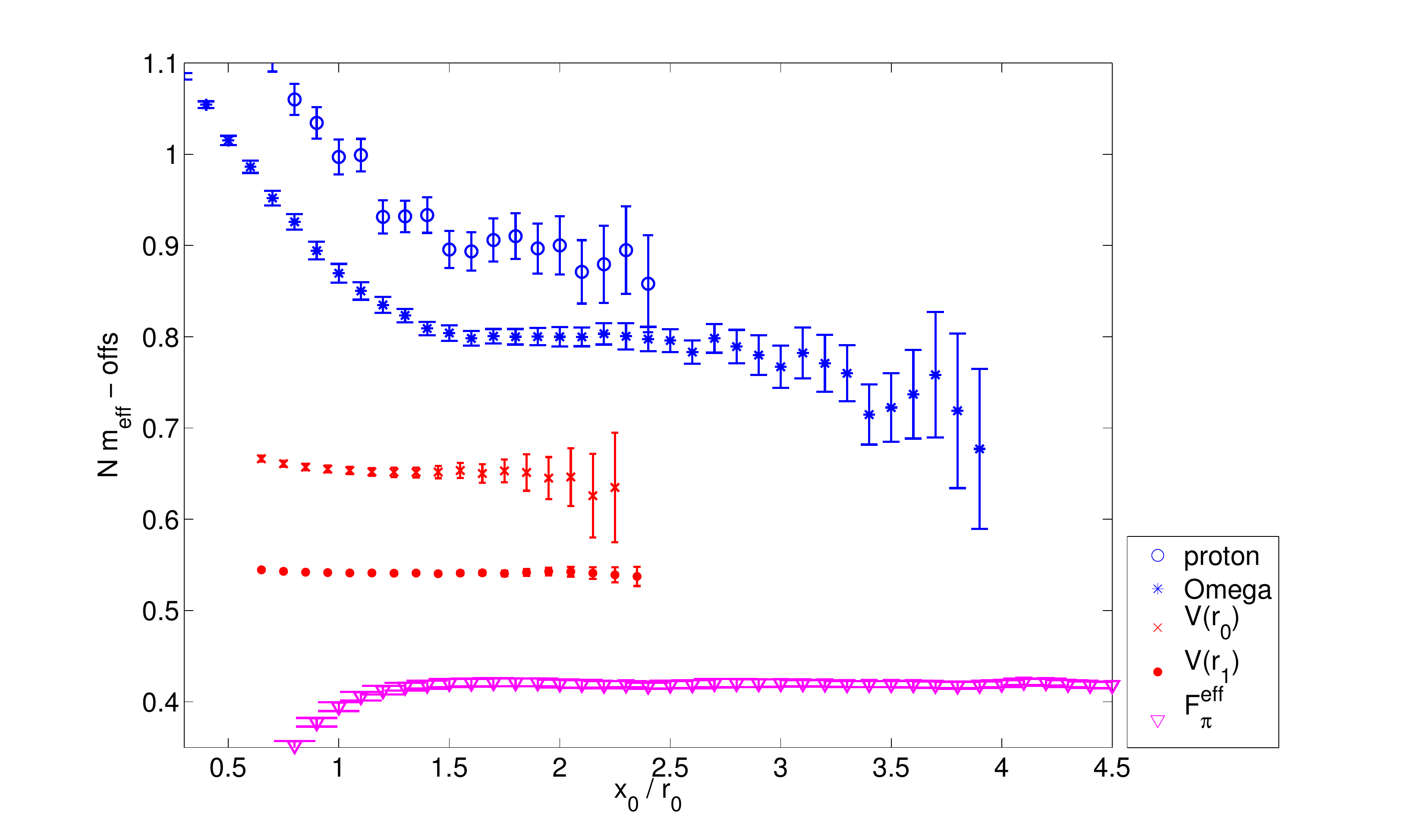}
   \vspace*{-2mm}
\caption{Effective masses for 
$m_\prot$ \cite{Jager:2013kha},
$m_\Omega$ \cite{Capitani:2011fg},
$V(\approx r_0)$, $V(\approx r_1)$~\cite{Fritzsch:2012wq} and $\fpi$~\cite{Lottini:2013rfa} on CLS ensemble N6 (see \cite{Lottini:2013rfa}).
All effective ``masses'' have been scaled 
such that the errors in the graph reflect
directly the errors of the determined scales. They have been shifted vertically. Figure from \cite{Sommer:2014mea}.
\label{f:plateaux} 
}
\end{figure*}

In the previous section the energy scales from the perturbative end
down to the implicitly defined hadronic scale $\Lmax$ has been covered
for coupling and quark mass running. It remains to connect
$\Lmax$ to a measurable quantity to bring in MeV units which we call
`scale setting'.
After the general discussion in section \ref{sec_hadronic} in principle,
if lattice QCD had all dimensionless scale ratios right, any observable would
be suitable and the mass of the stable proton would appear particularly natural.
In reality, however, many more mundane practical considerations are
essential \cite{Sommer:2014mea} to not give away too much precision in this step.

An obvious demand is that the quantity used for scale setting should be computable
on the lattice with good statistical precision (for a given computational effort)
and that it should have small systematic errors like cutoff and finite volume effects.
The latter would otherwise contaminate all quantities cited in MeV even if in lattice
units they themselves are well controlled. Another point is that it is desirable to use
a scale that is not very sensitive to the precise quark content and to the tuning of the masses
of included sea quarks to their physical values. This is obviously only possible to
a limited precision and in addition, for algorithmic reasons, the up/down quarks
in simulations are most of the time heavier than in Nature and a chiral extrapolation
is invoked. This step is theory guided by chiral perturbation theory, but it is clearly good
to keep the general scale as independent of this as possible. In addition the effective
theory with $\Nf<6$ is an approximation only and scale setting should be 
carried out in the sector of the theory that is
robust against these presently still unavoidable small errors.

With this said, estimates for a number of quantities that are considered by the community
are displayed in \fig{f:plateaux}. The plot includes the effective masses of the proton and the
$\Omega$ particle made of three strange quarks (still quenched for $\Nf=2$). We see the relatively
large errors combined with short plateaus, where the situation for the $\Omega$ is somewhat
superior. The static quark potential $V(r_0)$ is closely related to $r_0$ in (\ref{r0def}) and
$V(r_1)$ is a similar quantity where 1.65 is replaced by 1.00.
The lowest line with a very convincing plateau with small errors represents an effective
matrix element corresponding to the decay constant $f_{\pi}$.
A further bonus of such quantities is that they have been confirmed to be only weakly
coupled to the slow modes responsible for topological freezing. Therefore a reliable
error estimation is still possible down to the smallest lattice spacings $a\approx 0.05$ fm entering here.
A small draw back associated with decay constants has to be mentioned too: to cite
a physical value for them based on physical decays,
 separate experimental input for the relevant CKM matrix elements has to be used.

What has been said about $f_{\pi}$ is also true for the Kaon decay constant $\fK$.
It is our preferred scale setting quantity in this study, because of two more bonuses.
The chiral extrapolation from larger pion masses in the range of 
around $500\, \MeV$ to $270\, \MeV$
down to the physical value
is simpler for $\fK$. In this case the partially quenched variant of
chiral perturbation theory (pqChPT) is used.
A purely technical point to prefer the K sector in our $\Nf=2$ 
simulations is that the strange
quark mass can here be varied after the run without having to generate new configurations.
Only a few extra inversions for the strange quark propagator are needed which is
a relatively small effort.

\begin{table*}
\small
\centering
\begin{tabular}{cccccccccc}
\toprule
      $L/a$ & $\beta$  & $\kappa_{\rm sea}$ & $am$            & $\gbar^2(L)$  & $\delta[\gbar^2]$  &   $\Lmax/a$ \\\midrule
        $4$ & $5.2000$ &  $0.134700$        & $-0.03745(41)$  & $3.730(11)$    \\
        $4$ & $5.2000$ &  $0.133780$        & $-0.00086(35)$  & $3.797(11)$    \\
        $4$ & $5.2000$ &  --                & $\to 0$         & $3.798(11)$ &   $-0.686$   &   $  5.11(3)(13)$    \\
        $6$ & $5.2000$ &  $0.135600$        & $-0.01322(26)$  & $4.810(32)$    \\
        $6$ & $5.2000$ &  $0.135200$        & $+0.00289(24)$  & $4.984(33)$    \\
        $6$ & $5.2000$ &  --                & $\to 0$         & $4.954(33)$ &   $+0.470$   &   $  5.33(4)(11)$    \\\cmidrule(lr){1-6}
        $6$ & $5.2638$ &  $0.135673$        & $+0.00012(19)$  & $4.550(25)$ &   $+0.066$   &   $  5.89(4)(2)$     \\
        $8$ & $5.4689$ &  $0.136575$        & $+0.00046(11)$  & $4.526(32)$ &   $+0.042$   &   $  7.91(7)(1)$     \\
       $10$ & $5.6190$ &  $0.136700$        & $+0.00038(8)$   & $4.531(51)$ &   $+0.037$   &   $  9.87(14)(2)$    \\\cmidrule(lr){1-6}\addlinespace[0.005cm]\cmidrule(lr){1-6}
       $12$ & $5.7580$ &  $0.136623$        & $+0.00067(7)$   & $4.501(91)$ &   $+0.017$   &   $ 11.94(31)(1)$    \\
       $16$ & $5.9631$ &  $0.136422$        & $-0.00096(4)$   & $4.40(10)$  &   $+0.084$   &   $ 16.40(50)(6)$    \\
\bottomrule
\end{tabular}
\caption{\footnotesize Values of $\Lmax/a$ after correcting the simulated values $L/a$ to match the target $\gbar$ from \cite{Fritzsch:2012wq}. The data at the two largest $\beta$-values
are from \cite{Blossier:2012qu}. The second error
         on the final result is the systematic one.}
\label{tab:gbsq-L-to-L1}
\end{table*}

\subsection{Scale setting for $\nf=2$}
\label{s:s2}

%
%
We start with a compilation of simulation data
in tables \ref{t:L1inter} and \ref{tab:gbsq-L-to-L1}.
The non-integer values for $\Lmax/a$ derive from interpolations, see \cite{Fritzsch:2012wq}
for more details.
\begin{table*}[t!]
\small
\begin{center}
\begin{tabular}{@{\extracolsep{0.2cm}}ccccc}
\toprule
$\beta$  & $\Lmax/a$ & $\Lmax\fk$ & $r_0/\Lmax$ & $\mbar_\strange/\fk$\\
\midrule
$5.2$    &$5.367(82)  $&$0.318(6)(3)$ &$1.155(22)$ &$0.530(12)(6)$\\
$5.3$    &$6.195(51)  $ &$0.320(5)(4)$ &$1.169(15)  $ &$0.577(11)(7)$ \\
$5.5$    &$8.280(80)$ &$0.316(4)(2)$ &$1.213(17)$ &$0.617(11)(5)$ \\
\midrule
cont. & & $0.315(8)(2)$&$1.252(33)$&$0.678(12)(5)$\\
\bottomrule
\end{tabular}
\end{center}
\caption{\label{t:L1inter}\footnotesize Values of $\Lmax/a$, $\Lmax
   \fk$, $r_0/\Lmax$ and
   $\mbar_\strange/\fk$ together with the values extrapolated to the
      continuum limit. The running mass in the Schr\"odinger Functional scheme 
      $\mbar_\strange$ is given at the renormalization scale $\Lmax$.}
\end{table*}

The main difficulty and source of a systematic error
is the extrapolation to the proper quark masses,
the ``physical point''. Once we decide to set the
scale through $\fk$, this point is naturally defined
by 
\bes
    R_\mathrm{K} =  R_\mathrm{K}^\mathrm{phys}\,, \label{e:rkpiphys}
    \quad
    R_\mathrm{\pi} =  R_\mathrm{\pi}^\mathrm{phys}\,, 
\ees
where 
\bes  
   R_\mathrm{K} = {\mk^2 \over \fk^2}\,, \quad
   R_\mathrm{\pi} = {\mpi^2 \over \fk^2} \,,
\ees
and $R_\mathrm{K}^\mathrm{phys},\; R_\mathrm{\pi}^\mathrm{phys}$
are the values of these ratios in Nature. In an attempt to
minimize uncertainties, we take the physical masses and decay
constants to be the ones in the isospin symmetric limit with QED 
effects removed as discussed in \cite{Flag1}. We use
\bes
  \mpiphys=134.8\,\MeV\,,
  \mkphys=494.2\,\MeV\,.
\ees

One can then, for each lattice spacing, carry out two different strategies
for extrapolating to the physical point.

In strategy 1 one keeps  
\bes
    R_\mathrm{K} \equiv {\mk^2 \over \fk^2}  = R_\mathrm{K}^\mathrm{phys}\,, \label{e:Rkcond}
\ees    
fixed, as one varies the 
light (dynamical) quark mass.
The condition defines a curve in the 
quark mass plane spanned by $(m_\mathrm{d}=)m_\mathrm{u}$ and
$m_\mathrm{s}$. 
It is a very interesting one because along that curve
$\mk$ is constant up to terms of order
$m_\mathrm{u}^2$. In the ChPT expansion the mass-dependence 
is then small and in particular the coefficient of the chiral log,
$\mpi^2 \log \mpi^2 $, is small.
One expects that $\fk$ can be extrapolated rather 
easily to the physical point along this curve. 

The order $m_\mathrm{u}^2$ corrections are known 
in terms of one low energy constant, 
$\alpha_4$.
One just has to implement our condition
\eq{e:Rkcond}, which expresses $\mk$ in terms of $\mpi$, in the formulae of \cite{PQChPT:Steve}.

The predicted form is
\bes
  \label{e:fkstrat1}
  \fk &=& \fkphys\,[1 + \overline{L}_\mathrm{K}(y_1,\yk) \\ &&
  + (\alpha_4-\frac14)\,(y_1-\ypi) +\rmO(y^2), \nonumber
  \\
  \overline{L}_\mathrm{K}(y_1,\yk) &=&  L_\mathrm{K}(y_1,\yk) -
  L_\mathrm{K}(\ypi,\yk) \,, \label{e:fkstrat1b}
  \\
  L_\mathrm{K}(y_1,\yk) &=& - \frac12 y_1 \log(y_1) \nonumber
  \\ && 
  - \frac18y_1\log(2\yk/y_1 - 1) \,.
  \label{e:fkstrat1c}
\ees
The variables 
\bes
 y_1 &=& {\mpi^2\over 8\pi^2 \fk^2}
 \,,\quad \nonumber 
 \\\nonumber
 \ypi &=& {\mpiphys^2 \over 8\pi^2 \fkphys^2}=0.00958\,, 
 \\
 \yk &=& {\mkphys^2 \over 8\pi^2 \fkphys^2}=0.12875 \,,
 \nonumber
\ees
are proportional to (averages of) quark masses up to 
quadratic terms.
Because of \eq{e:Rkcond}, we have 
$y_3\equiv{\mk^2 / [8\pi^2 \fk^2]}= 2\yk-y_1 +\rmO(y^2)$
and $y_3$ does not appear in \eq{e:fkstrat1}.

Another option is strategy 2, where one keeps 
the (PCAC) strange quark mass fixed and uses
the expansion in just the up quark mass, i.~e. SU(2) chiral perturbation theory adopted to this
situation. 

\begin{figure}[bt!]
\begin{center}
\includegraphics[width=0.45\textwidth]{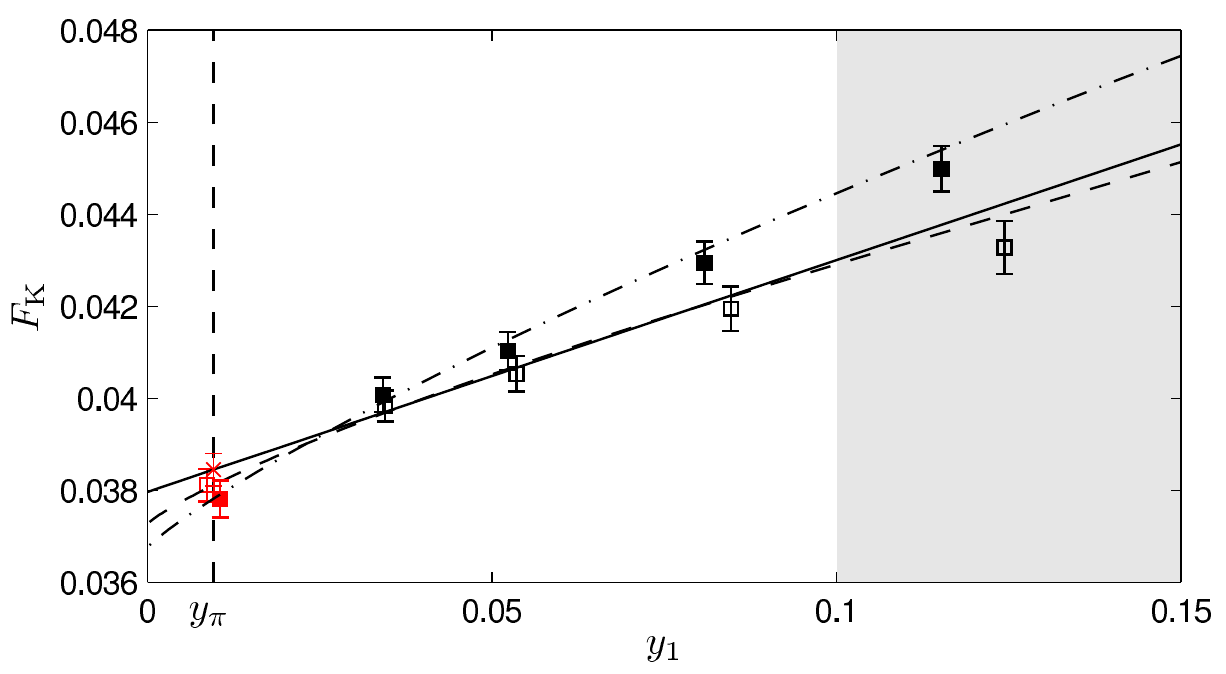}
\end{center}
  \caption{\footnotesize
    Physical point extrapolation of the 
    kaon decay constant in lattice     
    units. Open symbols and dashed lines
    correspond to strategy 1, whereas filled symbols and    
    dash-dotted lines represent strategy 2.
    Only data below $y_1=0.1$ enter the extrapolation. Figure from \cite{Fritzsch:2012wq}.
  }\label{f:fkextrap1}
\end{figure}

\Fig{f:fkextrap1} shows extrapolations with both strategies, which 
converge well at the physical point.

\begin{figure}[tb!]
\vspace{0pt}
\centerline{\includegraphics[width=0.45\textwidth]{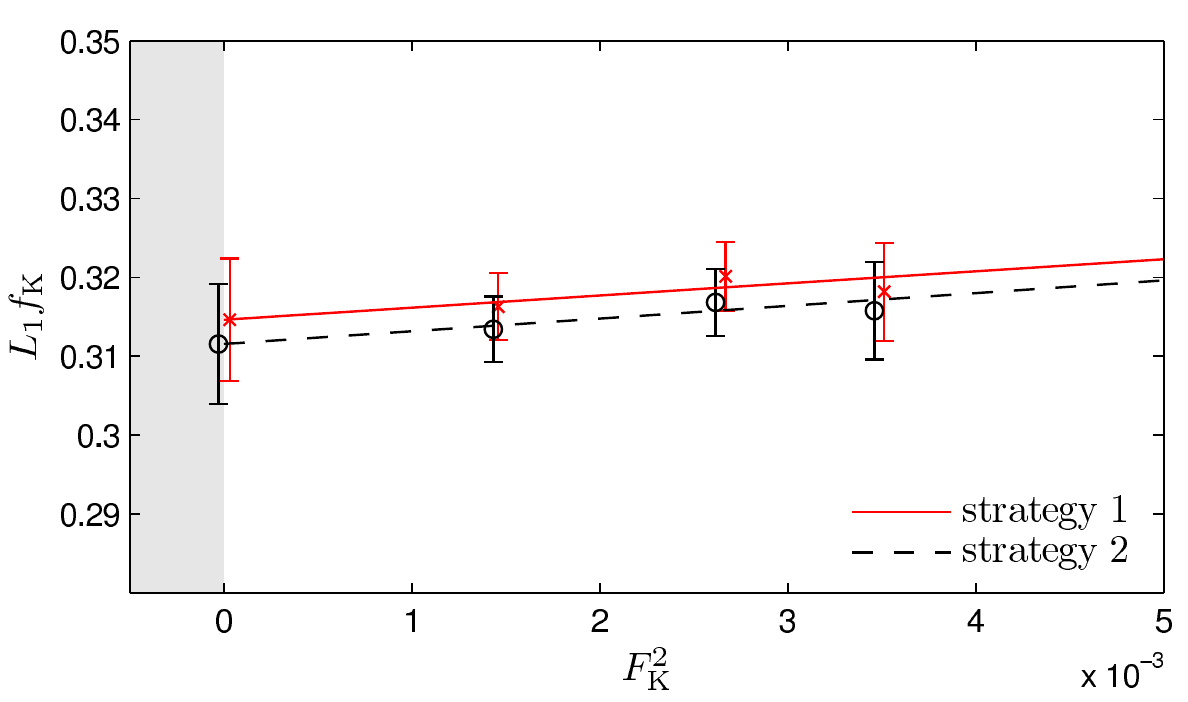}}
  \caption{
     Continuum extrapolation of $L_1=\Lmax$ in units of $\fk$. Even though the 
        data shows no cut-off effects, a linear extrapolation is used to 
        account for uncertainties from O$(a^2)$ effects hidden
        by the errors. The two strategies for the chiral extrapolation
        of $\fklat=a\fk$ agree well within statistics. 
        }\label{f:Lcont}
\end{figure}

It remains to perform a continuum extrapolation 
of the dimensionless combination $\fk \Lmax$ 
with the help of interpolations of the integer $\Lmax/a$ 
as a function of the bare coupling $g_0^2$.  Little
discretisation errors are seen in \fig{f:Lcont}
which leads to the continuum limit
\bes
\fk \Lmax = 0.315(8)(2) \,. \label{e:fklmax}
\ees

As in the previous section, the final results come from strategy 1 for the chiral extrapolation of $\fklat$. 
Strategy 2 is used to estimate
the systematic uncertainty in the second parenthesis; it is small compared to
the statistical errors. 
We then quote
\bes
\Lambda_\mathrm{SF}^{(2)}/\fk=0.84(6) .
\ees
Now, as a result of our analysis,
the error is dominated by the error on $\Lambda \Lmax$.
We translate to the  $\msbar$ scheme using
   $\Lambda_\msbar^{(2)}=2.382035(3)\Lambda_\mathrm{SF}^{(2)}$\cite{SF:LNWW,pert:1loop}
as well as to physical units 
\bes
\Lambda_\msbar^{(2)} = 310(20)\,\mathrm{MeV}\,,
\ees
where
\bes
\fkphys&=155\,\MeV\,
\ees
enters. 

As discussed previously with not all flavors of
QCD treated dynamically there is a small ambiguity in
the translation to MeV. We therefore also give the 
result
\bes
 r_0\,\Lambda_\msbar^{(2)}=0.789(52)\,, \label{e:r0lambda}
\ees 
based on
\bes
  r_0/\Lmax= 1.252(33)
\ees
in complete analogy to \eq{e:fklmax}.
Our result \eq{e:r0lambda} is
an unambiguous non-perturbative property of the 
two-flavor theory, sometimes called QCD-lite. 

\subsubsection{Strange quark mass}

The RGI mass
$M_\mathrm{s}$
is given in terms of the bare PCAC mass $m_\strange$ by
\bes
M_\mathrm{s}&=&\frac{M}{\mbar(L)} \mbar_\strange(L)
\\
&=&\frac{M}{\mbar(L)} \frac{\za}{\zp(L)} m_\strange \,,
\label{e:Ms}
\ees
where $m_\strange$ is the PCAC strange 
quark mass of strategy 2. 
The continuum value of the universal first factor 
$M/\mbar$ has been computed in Ref.~\cite{DellaMorte:2005kg} and discussed
in section \ref{runmass}.

Expressing $M_\mathrm{s}$ in units of $\fk$ we eliminate $\za$ and get
\bes
   {M_\strange\over \fkphys} &=&  {M \over \mbar(L)} \times 
    \frac{ m_\strange}{\fklatphys^\mathrm{bare}} \times \\
   && \quad  {1 \over \zp(L)} 
   [1+(\batil-\bptil) a m_\strange]\,   
   \label{e:mstrangergi} \ , \nonumber
\ees
with $\fklatphys^{\rm bare} = \fklatphys/\za$.
The second factor is $\rmO(a)$ improved, if we 
neglect a tiny correction proportional to the sea
quark mass, 
$(\babar-\bpbar)a m_\mathrm{sea}$. Note that 
$\babar,\bpbar=\rmO(g_0^4)$ are loop-suppressed {\em and}
$a m_\mathrm{sea}$ is very small.

Our final result is
\bes
\mbar_\strange/\fk&=&0.678(12)(5)\,, \\ M_\mathrm{s}/\fk &=& 0.887(19)(7)\,,
   \\ M_\mathrm{s}&=&138(3)(1)\,\mathrm{MeV} \,,
\ees
where we use $M/\mbar=1.308(16)$ at the scale $\Lmax$.
For reference, we also give the numbers  in the ${\rm
   \overline{MS\kern-0.05em}\kern0.05em}$  scheme. This 
conversion is the only part of the computation in which we need to take
recourse to perturbation theory, known in this case to four loops \cite{MS:4loop1,MS:4loop2,MS:4loop3,Czakon:2004bu}, which
differs from the two- and three-loop result by only a small amount. 
We use the same method as described in \cite{DellaMorte:2005kg}, but with the 
new value of $\Lambda_\msbar$ which leads us to
$\mbar^\msbar(2\,\mathrm{GeV})/M=0.740(12)$ and 
\bes
\mbar_\mathrm{s}^\msbar(2\,\mathrm{GeV})&=&\frac{M_\mathrm{s}}{\fk}\,
\frac{\mbar^\msbar(2\,\mathrm{GeV})}{M}\,  
\fkphys \\ &=& 102(3)(1)\,\mathrm{MeV} \ .
\label{e:msmsbar}
\ees

\subsection{Scale setting for $\nf=2+1$}
\label{s:2p1}

After the progress in simulation algorithms and the understanding
of how to get around the topological freezing, CLS has started 
large-scale QCD simulations with a strange quark in addition
to degenerate up and down quarks. The action $S_\mathrm{LW;SWsf}$
is used with non-perturbative $\csw$ \cite{Bulava:2013cta}.
The simulations started just about 1$\frac12$ years ago, but
have already reached a similar coverage of lattice spacings and 
pion masses as the $\nf=2$ simulations carried out before. 
A summary of the presently available ensembles is 
found in \tab{tab:ens} from \cite{Bruno:2014jqa}
where the details of the simulations are described.

\begin{table*}[tb]
\begin{center}
\small
\begin{tabular}{ccccllcccc}
\toprule
id &   $\beta$ &  $N_\mathrm{s}$  &  $N_\mathrm{t}$  &  $\kappa_u$ & $\kappa_s$ & $m_\pi$[MeV] &   $m_K$[MeV] &  $m_\pi L$\\
 \midrule
B105 & 3.40 & 32 & 64	& 0.136970 & 0.13634079		& 280 &460  & 3.9\\
H101 & 3.40 & 32 & 96	& 0.13675962 & 0.13675962	& 420 &420  & 5.8\\
H102 & 3.40 & 32 & 96	& 0.136865 & 0.136549339	& 350 &440  & 4.9\\
H105 & 3.40 & 32 & 96	& 0.136970 & 0.13634079		& 280 &460  & 3.9\\
C101 & 3.40 & 48 & 96	& 0.137030 & 0.136222041	& 220 &470  & 4.7 \\
D100 & 3.40 & 64 & 128	& 0.137090 & 0.136103607	& 130 &480  & 3.7 \\
\midrule                                                   
H200 & 3.55 & 32 & 96	& 0.137000 & 0.137000   & 420	&420  & 4.4 	\\
N200 & 3.55 & 48 & 128	& 0.137140 & 0.13672086	& 280&  460  & 4.4 	\\
D200 & 3.55 & 64 & 128	& 0.137200 & 0.136601748& 200 &	480  & 4.2		\\
\midrule		                                               
N300 & 3.70 & 48 & 128	& 0.137000 & 0.137000	 & 420	&420  & 5.1 	\\
N301 & 3.70 & 48 & 128	& 0.137005 & 0.137005	 & 410	&410  & 4.9 	\\
J303 & 3.70 & 64 & 192	& 0.137123 & 0.1367546608& 260	&470  & 4.1		\\
\bottomrule
\end{tabular}
\caption{\label{tab:ens}List of present CLS ensembles with up, down and strange sea quarks and the action of
\cite{Bulava:2013cta} . The numbers for
$m_\pi$ and $m_\mathrm{K}$ are rounded and use 
$\sqrt{8t_0}=0.4144\,\fm$. The
lattice spacings are roughly $a = 0.086\,\fm$, 
$0.064\,\fm$ and $0.05\,\fm$ for $\beta=3.4$, $3.55$ and $3.7$, respectively. Table from \cite{Bruno:2014jqa}.}
\end{center} 
\end{table*}

With a dynamical strange quark, the choice of the 
curve in the quark mass plane that one chooses to 
approach the physical point (cf. \sect{s:s2}) is much 
more important since each choice of the strange quark mass
means a new simulation. Various considerations enter into 
a choice of a trajectory. Strategy 2 of \sect{s:s2} 
seems a natural one, but one has to keep the renormalized 
strange mass fixed and in the non-degenerate case, the
renormalization of the quark mass contains a mixture of
the flavor singlet and the flavor non-singlet mass terms,
which renormalize differently. Similarly the $\rmO(a)$ 
terms become more complicated\cite{Bhattacharya:2005rb}.
As a result of this it is technically simpler to keep
the trace of the quark mass matrix constant as one changes 
the light quark mass\cite{Bietenholz:2010jr}. This condition is similar but not the 
same as strategy 1. The most important simplification is
that up to a (supposedly small) $\rmO(a)$ term, the
condition of a fixed trace of the renormalized mass matrix 
is equivalent to 
\bes
   \sum_{f=1}^3 \frac{1}{\kappa_{f}}  = \mbox{const} \,.
\label{eq:chitra}
\ees
It can thus be followed without any tuning errors. 
A non-trivial point is of course to choose the right value
of the trace,
the one which leads to a trajectory through the physical point. 
Slightly wrong choices and subsequent corrections
are unavoidable. We skip this issue here.

The scale setting will proceed in analogy to 
\sect{s:s2} through the decay constants 
$\fpi,\fk$. With open boundary conditions 
translation invariance in time is lost. Boundary-effects of correlation functions exist and have to 
be taken into account. The theoretical analysis of these 
effects is the same as with SF boundary conditions
in a large volume \cite{Guagnelli:1999zf}. Numerical 
aspects are presently being
studied in detail \cite{Bruno:2014lra,Bruno:2014jqa}. 
A short summary is that the boundary conditions
do not present an obstacle for the extraction of the 
hadronic matrix elements such as decay constants. In fact 
for some cases they may be advantageous compared to
the conventional torus, where particles may
propagate around the periodic time.
With open boundary conditions such effects are avoided. 

\begin{figure}[bt!]
\begin{center}
\includegraphics[width=0.45\textwidth]{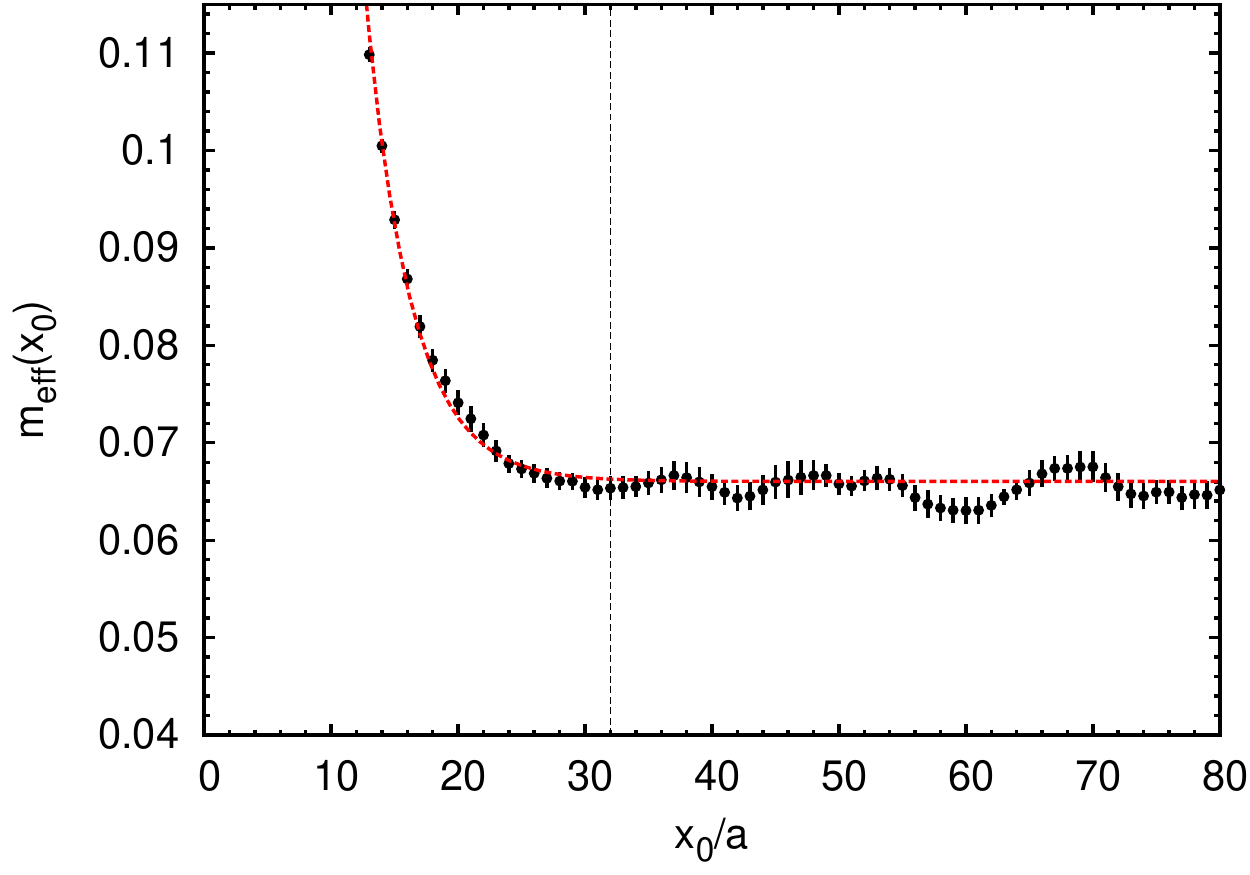}
\end{center}
  \caption{\footnotesize
    Effective pion mass of the D200 lattice (see \tab{tab:ens}).
    Graph from from \cite{Bruno:2014jqa}.
  }\label{f:D200meff}
\end{figure}

We show an example of an open boundary condition 
correlation function 
in \fig{f:D200meff}, from \cite{Bruno:2014jqa}. 
Apart from the  boundary effects the effective mass plot exhibts
significant wiggles at large time,
but these are of a purely statistical nature. Correlations between neighboring
time-slices are very strong, but over larger distances also anticorrelations are present and in the end the fitted two-state curve is statistically
compatible with the data points.

For the extraction of pseudo-scalar decay constants, 
just as for their mass,
one has various possibilities to combine correlation functions. 
There numerical study revealed that the details do not matter
too much and given a certain number of decorrelated configurations 
the statistical precision of these quantities is comparable to the 
one with periodic boundary condition ensembles \cite{Bruno:2014lra}.

In summary, the 2+1 simulations advance very fast and the 
scale setting is expected to be available rather soon.

\section{Outlook}
Let us first go back to 
\fig{f:nfdep}. Here the present knowledge of the running of the 
SF coupling is summarized in an easily accessible way.
For $\Nf$=3 the range of couplings is presently restricted.
However, within this range the precision is already much better
than for smaller flavor numbers. Still, the ALPHA collaboration
is reducing the errors further. 
In parallel, work is now progressing on extending 
the range towards small energies. As explained earlier,
in the lower energy region it becomes advantageous to use
the flow coupling.  Now that the reduction
of cutoff effects of flow observables
by Symanzik improvement is understood \cite{Ramos:2014kka}, we are ready to use it. The first preparation,
the determination of the critical lines for 
different $L/a$ in the necessary region of larger $g_0$
as compared to \fig{f:mcrit} is already far advanced. 
The step scaling functions of $g_\mathrm{GF}$ will be 
evaluated soon. 

As we have reported in the previous section, 
also the necessary large volume simulations are far advanced.
Hence, we foresee to soon present the three flavor 
$\Lambda^{(3)}_\msbar$ with a precision that is at least comparable to 
the present FLAG average \cite{Aoki:2013ldr}.
At this point, a perturbative relation of 
$\Lambda_\msbar^{(3)}$ to  $\Lambda^{(5)}_\msbar$ 
can be used and one obtains an estimate of 
$\alpha^{(5)}_\msbar(M_\mathrm{Z})$ with a precision
of the non-lattice PDG average \cite{pdg2012} or better.
The additional uncertainty introduced by this step can roughly be divided 
into two pieces (for a precise description of the 
decoupling of heavy quarks on the non-perturbative level
we refer to \cite{Bruno:2014ufa}). There are power-suppressed
$\rmO((M_\mathrm{charm}/\Lambda_\msbar^{(4)})^{-2})$ terms due to the 
neglect of higher dimensional operators in the effective Lagrangian 
when we treat the low energy theory by just three flavor QCD.  
Such effects are entirely due to charm quark loops and
are therefore suppressed by a factor of the number of colors in the
large $N$ expansion on top of a perturbative suppression, i.~e.
by  $\alpha/N$. 
In a (quite realistic, we would say) model it has recently 
been shown that indeed these power-suppressed terms are {\em very} 
small. We can safely neglect them within the envisaged accuracy.
What remains is the matching of QCD with 3 flavors to 
QCD with 4 flavors. In terms of the $\Lambda$-parameters this is
the relation
\bes 
 \Lambda_\msbar^{(3)} = P_{3,4}(M_\mathrm{charm}/\Lambda_\msbar^{(4)})\,\Lambda^{(4)}_\msbar\,,
 \ees 
which has a perturbative expansion in terms of 
the coupling  $\alpha(\mu)$ at the scale 
$\mu=M_\mathrm{charm}$. In the $\msbar$ scheme the relation is known
to four loops and the resulting perturbative uncertainty
looks very small \cite{Chetyrkin:2005ia}, see \cite{Bruno:2014ufa}
for the discussion of $P(M_\mathrm{charm}/\Lambda_\msbar)$. 

Nevertheless, perturbation theory at the scale $\mu=M_\mathrm{charm}$
is worrying per se. Therefore the ALPHA collaboration 
also foresees a further step to carry out an adapted
version of the full programme with four dynamical quark flavors. 
A first step is to bring Symanzik $\rmO(a)$ improvement under
control with a heavy charm quark. We plan to carry out
the steps at low and intermediate energy in a massive renormalization scheme \cite{lat14:felix}. Concerning improvement, this scheme does not need to be defined exactly with the charm mass at its physical value, 
but it is sufficient to have it fixed and close to it such 
that in an expansion in 
$a m_\mathrm{charm} - a m_\mathrm{charm}^\mathrm{phys}$
one can safely neglect higher order terms. 
This is not an easy undertaking, but first steps are 
promising \cite{lat14:felix}.
We can hence foresee to have in 
the near future a full four flavor non-perturbative determination of the $\Lambda$ parameter.
\\[2ex]

{\bf Acknowledgements}
\\[1ex]
We would like to acknowledge the fruitful and pleasant
collaboration with 
{  Mattia Bruno}, Michele Della Morte, {  Patrick
Fritzsch}, Jochen Heitger, Roland Hoffmann, Andreas J\"uttner, { 
Francesco Knechtli}, {  Tomasz Korzec}, Bj\"orn Leder,
Stefano Lottini, Marina Marinkovic, Harvey Meyer, {  Alberto Ramos},
Juri Rolf, {  Stefan Schaefer}, {  Stefan Sint}, {  Hubert Simma}, { 
Felix Stollenwerk}, Shinji Takeda, Fatih Tekin, Francesco Virotta, Ines
Wetzorke and Oliver Witzel. The project is a completion of
the ground-breaking work done in collaboration with Martin L\"uscher and
Peter Weisz whom we would like to thank very much for sharing 
their insights and support over the years. We thank 
Karl Jansen and Stefan Schaefer
for reading and helping to improve an earlier version
of this article.
\\
We are grateful for the support of the 
Deutsche Forschungsgemeinschaft (DFG)
in the SFB/TR~09 ``Computational Particle Physics'' and 
we have profited from the scientific exchange 
in the SFB. 
\\
Lastly, the results described here are also
due to a lot of support for computational ressources.
We gratefully acknowledge the Gauss Centre for Supercomputing (GCS)
for providing computing time through the John von Neumann Institute for
Computing (NIC) on the GCS share of the supercomputer JUQUEEN at J\"ulich
Supercomputing Centre (JSC). GCS is the alliance of the three national
supercomputing centres HLRS (Universit\"at Stuttgart), JSC (Forschungszentrum
J\"ulich), and LRZ (Bayerische Akademie der Wissenschaften), funded by the
German Federal Ministry of Education and Research (BMBF) and the German State
Ministries for Research of Baden-W\"urttemberg (MWK), Bayern (StMWFK) and
Nordrhein-Westfalen (MIWF).
We acknowledge PRACE for awarding us access to resource JUQUEEN in Germany at J\"ulich. We thank the HLRN for time on the supercomputers
Konrad and Gottfried and DESY for its support of the PAX cluster in Zeuthen.






\end{document}